\documentclass[twocolumn, trackchanges]{aastex701}

\usepackage{graphicx}
\usepackage{subcaption}
\usepackage{booktabs}
\usepackage{multirow}
\usepackage{amsmath}

\begin{document}

\title{SDSS+JWST Census of Stellar and Nebular Dust Attenuation at $z\sim0$--$7$:\\
Mass Dependence and Redshift Evolution}

\author[0000-0002-0846-7591]{Jie Song}
\affiliation{Department of Astronomy, University of Science and Technology of China, Hefei 230026, China}
\affiliation{School of Astronomy and Space Science, University of Science and Technology of China, Hefei 230026, China}
\affiliation{Institute for Cosmic Ray Research, The University of Tokyo, 5-1-5 Kashiwanoha, Kashiwa, Chiba 277-8582, Japan}
\email{jiesong@mail.ustc.edu.cn}

\author[0000-0002-1049-6658]{Masami Ouchi}
\affiliation{Institute for Cosmic Ray Research, The University of Tokyo, 5-1-5 Kashiwanoha, Kashiwa, Chiba 277-8582, Japan}
\affiliation{Kavli Institute for the Physics and Mathematics of the Universe (WPI), University of Tokyo, Kashiwa, Chiba 277-8583, Japan}
\affiliation{Department of Astronomical Science, The Graduate University for Advanced Studies, SOKENDAI, 2-21-1 Osawa, Mitaka, Tokyo, 181-8588, Japan}
\affiliation{National Astronomical Observatory of Japan, 2-21-1 Osawa, Mitaka, Tokyo, 181-8588, Japan}
\email{ouchims@icrr.u-tokyo.ac.jp}

\author[0009-0004-4332-9225]{Tomokazu Kiyota}
\affiliation{Department of Astronomical Science, The Graduate University for Advanced Studies, SOKENDAI, 2-21-1 Osawa, Mitaka, Tokyo, 181-8588, Japan}
\affiliation{National Astronomical Observatory of Japan, 2-21-1 Osawa, Mitaka, Tokyo, 181-8588, Japan}
\email{tomokazu.kiyota@grad.nao.ac.jp}

\author[0000-0002-9888-6895]{Chenghao Zhu}
\affiliation{Institute for Cosmic Ray Research, The University of Tokyo, 5-1-5 Kashiwanoha, Kashiwa, Chiba 277-8582, Japan}
\affiliation{Department of Physics, Graduate School of Science, The University of Tokyo, 7-3-1 Hongo, Bunkyo, Tokyo 113-0033, Japan}
\email{zchao@icrr.u-tokyo.ac.jp}

\author[0000-0002-7660-2273]{Xu Kong}
\affiliation{Department of Astronomy, University of Science and Technology of China, Hefei 230026, China}
\affiliation{School of Astronomy and Space Science, University of Science and Technology of China, Hefei 230026, China}
\affiliation{Institute of Deep Space Sciences, Deep Space Exploration Laboratory, Hefei 230026, China}
\email{xkong@ustc.edu.cn}

\author[0000-0002-0984-7713]{Yurina Nakazato}
\affiliation{Center for Computational Astrophysics, Flatiron Institute, 162 5th Avenue, New York, NY 10010}
\email{ynakazato@flatironinstitute.org}

\begin{abstract}
We present the demography of dust attenuation, including its mass dependence and redshift evolution, using spectroscopic samples of 34,182 SDSS galaxies at $z\sim0.1$ and 863 JWST/JADES galaxies at $z\sim1.5$--$7$. We find that, on average, ${\rm H\alpha}/{\rm H\beta}$ ratios are comparable to the Case B recombination value at $M_\ast \lesssim 10^9 M_\odot$, and increase beyond $M_\ast \sim 10^9 M_\odot$ both at $z\sim0.1$ and $1.5$--$7$. We derive the nebular attenuation $A_{\rm V, nebular}$ from Balmer decrements and the stellar attenuation $A_{\rm V, stellar}$ from rest-frame UV--optical spectra with supplementary \textit{GALEX} data, via comparisons with stellar-population models and multiple attenuation curves in a consistent manner across cosmic time. We find no significant redshift evolution of $A_{\rm V, nebular}$ and $A_{\rm V, stellar}$ at fixed $M_\ast$ over $z\sim0$--$7$, forming a universal extinction relation, and both rise from $0.2$--$0.4$ at $M_\ast \lesssim 10^9 M_\odot$ to $\sim1$ at $M_\ast \sim 10^{11} M_\odot$. Interestingly, at $M_\ast \gtrsim 10^9 M_\odot$, $A_{\rm V, nebular}$ rises more steeply than $A_{\rm V, stellar}$. This correlation holds within an uncertainty of $\sim\pm0.2$ for various combinations of attenuation curves (Calzetti, SMC, and Milky Way). These results indicate that $M_\ast \sim 10^9 M_\odot$ is a transition mass in dust attenuation, whose low-mass behavior reflects dust widely distributed by feedbacks. These mass-dependent extinction results address the long-standing issue of appropriate choice of the stellar-to-nebular color excess ratio, $f\equiv E(B-V)_{\rm stellar}/E(B-V)_{\rm nebular}=1.0$ or $0.44$, and suggest that galaxy $M_\ast$ determines $f$ from $\sim1.0$ to $\sim0.44$ across low- to high-mass galaxies.

\end{abstract}


\section{Introduction} \label{sec:1}
Dust attenuation is one of the most fundamental physical processes in extragalactic astronomy, fundamentally affecting the derivation of galaxy physical properties, such as stellar masses ($M_{\ast}$) and star formation rates (SFRs), across all cosmic epochs (e.g., \citealt{1998ARA&A..36..189K, 2000ApJ...533..682C, 2000ApJ...539..718C, 2012ARA&A..50..531K, 2017MNRAS.466..105A, 2018ARA&A..56..673G}). Interstellar dust grains absorb and scatter ultraviolet (UV) and optical photons emitted by stars and ionized gas and subsequently re-radiate the absorbed energy at far-infrared wavelengths (e.g., \citealt{2006A&A...451..417D, 2011piim.book.....D, 2013ApJ...779...32V, 2018ARA&A..56..673G}). This process biases virtually every observable quantity derived from short-wavelength light. Accurately characterizing dust attenuation is therefore essential for understanding the overall formation and evolution of galaxies.

Numerous previous studies have demonstrated that galactic dust attenuation is affected by the total dust content, grain size distribution, and chemical composition (e.g., \citealt{1977ApJ...217..425M, 2001ApJ...548..296W, 2011piim.book.....D, 2013MNRAS.432..637A, 2020MNRAS.491.3844A, 2020ARA&A..58..529S, 2026A&A...705A..75M}). 
Furthermore, the geometry between dust and stars is also believed to be an important factor governing dust attenuation, as dust is not necessarily uniformly distributed within galaxies (e.g., \citealt{1997AJ....113..162C, 2000ApJ...539..718C, 2017MNRAS.471.3152P, 2018ApJ...869...70N, 2019PASJ...71....8K, 2024MNRAS.527.7337V, 2026arXiv260207347N}). This non-uniform distribution has been examined in a series of observational studies (e.g., \citealt{1994ApJ...429..582C, 1997AJ....113..162C, 2000ApJ...533..682C, 2013ApJ...777L...8K, 2014ApJ...788...86P, 2015ApJ...806..259R, 2026arXiv260311338K}), which have revealed that the attenuation experienced by the stellar continuum, $A_{\rm V, stellar}$, differs systematically from that experienced by the ionized gas, $A_{\rm V,nebular}$, motivating the now-standard practice of treating $A_{\rm V, stellar}$ and $A_{\rm V, nebular}$ as distinct quantities.

The stellar continuum attenuation can be derived from spectral energy distribution (SED) fitting, the rest-frame UV slope, or the infrared excess (e.g., \citealt{2004MNRAS.349..769K, 2018MNRAS.476.3991M, 2018ARA&A..56..673G, 2025arXiv251000235W}). In previous works, $M_{\ast}$ of a galaxy has been identified as a robust proxy of the dust attenuation affecting its UV emission over a broad redshift range (e.g., \citealt{2009ApJ...698L.116P, 2013ApJ...777L...8K, 2014MNRAS.439.1337O, 2015ApJ...807..141P}), reflecting the expectation that the dust content of a galaxy grows together with its assembled $M_{\ast}$. Although some studies have found that high-redshift galaxies host less dust than their local counterparts of similar mass (e.g., \citealt{2022ApJ...928...68S, 2025A&A...693A.190J, 2026arXiv260401089B}), many other works have shown that the $A_{V,\rm stellar}$--$M_\ast$ relation does not evolve significantly with redshift at fixed $M_{\ast}$, at least out to $z \sim 3$ (e.g., \citealt{2015ApJ...807..141P, 2017ApJ...850..208W, 2018MNRAS.476.3991M}). For instance, using a sample at $0 < z < 2.5$, \cite{2017ApJ...850..208W} found that the obscured fraction of star formation is approximately $50\%$ at $\log(M_*/M_\odot) = 9.4$ and exceeds $90\%$ above $\log(M_*/M_\odot) = 10.5$ at all redshifts, providing evidence that the $A_{V,\rm stellar}$--$M_*$ relation is robust across cosmic noon. However, less consensus has been reached at higher redshifts (e.g., \citealt{2020MNRAS.491.4724F, 2023MNRAS.518.6142A, 2026arXiv260204765W}).

For $A_{V,\rm nebular}$, the most common approach is to infer it from the Balmer decrement (hereafter BD) $\rm H\alpha/H\beta$ under the assumption of Case~B recombination. As with $A_{V,\rm stellar}$, a number of studies have established that $M_{\ast}$ is the dominant galaxy property setting $A_{V,\rm nebular}$ (e.g., \citealt{2004MNRAS.351.1151B, 2010MNRAS.409..421G, 2022ApJ...926..145S, 2023ApJ...954..157S, 2024A&A...691A.305S, 2025arXiv251000235W}). For example, \cite{2010MNRAS.409..421G} found that, after controlling for $M_{\ast}$, no residual correlation with SFR or metallicity remains using a sample from Sloan Digital Sky Survey (SDSS, \citealt{2009ApJS..182..543A}). Although numerous studies have demonstrated that the dust content of galaxies correlates with various physical properties (e.g., \citealt{2018ApJ...853..179T, 2019MNRAS.490.1425L, 2021ApJ...914...19S}), the $A_{V,\rm nebular}$--$M_\ast$ relation has been found to remain nearly invariant from $z \sim 7$ to $z \sim 0$ at fixed $M_{\ast}$.

The difference between $A_{V,\rm nebular}$ and $A_{V,\rm stellar}$, which encodes the spatial distribution of stars relative to dust within galaxies, is considerably more complex. The canonical empirical relationship between $A_{V,\text{stellar}}$ and $A_{V,\text{nebular}}$ originates from the work of \citet{1997AJ....113..162C}, who found $f=0.44$, where $f$ is defined as:
\begin{equation} 
    f  \equiv E(B-V)_{\text{stellar}}/E(B-V)_{\text{nebular}}
    \label{eq:1}
\end{equation}
This is still widely adopted in some recent studies (e.g., \citealt{2023ApJ...952..133M}). A physical explanation for this differential attenuation is provided by the two-component dust model \citep{2000ApJ...539..718C, 2011MNRAS.417.1760W, 2013MNRAS.432.2061C}, which remains the standard theoretical framework for interpreting $f$ to this day. In the model of \cite{2000ApJ...539..718C}, the interstellar medium (ISM) is composed of a diffuse ISM component and dense molecular birth clouds. Stars younger than $\sim 10^{7}\,\rm yr$ remain embedded within optically thick birth clouds and experience attenuation from both components, while older stars have migrated out of their natal clouds and are attenuated only by the diffuse ISM. Because nebular emission lines arise exclusively from \ion{H}{2} regions surrounding massive O and B stars still inside birth clouds, the nebular emission is always more attenuated than the integrated stellar continuum, naturally producing $f < 1$ without requiring any special dust geometry.

However, analyses of large statistical samples in the local universe have demonstrated that $f$ is not a universal constant, indicating that the differential attenuation between the nebular and stellar components varies systematically with galaxy physical properties (e.g., \citealt{2011MNRAS.417.1760W, 2017ApJ...847...18Z, 2019PASJ...71....8K, 2020ApJ...888...88L, 2021ApJ...917...72L}). In particular, early studies using two-component models demonstrated that the relative contributions of young and old stellar populations, which are tightly regulated by the specific SFR, strongly dictate the integrated differential dust attenuation \citep{2004MNRAS.349..769K}. For example, by combining Wide-field Infrared Survey Explorer (WISE, \citealt{2010AJ....140.1868W}), SDSS, and Galaxy Evolution Explorer (GALEX, \citealt{2005ApJ...619L...1M}) photometry for more than $50{,}000$ local galaxies at $0.02 < z < 0.10$, \cite{2019PASJ...71....8K} revealed that $f$ decreases with $M_\ast$ and increases with sSFR.

At higher redshift, the conclusions regarding differential dust attenuation become increasingly complex, with a diverse range of $f$ values reported in the literature (e.g., \citealt{2013ApJ...777L...8K, 2013ApJ...779..135W, 2014ApJ...788...86P, 2015ApJ...807..141P, 2025arXiv251000235W, 2025ApJ...988L..20L, 2026ApJ...997..319T, 2026arXiv260311338K}). Utilizing a sample of 473 galaxies from the 3D-HST survey at $0.7 < z < 1.5$, \cite{2013ApJ...779..135W} reported a result consistent with the local \cite{1997AJ....113..162C} relation, while \cite{2016A&A...586A..83P} reported a value of $f \approx 0.93$ at $z \sim 1$ also based on a 3D-HST sample. However, in another study using a sample from the COSMOS field, \cite{2013ApJ...777L...8K} found $f \approx 0.7\text{--}0.83$ at $z \sim 1.6$. Moreover, several investigations have revealed that this differential attenuation continues to exhibit dependencies on galaxy $M_\ast$ within this redshift regime (e.g., \citealt{2014ApJ...788...86P, 2015ApJ...806..259R, 2026arXiv260311338K}).

This variable $f$ implies that, although both $A_{V,\rm nebular}$ and $A_{V,\rm stellar}$ correlate with $M_\ast$, their dependences on $M_\ast$ are not identical. A more careful and systematic characterization of how $A_{V,\rm nebular}$, $A_{V,\rm stellar}$, and their difference depend on $M_\ast$, and how each evolves with redshift, is therefore necessary. Although significant progress has been made in previous studies, a number of important questions remain open. First, measuring $A_{V,\rm nebular}$ requires the BD. At high redshift, the $\rm H\alpha$ emission line is redshifted into the observed-frame near-infrared, a wavelength regime that was largely inaccessible for statistical galaxy samples prior to the advent of the \textit{James Webb Space Telescope} (JWST). Detailed studies of these questions at $z > 3$ are therefore still very limited (e.g., \citealt{2025arXiv251000235W, 2026arXiv260311338K, 2026ApJ...997..319T, 2026ApJ...999...15R}). Second, prior work has focused on different redshift ranges and adopted a wide variety of methods to derive $A_{V,\rm stellar}$, often in conjunction with different assumptions about the stellar population templates (e.g., \citealt{2013ApJ...777L...8K, 2014ApJ...788...86P, 2016A&A...586A..83P}). This methodological heterogeneity makes direct, quantitative comparisons across studies difficult. A systematic study characterizing $A_{V,\rm nebular}$, $A_{V,\rm stellar}$, and their difference across a wide range of redshifts using a self-consistent methodology has hitherto been lacking.

In this paper, we present a systematic census of nebular and stellar dust attenuation from $z \sim 0$ to $z \sim 7$, utilizing a combined sample of 34,182 SDSS+GALEX star-forming galaxies at $z \sim 0.1$ and 863 JWST galaxies at $z \sim 1.5\text{--}7$, analyzed with a fully self-consistent methodology. We derive $A_{V,\text{stellar}}$ from SED fitting with the \texttt{Bagpipes} code \citep{2018MNRAS.480.4379C, 2019MNRAS.490..417C}, adopting identical priors for both samples to ensure uniformity. Simultaneously, $A_{V,\text{nebular}}$ is estimated from the observed BD, facilitating a self-consistent comparison across cosmic time. We find that both $A_{V,\rm nebular}$ and $A_{V,\rm stellar}$ correlate positively with $M_{\ast}$ and show no significant redshift evolution. Moreover, $A_{V,\rm nebular}$ increases more steeply with $M_\ast$ than $A_{V,\rm stellar}$ for galaxies with $\log(M_*/M_\odot) \gtrsim 9$. Consequently, $f$ approaches 1.0 at $\log(M_*/M_\odot) \lesssim 9$ and decreases monotonically toward the canonical value of 0.44 above this threshold, with no statistically significant redshift evolution from $z \sim 7$ to $z \sim 0$ at fixed $M_{\ast}$.

The structure of this paper is as follows. In Section~\ref{sec:2}, we detail the dataset construction and sample selection criteria. Section~\ref{sec:3} outlines our procedures for deriving the nebular and stellar dust attenuation, respectively. We present our main results in Section~\ref{sec:4}. Section~\ref{sec:5} explores the sensitivity of our findings to the assumed dust attenuation curve and provides a possible physical interpretation of our results. Finally, a summary of our conclusions is given in Section~\ref{sec:6}. Throughout this work, we assume a \cite{2001MNRAS.322..231K} initial mass function and adopt a flat $\Lambda$CDM cosmology characterized by $H_0 = 70\,\rm km\,s^{-1}\,Mpc^{-1}$, $\Omega_m = 0.3$, and $\Omega_\Lambda = 0.7$.

\section{Dataset and sample selection} \label{sec:2}
\subsection{SDSS-GALEX} \label{sec:2.1}
We constructed our low-redshift sample by leveraging the synergy between two large surveys: SDSS DR8 \citep{2000AJ....120.1579Y, 2011ApJS..193...29A}, which provides optical imaging and spectroscopy, and GALEX GR6 \citep{2007ApJS..173..342M}, which offers the ultraviolet coverage needed to constrain young stellar populations.

Our parent optical sample is drawn from the MPA-JHU DR8 Value-Added Catalog\footnote{\url{https://www.sdss4.org/dr17/spectro/galaxy_mpajhu/}} \citep{2003MNRAS.341...33K, 2004MNRAS.351.1151B, 2004ApJ...613..898T}. We restricted the redshift range to $0.05 < z < 0.10$ (hereafter, we adopt $z\sim 0.1$ to represent this redshift range) to minimize fiber aperture effects. To guarantee the reliability of the derived physical properties, we excluded objects with uncertain redshifts or unreliable emission-line measurements by requiring \texttt{z\_warning=0} and \texttt{reliable=1}. To ensure robust estimates of nebular dust extinction via BD, we further imposed a signal-to-noise ratio (S/N) cut of $\text{S/N} > 5$ for both the $\rm H\alpha$ and $\rm H\beta$ emission lines. 

To isolate star-forming galaxies and remove contamination from active galactic nuclei (AGN), we used the BPT diagram \citep{1981PASP...93....5B} and selected star-forming galaxies following the separation line defined by \citet{2003MNRAS.346.1055K}. These criteria yielded an initial optical sample of 76,758 galaxies. For these objects, we adopted the emission-line flux measurements from the MPA-JHU catalog, while the corresponding optical spectra and broadband photometry were retrieved from the SDSS Science Archive Server\footnote{\url{http://dr8.sdss.org/home}} and the SDSS CasJobs site\footnote{\url{https://skyserver.sdss.org/CasJobs/default.aspx}}, respectively.

The wavelength coverage of the SDSS spectra spans 3800--9200~\AA. Some previous studies have suggested that optical data alone may not fully break the degeneracy between age, dust, and metallicity \citep[e.g.,][]{2012MNRAS.422.3285P, 2016ApJS..227....2S}. Other works, however, have shown that stellar population properties derived from optical-only spectral fitting are broadly consistent with those obtained from joint optical--UV analyses \citep[e.g.,][]{2016MNRAS.458..184L}. To be conservative, we incorporated UV data to obtain more robust results in this work. To this end, we retrieved UV photometry in the FUV and NUV bands via the GALEX CasJobs server\footnote{\url{https://galex.stsci.edu/casjobs/default.aspx}}. Following the matching methodology recommended by \citet{2009ApJ...694.1281B}, we cross-matched each SDSS source with the nearest GALEX detection within a search radius of 3\arcsec. This process yielded a final low-redshift sample of 34,182 star-forming galaxies with UV detections.

\subsection{JADES} \label{sec:2.2}
To investigate dust attenuation properties in the high-redshift universe, we drew our sample from the JWST Advanced Deep Extragalactic Survey (JADES, \citealt{2023arXiv230602465E}). JADES is a comprehensive survey targeting the GOODS-S and GOODS-N fields, providing deep near-infrared imaging and spectroscopy with high sensitivity. For a detailed description of the survey design, data reduction, and data products, we refer the reader to the JADES data release papers \citep{2024A&A...690A.288B, 2024ApJ...964...71H, 2023ApJS..269...16R, 2025ApJS..281...50E, 2025ApJS..277....4D, 2025arXiv251001033C, 2025arXiv251001034S, 2026arXiv260115956R}.

The spectroscopic observations used in this work were obtained with the NIRSpec Micro-Shutter Assembly in Multi-Object Spectroscopy mode \citep{2022A&A...661A..80J, 2022A&A...661A..81F, 2023PASP..135c8001B}. The survey strategy employs both the low-resolution prism (PRISM/CLEAR; $R \sim 100$) and the medium-resolution gratings (G140M/F070LP, G235M/F170LP, and G395M/F290LP; $R \sim 1000$). We retrieved the fully reduced spectroscopic data from the Mikulski Archive for Space Telescopes (MAST)\footnote{\url{https://archive.stsci.edu/hlsp/jades}}.

Our high-redshift sample is constructed from the spectroscopic catalogs of JADES DR4 \citep{2025arXiv251001034S}. To deblend the $\rm H\alpha$ emission line from the [\ion{N}{2}] doublet, we used emission-line fluxes derived from the medium-resolution grating observations. We restricted our analysis to galaxies in the redshift range $1.5 < z < 7.0$. Consistent with our low-redshift selection criteria, we required $\text{S/N} > 5$ for both the $\rm H\alpha$ and $\rm H\beta$ emission lines. To account for slit-loss effects, we also retrieved the corresponding photometric catalogs from MAST and required that all selected galaxies have valid NIRCam observations.

To remove potential contamination from AGN, we performed a careful exclusion procedure based on the following criteria. First, for galaxies with detections of both [\ion{O}{3}] and [\ion{N}{2}], we rejected sources falling in the AGN region of the BPT diagram \citep{1981PASP...93....5B} following the criterion of \citet{2003MNRAS.346.1055K}. Second, we cross-checked our sample against the literature and excluded known broad-line and narrow-line AGNs reported in these fields \citep{2025arXiv250403551J, 2025A&A...697A.175S}. Finally, we cross-matched our sources with deep X-ray catalogs in the GOODS-S and GOODS-N fields \citep{2016ApJS..224...15X, 2022ApJ...941..191L} and rejected any objects with X-ray counterparts. After applying these criteria, our final high-redshift sample consists of 863 star-forming galaxies.

\section{Analysis} \label{sec:3}

\begin{table}
\centering
\caption{The SED fitting parameters and their priors used in this work}
\label{tab:1}
\begin{tabular}{ccc}
\hline\hline
Parameter & Limits & Prior \\
\hline 
$z$ & ($z_{\rm spec} - 0.01$, $z_{\rm spec} + 0.01$) & Uniform \\
$\log(M_{\rm \ast, form}/M_{\odot})$ & (4, 13) & Uniform\\
$Z/Z_{\odot}$ & (1e$-$4, 10) & Logarithmic \\
log U & ($-$4, 0) & Uniform \\
$A_{\rm V}$ & (0, 4) & Uniform \\
\hline 
\end{tabular}
\end{table}

\begin{figure}[htbp] 
    \centering
    \includegraphics[width=1\linewidth]{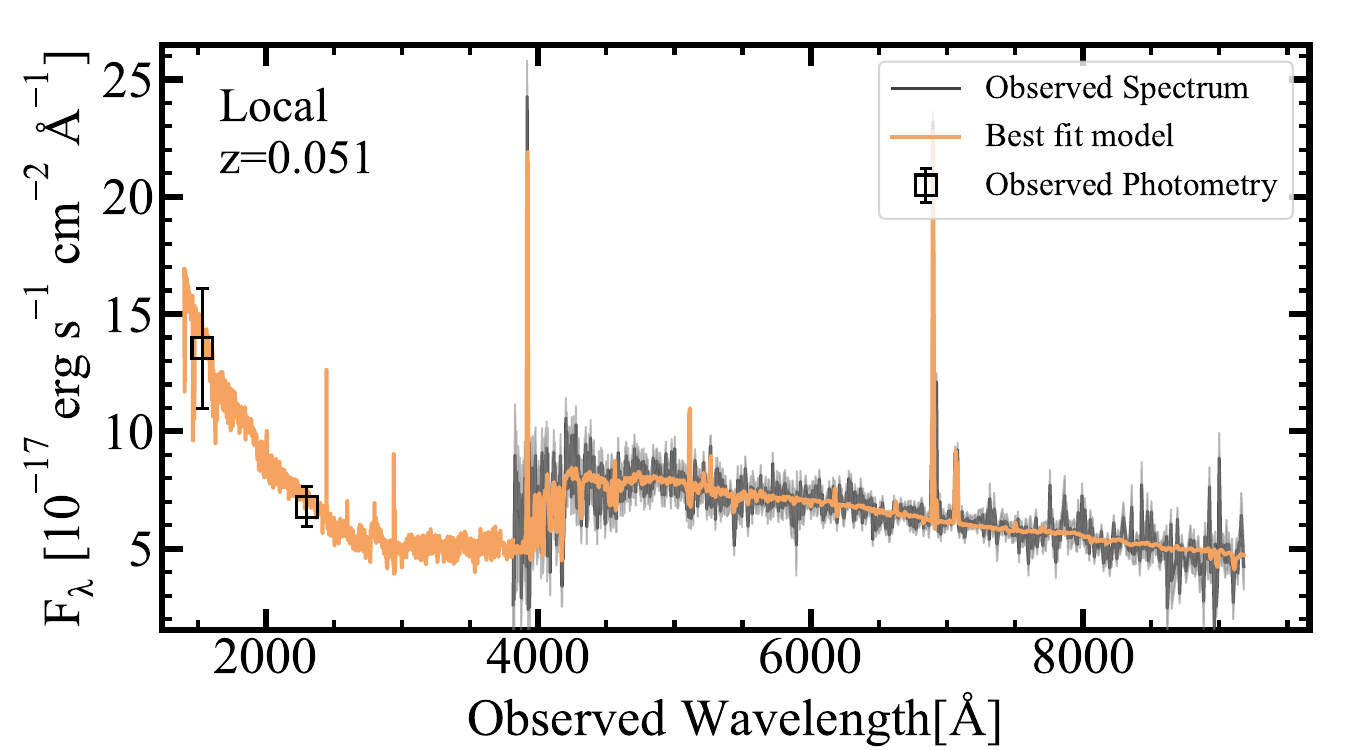} \\ 
    \includegraphics[width=1\linewidth]{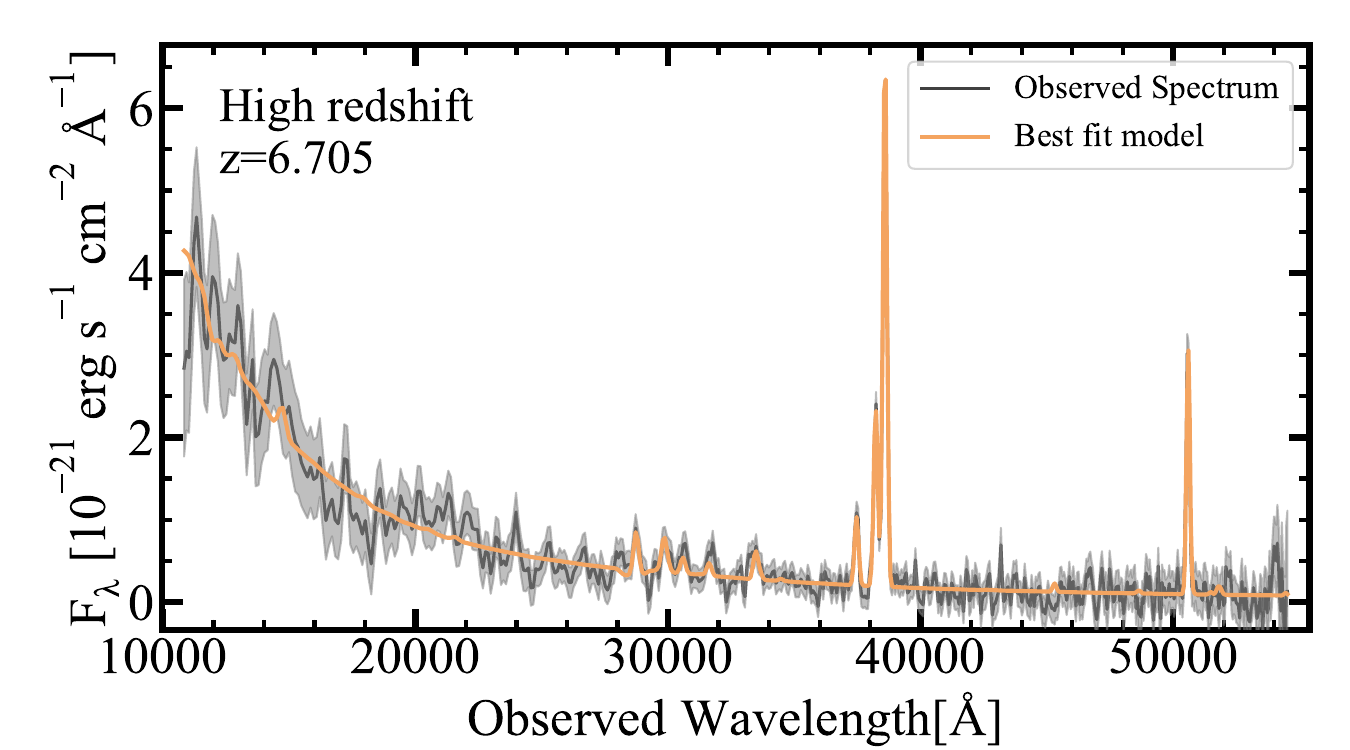}
    \caption{Top panel: SED fitting example from the SDSS sample; bottom panel: SED fitting example from the JADES sample. The black solid line and gray shaded region show the observed spectrum and its uncertainty, respectively, and the yellow solid line shows the best-fit model. The black squares with error bars denote the observed GALEX UV photometry and associated uncertainties.}
    \label{Fig:1}
\end{figure}

\subsection{$A_{\rm V,stellar}$ Measurements} \label{sec:3.1}
In this work, we used {\tt Bagpipes} to measure the $M_{\ast}$ and $A_{\text{V,stellar}}$ of the SDSS and JADES galaxies via SED fitting. Before proceeding with the SED fitting analysis, and following the method adopted by the MPA-JHU group, the observed data were corrected for Galactic extinction using the \citet{2011ApJ...737..103S} dust maps and the \citet{1994ApJ...422..158O} attenuation curve.

For the SDSS sample, following the methodology of \citet{2019MNRAS.483.2382W}, we simultaneously fit the GALEX UV photometry and the SDSS optical spectra. During the fitting process, we assumed a non-parametric star formation history (SFH) with a Student-$t$ distribution acting as a continuity prior \citep[e.g.,][]{2019ApJ...876....3L, 2022ApJ...927..170T}. The SFH is divided into six time bins: the two most recent bins are fixed at 0--3~Myr and 3--10~Myr, while the remaining four are logarithmically spaced in time back to the beginning of the Universe. To account for dust, we tested various attenuation and extinction laws, including the \citet{2000ApJ...533..682C} curve (hereafter CAL), the Small Magellanic Cloud attenuation curve (hereafter SMC) of \cite{2003ApJ...594..279G}, and the Milky Way extinction curve (hereafter MW) of \cite{1989ApJ...345..245C}. All other SED fitting parameters and adopted priors are summarized in Table~\ref{tab:1}. To recover the total physical properties of each galaxy, we applied an aperture correction to the fitting results based on the difference between the total and fiber-based $r$-band magnitudes \citep[e.g.,][]{2005PASP..117..227K, 2019PASJ...71....8K}.

For the JADES sample, we fit the full JWST/NIRSpec PRISM spectra, masking rest-frame wavelengths below 1400~\AA\ to exclude the $\rm Ly\alpha$ damping region \citep{2025A&A...693A..60H} and short-wavelength instrumental effects. Following the method described in \citet{2025NatAs...9..458M}, we also applied a slit-loss correction to the JADES spectra by scaling them according to the difference between the photometric and spectroscopic flux densities for each source. The priors adopted in the SED fitting are the same as those used for the SDSS sample. Figure~\ref{Fig:1} shows two representative examples of our SED fitting results for the SDSS (top panel) and JADES (bottom panel) samples. As shown in Figure \ref{Fig:1}, the spectral features are well reproduced by our fits.

\begin{figure}[htbp] 
    \centering
    \includegraphics[width=\linewidth]{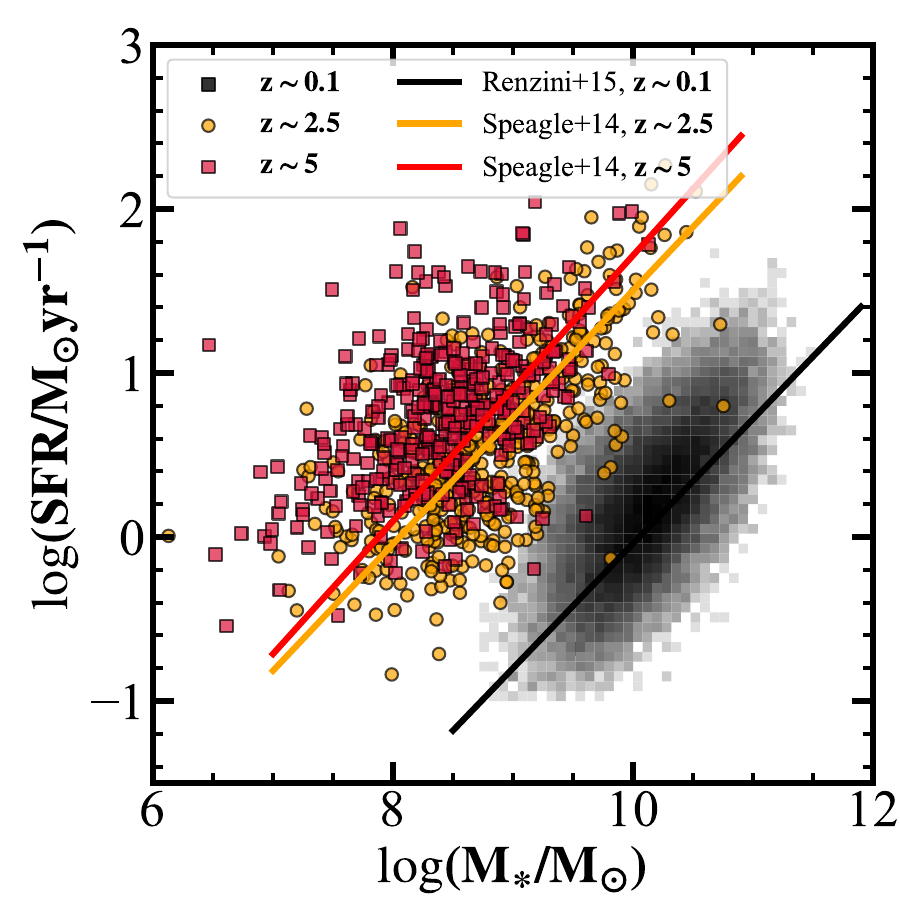}
    \caption{Distribution of our sample in the SFR--$M_\ast$ plane. Black two-dimensional histogram show the result for our local sample, while orange and red symbols show the result for galaxies at $1.5 < z < 3.5$ and $3.5 < z < 7$, respectively. For comparison, SFMS at $z \sim 0.1$ from \citet{2015ApJ...801L..29R} and at $z \sim 2.5$ and $z \sim 5$ from \citet{2014ApJS..214...15S}, are shown as black, orange, and red solid lines, respectively.}
    \label{Fig:2}
\end{figure}

\begin{rotatetable*}
\setlength{\tabcolsep}{2.5pt} 
\begin{deluxetable*}{lcccccccccccccccc} 
\tabletypesize{\scriptsize}
\centerwidetable 
\tablecaption{Part of The Final Catalogue \label{tab:2}}
\tablehead{
\colhead{ID} & \colhead{RA} & \colhead{DEC} & \colhead{$z$} & 
\multicolumn{4}{c}{CAL Curve} & \multicolumn{4}{c}{MW Curve} & \multicolumn{4}{c}{SMC Curve} \\
\cmidrule(lr){5-8} \cmidrule(lr){9-12} \cmidrule(lr){13-16}
\colhead{} & \colhead{(deg)} & \colhead{(deg)} & \colhead{} & 
\colhead{$\log M_\ast$} & \colhead{$A_{\rm V,stellar}$} & \colhead{$A_{\rm V,nebular}$} & \colhead{SFR} & 
\colhead{$\log M_\ast$} & \colhead{$A_{\rm V,stellar}$} & \colhead{$A_{\rm V,nebular}$} & \colhead{SFR} & 
\colhead{$\log M_\ast$} & \colhead{$A_{\rm V,stellar}$} & \colhead{$A_{\rm V,nebular}$} & \colhead{SFR} \\
\colhead{(1)} & \colhead{(2)} & \colhead{(3)} & \colhead{(4)} & \colhead{(5)} & \colhead{(6)} & \colhead{(7)} & \colhead{(8)} & \colhead{(9)} & \colhead{(10)} & \colhead{(11)} & \colhead{(12)} & \colhead{(13)} & \colhead{(14)} & \colhead{(15)} & \colhead{(16)}
}
\startdata
\tableline
\multicolumn{16}{c}{SDSS Sample} \\
\tableline
299493800143775744 & 146.63167 & -0.98828 & 0.053 & $9.50 \pm 0.02$ & $0.80 \pm 0.04$ & $0.36 \pm 0.23$ & $0.21 \pm 0.01$ & $9.46 \pm 0.02$ & $0.76 \pm 0.04$ & $0.33 \pm 0.21$ & $0.21 \pm 0.01$ & $9.51 \pm 0.02$ & $0.58 \pm 0.03$ & $0.27 \pm 0.17$ & $0.20 \pm 0.01$ \\
299529809149585408 & 146.35204 & -0.33269 & 0.052 & $10.39 \pm 0.01$ & $1.14 \pm 0.03$ & $1.81 \pm 0.08$ & $5.19 \pm 0.11$ & $10.30 \pm 0.01$ & $0.89 \pm 0.03$ & $1.65 \pm 0.07$ & $4.56 \pm 0.09$ & $10.29 \pm 0.01$ & $1.07 \pm 0.03$ & $1.34 \pm 0.06$ & $3.50 \pm 0.05$ \\
299533107684468736 & 146.39902 & -0.35038 & 0.054 & $9.74 \pm 0.01$ & $1.43 \pm 0.02$ & $1.61 \pm 0.13$ & $1.43 \pm 0.05$ & $9.70 \pm 0.01$ & $1.25 \pm 0.02$ & $1.46 \pm 0.12$ & $1.27 \pm 0.04$ & $9.64 \pm 0.01$ & $1.18 \pm 0.02$ & $1.19 \pm 0.10$ & $1.01 \pm 0.02$ \\
299572690103068672 & 145.32060 & -0.25286 & 0.052 & $10.23 \pm 0.00$ & $0.85 \pm 0.02$ & $0.95 \pm 0.11$ & $1.07 \pm 0.02$ & $10.16 \pm 0.01$ & $0.53 \pm 0.01$ & $0.86 \pm 0.10$ & $1.00 \pm 0.02$ & $10.13 \pm 0.00$ & $0.52 \pm 0.01$ & $0.70 \pm 0.08$ & $0.87 \pm 0.01$ \\
299604301062367232 & 145.89355 & 1.02842 & 0.053 & $10.21 \pm 0.01$ & $0.71 \pm 0.02$ & $1.26 \pm 0.18$ & $1.40 \pm 0.03$ & $10.15 \pm 0.01$ & $0.68 \pm 0.02$ & $1.14 \pm 0.16$ & $1.28 \pm 0.03$ & $10.18 \pm 0.01$ & $0.79 \pm 0.02$ & $0.93 \pm 0.13$ & $1.07 \pm 0.02$ \\
299491051834468352 & 146.63167 & -0.98828 & 0.053 & $9.44 \pm 0.01$ & $0.61 \pm 0.04$ & $0.52 \pm 0.22$ & $0.26 \pm 0.01$ & $9.43 \pm 0.01$ & $0.55 \pm 0.04$ & $0.47 \pm 0.20$ & $0.25 \pm 0.01$ & $9.44 \pm 0.01$ & $0.48 \pm 0.03$ & $0.38 \pm 0.16$ & $0.23 \pm 0.01$ \\
\tableline
\multicolumn{16}{c}{JADES Sample} \\
\tableline
goods-s-mediumhst\_2430 & 53.12819 & -27.78769 & 5.481 & $7.40 \pm 0.15$ & $0.32 \pm 0.11$ & $0.21 \pm 0.52$ & $4.00 \pm 1.20$ & $7.33 \pm 0.10$ & $0.24 \pm 0.08$ & $0.19 \pm 0.48$ & $3.94 \pm 1.08$ & $7.31 \pm 0.10$ & $0.15 \pm 0.05$ & $0.16 \pm 0.39$ & $3.82 \pm 0.83$ \\
goods-s-deephst\_2923 & 53.15407 & -27.82094 & 3.014 & $7.80 \pm 0.07$ & $0.11 \pm 0.06$ & $0.00 \pm 0.50$ & $0.82 \pm 0.16$ & $7.87 \pm 0.05$ & $0.03 \pm 0.03$ & $0.00 \pm 0.46$ & $0.82 \pm 0.14$ & $7.85 \pm 0.06$ & $0.04 \pm 0.02$ & $0.00 \pm 0.37$ & $0.82 \pm 0.11$ \\
goods-s-mediumjwst\_3049 & 53.07061 & -27.90819 & 3.469 & $8.38 \pm 0.09$ & $0.48 \pm 0.06$ & $0.29 \pm 0.33$ & $3.43 \pm 0.52$ & $8.35 \pm 0.09$ & $0.39 \pm 0.05$ & $0.26 \pm 0.30$ & $3.36 \pm 0.46$ & $8.35 \pm 0.09$ & $0.24 \pm 0.03$ & $0.22 \pm 0.24$ & $3.22 \pm 0.35$ \\
goods-s-deephst\_3184 & 53.15010 & -27.81971 & 3.466 & $8.57 \pm 0.05$ & $0.50 \pm 0.02$ & $0.10 \pm 0.27$ & $2.97 \pm 0.26$ & $8.81 \pm 0.01$ & $0.11 \pm 0.01$ & $0.09 \pm 0.25$ & $2.95 \pm 0.23$ & $8.89 \pm 0.03$ & $0.13 \pm 0.01$ & $0.07 \pm 0.20$ & $2.91 \pm 0.18$ \\
goods-s-mediumjwst\_3191 & 53.06071 & -27.90770 & 2.567 & $8.95 \pm 0.03$ & $0.34 \pm 0.12$ & $0.00 \pm 0.63$ & $0.53 \pm 0.14$ & $8.93 \pm 0.03$ & $0.24 \pm 0.10$ & $0.00 \pm 0.57$ & $0.53 \pm 0.13$ & $8.91 \pm 0.03$ & $0.31 \pm 0.09$ & $0.00 \pm 0.47$ & $0.53 \pm 0.10$ \\
goods-s-deephst\_3334 & 53.15138 & -27.81917 & 6.705 & $7.45 \pm 0.15$ & $0.04 \pm 0.04$ & $0.00 \pm 0.59$ & $1.07 \pm 0.39$ & $7.44 \pm 0.17$ & $0.08 \pm 0.05$ & $0.00 \pm 0.53$ & $1.07 \pm 0.35$ & $7.45 \pm 0.14$ & $0.02 \pm 0.02$ & $0.00 \pm 0.43$ & $1.07 \pm 0.29$ \\
\enddata
\tablecomments{(1) Source ID, corresponding to the \texttt{SPECOBJID} from the MPA-JHU DR8 catalog for the SDSS sample, and the \texttt{Unique\_ID} from the JADES DR4 catalog for the JADES sample; (2) R.A. in decimal degrees; (3) Decl. in decimal degrees; (4) Spectroscopic redshift; (5)--(8) Stellar mass ($\log M_*/M_\odot$), $A_{V, \rm stellar}$, $A_{V, \rm nebular}$, and SFR derived assuming a CAL attenuation curve; (9)--(12) Same as (5)--(8), but assuming a MW extinction curve; (13)--(16) Same as (5)--(8), but assuming a SMC extinction curve.}
\end{deluxetable*}
\end{rotatetable*}

\subsection{$A_{\rm V,nebular}$ Measurements} \label{sec:3.2}
We derived $A_{\text{V,nebular}}$ for each galaxy from BD. The relevant relations are
\begin{equation}
    E(B-V)_{\text{nebular}} = \frac{2.5}{\kappa(\text{H}\beta) - \kappa(\text{H}\alpha)} \log \left( \frac{R_{\text{obs}}}{R_{\text{int}}} \right)
    \label{eq:2}
\end{equation}
\begin{equation}
    A_{V,\text{nebular}} = R_V \times E(B-V)_{\text{nebular}}
    \label{eq:3}
\end{equation}
where $\kappa(\text{H}\alpha)$ and $\kappa(\text{H}\beta)$ are the values of the attenuation curve at the wavelengths of H$\alpha$ and H$\beta$, respectively, $R_{\rm obs} \equiv F_{\text{H}\alpha}/F_{\text{H}\beta}$ is the observed BD, and $R_{\text{int}}$ is the intrinsic BD assumed for the ionized gas.

For the SDSS sample, we adopted the widely used assumption of Case~B recombination with an electron density of $n_e = 10^2~\text{cm}^{-3}$ and a temperature of $T_e = 10^4~\text{K}$, yielding $R_{\text{int}} = 2.86$ \citep[e.g.,][]{1989agna.book.....O, 2004MNRAS.351.1151B}. However, recent studies have shown that this assumption, calibrated in the local Universe, may not be appropriate at high redshift \citep[e.g.,][]{2023ApJ...954..157S, 2024A&A...691A.305S, 2024MNRAS.534..523C}. For the JADES sample, we therefore adopted the conditions suggested by \citet{2023ApJ...948...83R}, characterized by $T_e = 15{,}000~\text{K}$ and $n_e = 100~\text{cm}^{-3}$, which correspond to $R_{\text{int}} = 2.79$.

For $A_{\text{V,nebular}}$, we also considered multiple dust attenuation curves, including the CAL, SMC, and MW curves. We note that a subset of galaxies in our sample exhibit observed BD below the theoretical intrinsic value. While this could reflect departures from Case~B recombination \citep[e.g.,][]{2024ApJ...974..180Y, 2024MNRAS.529.3301T, 2025MNRAS.540..190M}, it is more likely attributable to measurement uncertainty in galaxies whose intrinsic BD lie close to the theoretical floor, causing the observed ratio to scatter below the theoretical value \citep[e.g.,][]{2024A&A...691A.201L}. For these galaxies, we set $E(B-V)_{\text{nebular}}=0$ to represent their intrinsically low dust attenuation. 

We then applied an aperture correction to the dust-corrected $\rm H\alpha$ flux based on the difference between the total photometric flux of each galaxy and the synthetic flux derived from the spectra, following the method described in \citet{2019PASJ...71....8K} and \citet{2024ApJ...977..133C}. From the aperture-corrected $\rm H\alpha$ flux, we derived SFR using the calibration of \citet{1998ARA&A..36..189K}. The results derived here, together with those obtained from the SED fitting described above, will be made available online. Representative examples of our derived parameters are presented in Table \ref{tab:2}.

Although we have computed $A_{V,\rm stellar}$ and $A_{V,\rm nebular}$ under multiple assumed dust attenuation curves, numerous studies have demonstrated that the CAL law provides a robust description of the integrated attenuation for the stellar continuum in star-forming galaxies at both low (e.g., \citealt{2000ApJ...533..682C, 2016ApJ...818...13B, 2017ApJ...851...90B}) and high redshifts (e.g., \citealt{2026arXiv260409763R}). For the nebular component, the MW extinction curve with $R_V = 3.1$ has been similarly well-supported across cosmic time (e.g., \citealt{2015ApJ...806..259R, 2021MNRAS.506.3588R}). Furthermore, the combination of a CAL-like stellar attenuation curve and a MW-like nebular extinction curve remains the most widely adopted convention in the existing literature (e.g., \citealt{1997AJ....113..162C, 2019PASJ...71....8K}). Consequently, we adopt $A_{V,\rm stellar}$ from the CAL law and $A_{V,\rm nebular}$ from the MW curve as our fiducial measurements, and present the results derived from this combination in the following sections. A detailed discussion of the impact of alternative attenuation curves is provided in Section~\ref{sec:5.1}.

Figure~\ref{Fig:2} shows the distribution of our sample in the SFR--$M_\ast$ plane. Because our JADES sample spans a broad redshift range, we divide it into two redshift bins, $1.5 < z < 3.5$ (491 galaxies) and $3.5 < z < 7$ (372 galaxies), for the analyses that follow (hereafter, we adopt $z\sim 2.5$ and $z\sim 5$ to represent these two redshift ranges). For reference, we also overlay the star-forming main sequence (SFMS) at $z \sim 0.1$ from \citet{2015ApJ...801L..29R}, together with the relations at $z \sim 2.5$ and $z \sim 5$ from \citet{2014ApJS..214...15S} in Figure~\ref{Fig:2}. As is evident from this comparison, our sample predominantly follows the SFMS across all redshift bins.

\section{Galaxy Dust Properties} \label{sec:4}

\begin{figure*}[htbp] 
    \centering
    \includegraphics[width=\linewidth]{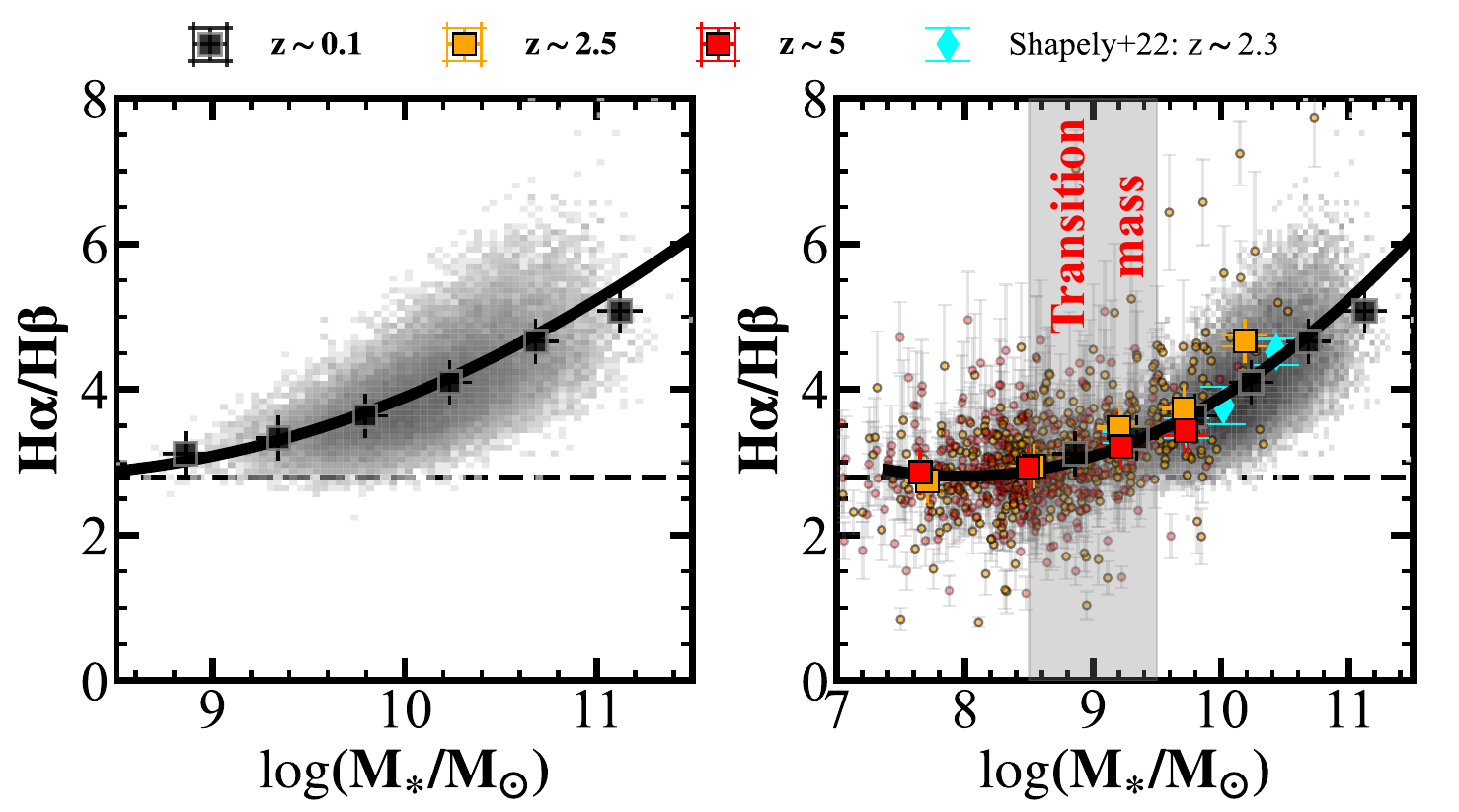}
    \caption{The distribution of BD as a function of $M_\ast$. The left panel shows the SDSS results, and the right panel shows the JADES results together with the SDSS results. The black, orange, and red symbols represent results at $0.05 < z < 0.1$, $1.5 < z < 3.5$, and $3.5 < z < 7$, respectively. Squares denote the median values in bins of $M_\ast$, while the error bars indicate the standard error within each bin. The black dashed line represents the theoretical value 2.79 under Case~B recombination. The best-fitting result is represented by the solid black line. Shaded gray region represents the transition mass at $\log(M_\ast/M_\odot)=9$. For comparison, we also show the results from \cite{2022ApJ...926..145S} at $z \sim 2.3$ as cyan diamonds. It can be seen from this figure, at fixed $M_\ast$, BD exhibits little to no evolution with redshift. }
    \label{Fig:3}
\end{figure*}

\begin{figure*}[htbp]
    \centering
    \includegraphics[width=\linewidth]{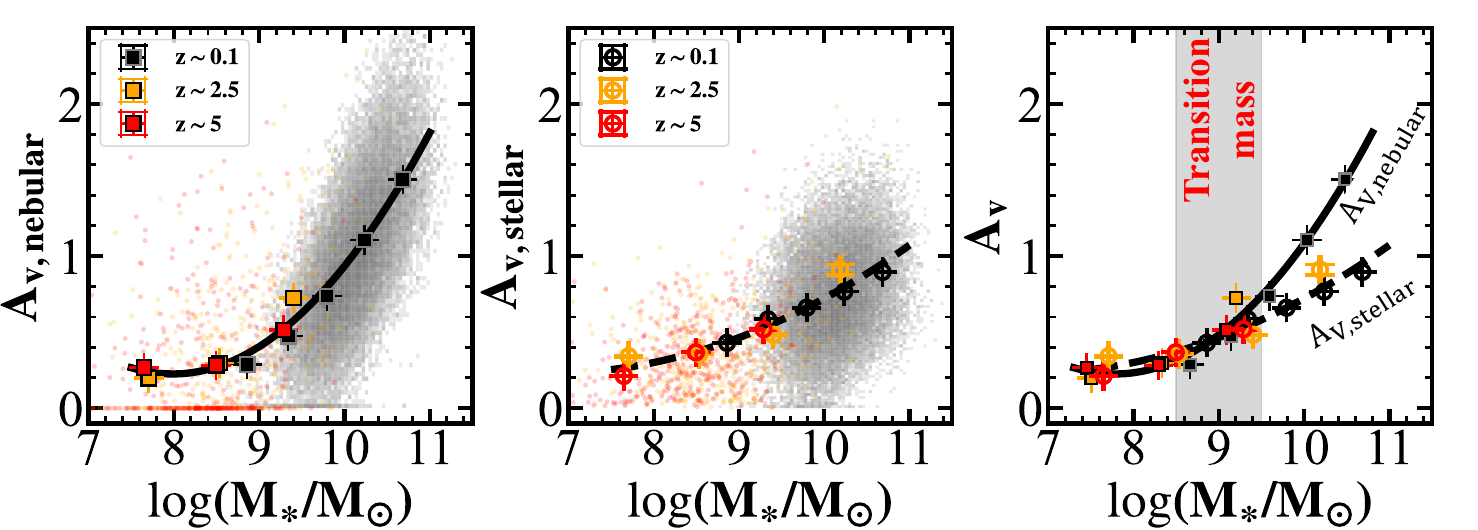}
    \caption{Dust attenuation as a function of $M_\ast$. The layout is similar to Figure~\ref{Fig:3}. The left panel presents $A_{V,\text{nebular}}$ versus $M_\ast$, while the middle panel shows $A_{V,\text{stellar}}$ as a function of $M_\ast$. The filled squares and open circles represent the mean $A_{V,\text{nebular}}$ and $A_{V,\text{stellar}}$ values, respectively, calculated within different $M_\ast$ bins. The error bars represent the standard error within each corresponding $M_\ast$ bin. The corresponding best-fit relations derived in Section \ref{sec:4.3} are indicated by the black solid and dashed lines. For ease of comparison, the right panel displays both quantities side by side.}
    \label{Fig:4}
\end{figure*}

\subsection{$A_{\rm V,nebular}$ and $A_{\rm V,stellar}$} \label{sec:4.1}
We first examine how galactic dust attenuation varies with $M_\ast$. The most direct approach to this question is the relation between the BD and $M_\ast$, because the BD is a directly observed quantity that does not rely on stellar-population modeling. Previous studies have established that the BD increases monotonically with $M_\ast$, and that this relation exhibits little to no evolution with redshift \citep[e.g.,][]{2010MNRAS.409..421G, 2013ApJ...763..145D, 2022ApJ...926..145S, 2023ApJ...954..157S, 2024A&A...691A.305S, 2025arXiv251000235W}. In this study, we revisit this relation by combining the SDSS and JADES samples, and the results are presented in Figure~\ref{Fig:3}. The left panel shows the SDSS results, while the right panel presents the results from JADES together with those from SDSS. The black two-dimensional histogram illustrates the result at $z \sim 0.1$, while orange and red circles represent the results at $z \sim 2.5$ and $z \sim 5$, respectively. Squares denote the mean values within each $M_\ast$ bin, with error bars representing the corresponding standard error. The black dashed line indicates the theoretical BD of $2.79$ adopted for the JADES sample. For comparison, the results at $z \sim 2.3$ from \cite{2022ApJ...926..145S} are also overlaid as cyan diamonds.

Figure~\ref{Fig:3} shows that the BD increases monotonically with $M_\ast$ across the entire redshift range, indicating that the correlation between the BD and $M_\ast$ is already established at $z \sim 7$. Moreover, at fixed $M_\ast$, the BD values measured in different redshift bins are quite similar, demonstrating that $M_{\ast}$ is the dominant parameter governing the BD. This result is in good agreement with previous findings \citep[e.g.,][]{2023ApJ...954..157S, 2024A&A...691A.305S, 2025arXiv251000235W}. Furthermore, when we combine the samples across all redshift bins, evidence for a characteristic transition mass in this relation emerges. At $\log(M_\ast / \text{M}_\odot) \lesssim 9$, the BD rises only slowly with $M_\ast$ and remains close to the theoretical Case~B value, whereas at $\log(M_\ast / \text{M}_\odot) \gtrsim 9$, the BD increases more rapidly with $M_\ast$. This transition mass is marked by the shaded gray region in Figure~\ref{Fig:3}.

Considering that the lack of redshift evolution of the BD--$M_*$ relation has been reported by many previous studies, we characterize the relation between $M_\ast$ and the $\rm H\alpha/H\beta$ ratio using a single polynomial function for all redshift bins. To prevent the fit from being dominated by the SDSS sample, we performed a weighted least-squares fit to the mean values in each $M_\ast$ bin. The resulting best-fit relation is
\begin{equation}
\begin{aligned}
    \mathrm{H}\alpha/\mathrm{H}\beta
    &= 0.26\,\log(M_*/M_\odot)^{2} \\
    &\quad - 4.25\,\log(M_*/M_\odot) + 19.75 ,
\end{aligned}
\label{eq:4}
\end{equation}

Additionally, we also investigated the mass dependence of each attenuation components. Figure~\ref{Fig:4} presents $A_{V,\rm nebular}$ (left panel) and $A_{V,\rm stellar}$ (middle panel) as functions of $M_\ast$, following a presentation similar to that of Figure~\ref{Fig:3}, except that filled squares and open circles denote the mean values of $A_{V,\rm nebular}$ and $A_{V,\rm stellar}$ within each mass bin, respectively. As noted in Section~\ref{sec:3.1}, a subset of galaxies in our sample exhibit an observed $\rm H\alpha/H\beta$ ratio below the theoretical BD, which causes their inferred $E(B-V)_{\rm nebular}$ to be set to zero. We retain these galaxies in the following analysis because this behavior may reflect intrinsically low dust attenuation scattered below the theoretical floor by measurement uncertainty, as discussed in Section~\ref{sec:3.1}.

Across the full redshift range, $A_{V,\rm nebular}$ exhibits a clear positive correlation with $M_\ast$. Similar to the result of BD, we also see indications of a characteristic transition mass: above $\log(M_*/M_\odot) \sim 9$, $A_{V,\rm nebular}$ rises rapidly with increasing $M_\ast$. This close parallel with the BD--$M_*$ relation is expected, since $A_{V,\rm nebular}$ is derived directly from the BD. The stellar attenuation $A_{V,\rm stellar}$ likewise correlates positively with $M_\ast$ at all redshifts, but the rate of increase appears slower than for $A_{V,\rm nebular}$, and the evidence for a transition mass is correspondingly weaker. In addition, at fixed $M_\ast$, both $A_{V,\rm nebular}$ and $A_{V,\rm stellar}$ appear to remain constant with redshift. We will discuss the redshift evolution in more detail in Section~\ref{sec:4.3}.

To more directly assess whether $A_{V,\rm nebular}$ and $A_{V,\rm stellar}$ share the same dependence on $M_{\ast}$, the right panel of Figure~\ref{Fig:4} displays the two quantities side by side. For clarity of presentation, the $A_{V,\rm nebular}$ points are shifted leftward by $0.2$ dex in $\log(M_*/M_\odot)$ to avoid overlap with the $A_{V,\rm stellar}$ points. In this combined view, the evidence for a transition mass becomes more pronounced. At $\log(M_*/M_\odot) \lesssim 9$, $A_{V,\rm nebular}$ and $A_{V,\rm stellar}$ are comparable in magnitude and exhibit similar dependences on $M_\ast$. However, $A_{V,\rm nebular}$ increases with $M_\ast$ at a markedly faster rate than $A_{V,\rm stellar}$ above $\log(M_*/M_\odot) \sim 9$, so that $A_{V,\rm nebular}$ becomes systematically larger than $A_{V,\rm stellar}$ at the high-mass end.

\begin{figure*}[htbp]
    \centering
    \includegraphics[width=\linewidth]{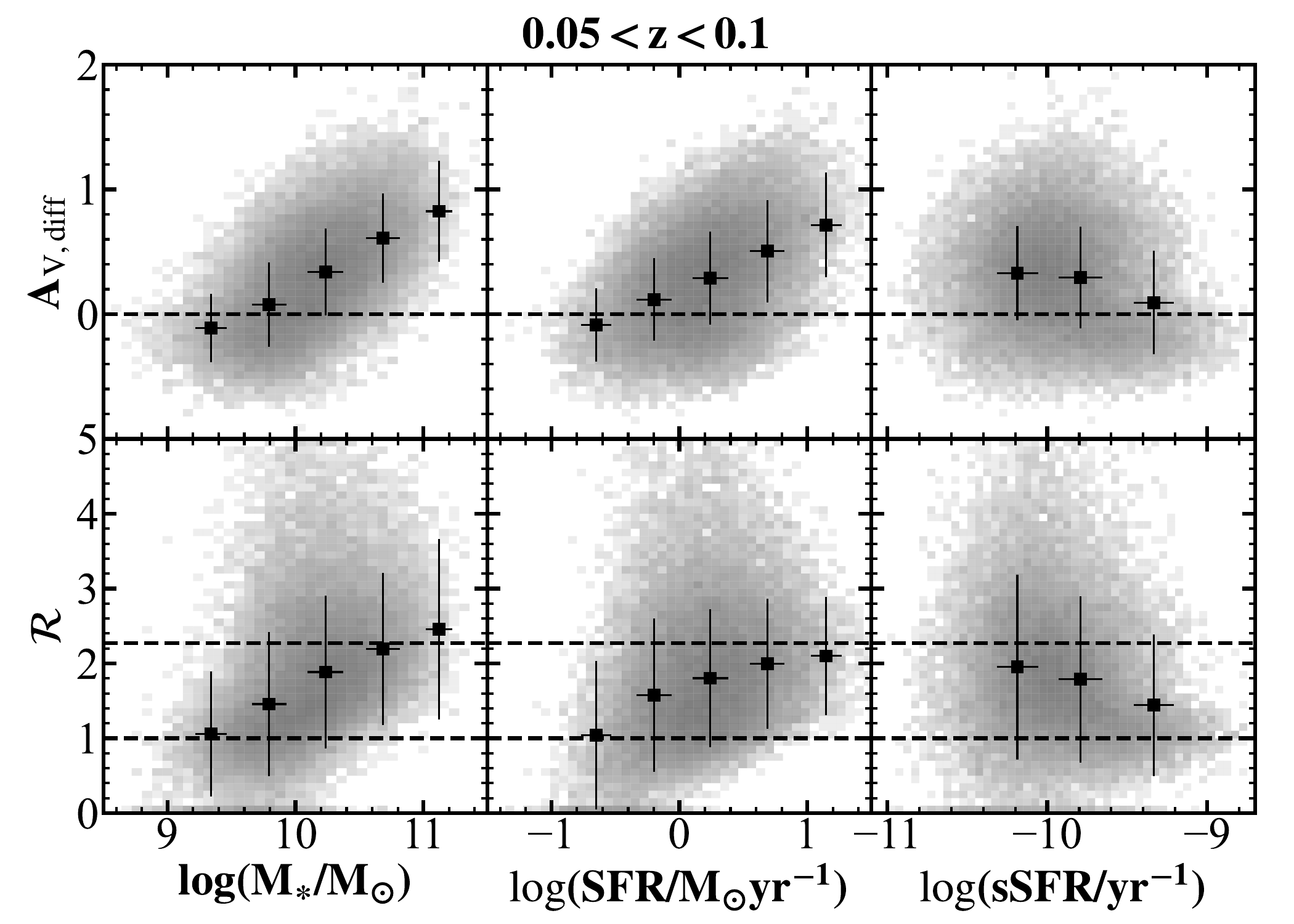}
    \caption{The difference in dust reddening between nebular and stellar components as a function of $M_\ast$ (left column), SFR (middle column), and sSFR (right column). The top row shows the results of $A_{\rm V, diff}$, while the bottom row presents the results of $\mathcal{R}$. Squares denote the values derived from the mean attenuation of the respective components in each bin of the relevant physical property, and the error bars show the corresponding $1\sigma$ dispersion. For ease of comparison, we indicate $A_{\rm V, diff} = 0$ with a dashed line in the top panel. In the lower panel, dashed lines denote the unity ratio ($\mathcal{R}=1$) and the canonical value of $\mathcal{R}=2.27$ in the local universe.}
    \label{Fig:5}
\end{figure*}

\begin{table*}[htbp]
    \centering
    \caption{Spearman correlation results across different redshift bins}
    \label{tab:3}
    \begin{tabular}{cccccccc}
        \toprule
        \multirow{2}{*}{Redshift} & \multirow{2}{*}{Parameter} & \multicolumn{2}{c}{$\log(M_*/M_\odot)$} & \multicolumn{2}{c}{$\log\rm (SFR/M_\odot yr^{-1})$} & \multicolumn{2}{c}{$\log\rm (sSFR/yr^{-1})$} \\
        \cmidrule(lr){3-4} \cmidrule(lr){5-6} \cmidrule(lr){7-8}
        & & $r$ & $p$ & $r$ & $p$ & $r$ & $p$ \\
        \midrule
        \multirow{2}{*}{$0.05 < z < 0.1$} 
        & $A_{V, \mathrm{diff}}$ &  0.568 & $< 10^{-5}$ &  0.422 & $< 10^{-5}$ & -0.146 & $< 10^{-5}$ \\
        & $\mathcal{R}$                    &  0.426 & $< 10^{-5}$ &  0.272 & $< 10^{-5}$ & -0.158 & $< 10^{-5}$ \\
        \midrule
        \multirow{2}{*}{$1.5 < z < 3.5$}   
        & $A_{V, \mathrm{diff}}$ &  0.311 & $< 10^{-5}$ &  0.339 & $< 10^{-5}$ & -0.134 & 0.003 \\
        & $\mathcal{R}$                    &  0.373 & $< 10^{-5}$       &  0.413 & $< 10^{-5}$       & -0.070 & 0.126 \\
        \midrule
        \multirow{2}{*}{$3.5 < z < 7$}     
        & $A_{V, \mathrm{diff}}$ & -0.096 & 0.066       &  0.224 & $< 10^{-3}$ &  0.213 & 0.001 \\
        & $\mathcal{R}$                    & 0.096 & 0.067       &  0.432 & $< 10^{-3}$       &  0.182 & $< 10^{-3}$ \\
        \bottomrule
    \end{tabular}
\end{table*}

\subsection{Nebular-to-Stellar Attenuation Difference} \label{sec:4.2}
\subsubsection{Galaxies at $z\sim 0.1$} \label{sec:4.2.1}

In the previous subsection, we examined how $A_{V,\rm nebular}$ and $A_{V,\rm stellar}$ depend on $M_\ast$. We now turn to their difference and investigate how it correlates with other galaxy physical properties. 
In this section, we consider both the difference in $A_{\rm V}$, which is defined as:
\begin{equation}
    A_{\rm V, diff} \equiv A_{\rm V,nebular} - A_{\rm V,stellar}
    \label{eq:5}
\end{equation}
and the ratio of $E(B-V)$. To avoid infinite values arising from setting $E(B-V)_{\rm nebular}$ to zero, we adopt the inverse of the conventional $f$-factor throughout the following sections, defining 
\begin{equation}
    \mathcal{R} \equiv E(B-V)_{\rm nebular}/E(B-V)_{\rm stellar}
    \label{eq:6}
\end{equation} 
Under this convention, the classic result of $f = 0.44$ from \citet{1997AJ....113..162C} corresponds to $\mathcal{R} = 2.27$.

Figure~\ref{Fig:5} presents $A_{V,\rm diff}$ and $\mathcal{R}$ as functions of $M_\ast$ (left), SFR (middle), and sSFR (right) for our SDSS sample. The top row shows the results for $A_{V,\rm diff}$, while the bottom row displays the corresponding trends for $\mathcal{R}$. Squares represent the values derived from the mean attenuation of the respective components in each bin of the relevant physical property, and the error bars indicate the $1\sigma$ dispersion within each bin.

As shown in Figure~\ref{Fig:5}, both $A_{V,\text{diff}}$ and $\mathcal{R}$ exhibit significant positive correlations with $M_\ast$ and the SFR, while being anti-correlated with the sSFR. The quantity $A_{V,\text{diff}}$ approaches 0 at $\log(M_\star/M_\odot) \sim 9$ and increases monotonically to nearly 1 mag at $\log(M_\star/M_\odot) \sim 11$. Correspondingly, $\mathcal{R}$ also increases monotonically from approximately 1 to the classic value of 2.27 observed in the local Universe over this mass range. These trends are mutually consistent, reflecting the underlying positive correlation between $M_*$ and the SFR, as well as the inverse relationship between $M_*$ and the sSFR. To quantify the significance of these relations, we computed the Spearman rank-order correlation coefficients for the full sample, as summarized in Table~\ref{tab:3}. In all cases, the associated $p$-values are below $10^{-5}$, indicating that the differential dust attenuation is statistically correlated with all three galaxy properties. 

These results are broadly consistent with many prior studies of local star-forming galaxies (e.g., \citealt{2011MNRAS.417.1760W, 2016ApJ...818...13B, 2017ApJ...847...18Z, 2019PASJ...71....8K, 2019ApJ...886...28Q, 2020ApJ...888...88L, 2021ApJ...917...72L}). For example, \cite{2017ApJ...847...18Z} derived $A_{V,\rm stellar}$ from stacked SDSS optical spectra and compared it to $A_{V,\rm nebular}$, finding that the nebular-to-continuum attenuation ratio increases systematically with $M_\ast$. Similarly, \cite{2019PASJ...71....8K} demonstrated that $\mathcal{R}$ increases with $M_\ast$, approaching unity at the low-mass end. Numerous other studies have likewise reported a positive correlation between the differential attenuation and the SFR (e.g., \citealt{2020ApJ...888...88L, 2021ApJ...917...72L}), in agreement with our findings. These findings are also consistent with the results presented in Figure~\ref{Fig:4}. At the high-mass end, $A_{V,\text{nebular}}$ increases more rapidly with $M_\ast$ than $A_{V,\text{stellar}}$, producing an increasing difference between the two components as $M_\ast$ increases.

\begin{figure*}[htbp]
    \centering
    \includegraphics[width=\linewidth]{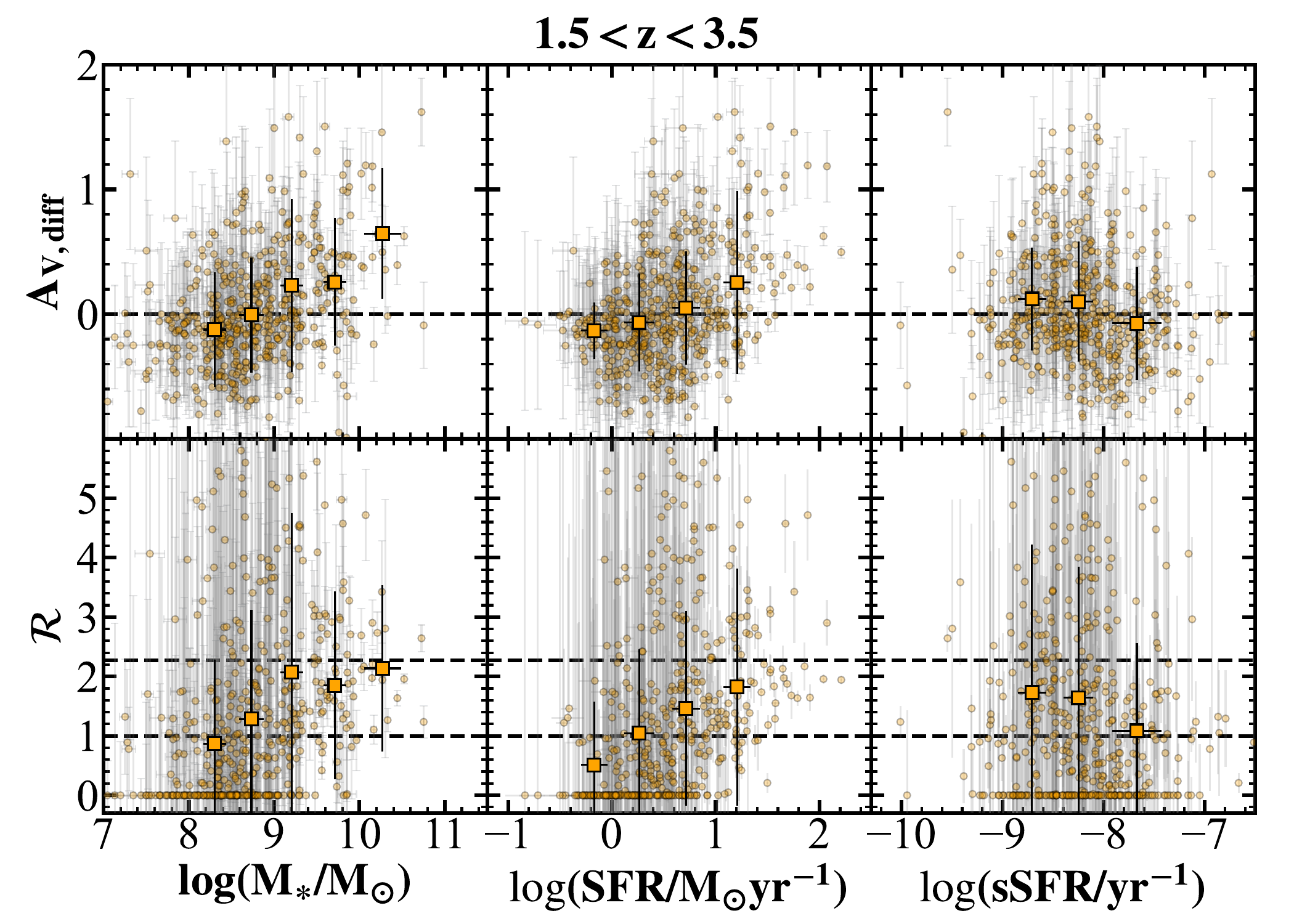}
    \caption{Similar to Figure \ref{Fig:5}, but showing the results for the sample at $1.5 < z < 3.5$. The background circles represent the measurements for individual galaxies. The squares denote the results in bins of each relevant physical property. The error bars indicate the $1\sigma$ dispersion within the corresponding physical property bins.}
    \label{Fig:6}
\end{figure*}

\begin{figure*}[htbp]
    \centering
    \includegraphics[width=\linewidth]{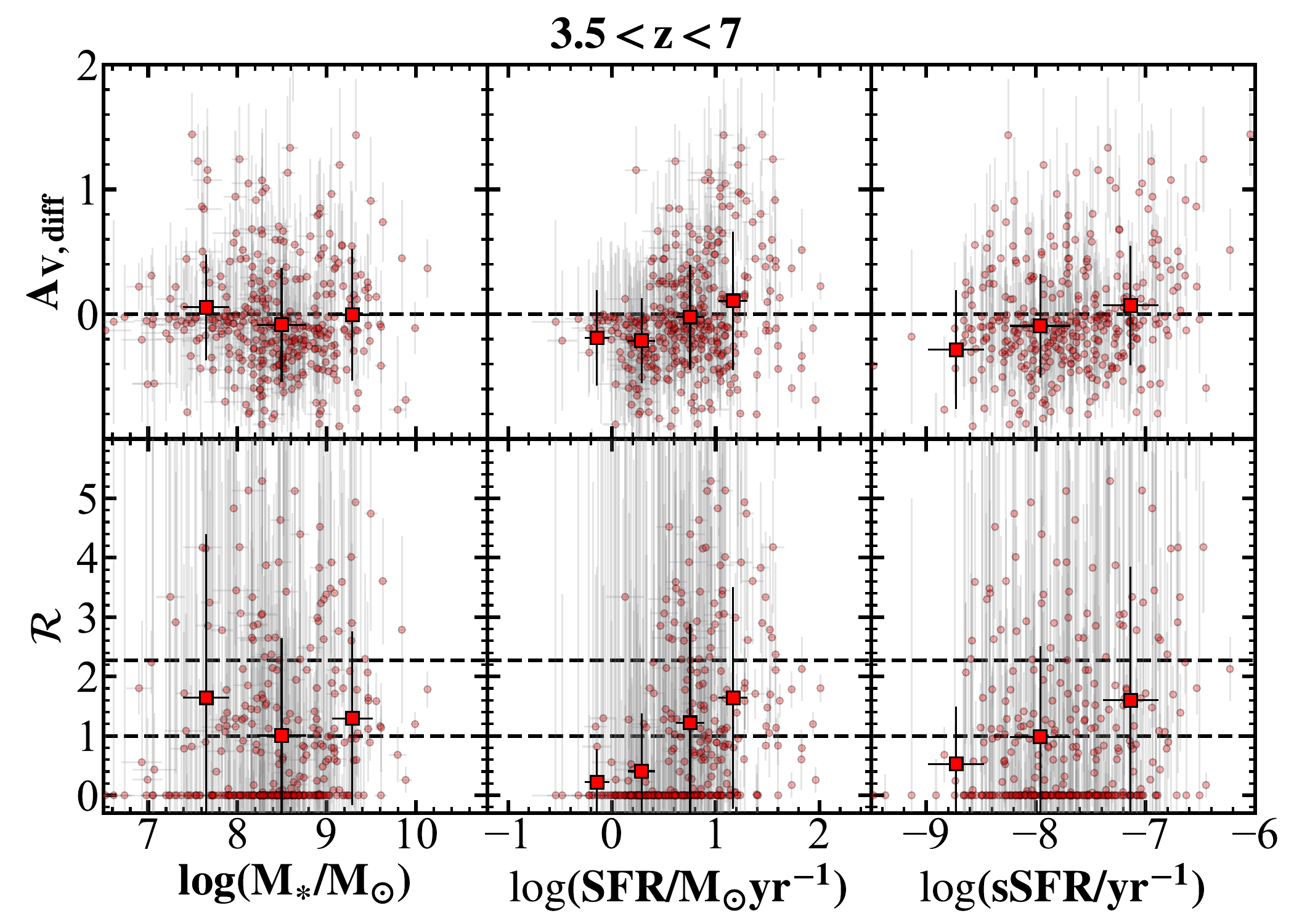}
    \caption{Similar to Figure \ref{Fig:6}, but showing the results for the sample at $3.5 < z < 7$.}
    \label{Fig:7}
\end{figure*}

\subsubsection{Galaxies at $z\sim 1.5-7$} \label{sec:4.2.2}
Several studies have established that the $\mathcal{R}$-factor derived in the local universe is smaller than that observed in typical high-redshift star-forming galaxies. In this section, we utilize the JADES data to further investigate this issue. Following the approach described in Section~\ref{sec:3.2}, we divide our high-redshift sample into two redshift bins: $1.5 < z < 3.5$ and $3.5 < z < 7$. The results for galaxies in these two redshift ranges are presented in Figures~\ref{Fig:6} and~\ref{Fig:7}, respectively, in a format similar to that of Figure~\ref{Fig:5}.

For the sample at $z \sim 2.5$, we find that the positive correlations of both $A_{V,\rm diff}$ and $\mathcal{R}$ with $M_\ast$ persist at high redshift, demonstrating that mass-dependent differential attenuation is not a phenomenon restricted to the local universe. Consistent with the local sample, both $A_{V,\rm diff}$ and $\mathcal{R}$ exhibit positive correlations with the SFR, alongside a negative correlation with the sSFR. The corresponding Spearman rank-order correlation coefficients are summarized in Table~\ref{tab:3}. Notably, the JADES sample probes a lower characteristic $M_\ast$ range ($\log(M_*/M_\odot) \sim 7.5\text{--}10.5$) than the SDSS sample ($\log(M_*/M_\odot) \sim 9\text{--}11.5$). Given the positive correlation between $A_{V,\rm diff}$ and $M_\ast$, this lower characteristic mass naturally accounts for the median $A_{V,\rm diff}$ values that lie closer to zero at high redshift relative to the SDSS sample, without requiring genuine redshift evolution.

Prior to the JWST era, several studies had already investigated differential dust attenuation at $z \sim 1\text{--}3$, and our results are in good agreement with their findings. For instance, using 3D-HST spectroscopy, \cite{2014ApJ...788...86P} reported that $A_{V,\rm diff}$ exhibits positive correlations with both $M_\ast$ and SFR at $z \sim 1.5$. Similarly, \cite{2015ApJ...806..259R} studied galaxies from the MOSDEF survey at $1.4 < z < 2.6$ and reached the same conclusion. A number of other studies covering this epoch reported single-valued $f$-factor estimates without examining an explicit dependence on galaxy properties. For instance, \cite{2013ApJ...777L...8K} measured $f \simeq 0.83$ at $z \sim 1.6$, while \cite{2019ApJ...871..128T} found an intermediate value of $f \simeq 0.74$ at $z \sim 2.0\text{--}2.7$. Additionally, \cite{2015ApJ...807..141P} obtained $f \sim 1$ at $z > 1.5$, and \cite{2016A&A...586A..83P} reported $f \sim 0.93$ at $z \sim 1$. These discrepancies are naturally understood when the different $M_\ast$ ranges probed by these studies are taken into account: samples dominated by lower-mass galaxies preferentially yield $f$ values closer to unity, whereas those with a higher characteristic mass tend to recover the canonical \cite{1997AJ....113..162C} value of $f \approx 0.44$.

Regarding the results at $z\sim5$, we find no statistically significant correlation between $A_{V,\rm diff}$ (or $\mathcal{R}$) and $M_\ast$, with the corresponding $p$-values of $0.066$ and $0.067$, respectively (as shown in Table~\ref{tab:3}). Furthermore, the median $A_{V,\rm diff}$ in this redshift bin is consistent with zero. This result likely reflects a combination of two sample-selection effects: the incompleteness of our sample at $\log(M_*/M_\odot) < 8.5$ and the scarcity of massive galaxies ($\log(M_*/M_\odot) > 10$) in the high-redshift JADES sample \citep{2025arXiv251006681C}. Nonetheless, $A_{V,\rm diff}$ retains a moderate positive correlation with the SFR ($p < 0.05$). Similar results are also found over the redshift range $1.5 < z < 3.5$, where both $A_{V,\mathrm{diff}}$ and $\mathcal{R}$ appear to exhibit stronger correlations with SFR, as shown in Table \ref{tab:3}. However, as discussed in the Introduction, many previous studies have found that the relation between galaxy dust attenuation and $M_\ast$ is redshift invariant. Therefore, here we mainly focus on the relation between $A_{V,\mathrm{diff}}$ and $M_\ast$. The physical quantity that exhibits the strongest correlation with $A_{V,\mathrm{diff}}$ will be investigated in more detail in future work.

Several other JWST-based studies have also examined differential dust attenuation at $z > 3$. \cite{2026arXiv260311338K} analyzed a JADES sample of 283 star-forming galaxies at $2.7 < z < 7$ and found a flat relation between $E(B-V)_{\rm nebular} - E(B-V)_{\rm stellar}$ and $M_*$ at $z > 4$, with values consistent with zero at $z > 5$, which is in good agreement with our results. In contrast, \citet{2025arXiv251000235W} reported an unusually high mean ratio of $\mathcal{R} \sim 14$ at $z > 3$. However, their analysis excluded galaxies with observed BD below $2.86$, thereby preferentially removing low-attenuation systems from the sample. This selection criterion introduces a systematic bias toward elevated $E(B-V)_{\rm nebular}$ values, inflating the inferred $\mathcal{R}$. Additional JWST constraints from \cite{2026ApJ...997..319T} and \cite{2026arXiv260311338K} report $f$-values in the range of $\sim\! 0.5\text{--}1.0$ at $z \sim 3\text{--}7$, which are broadly consistent with our findings despite the significant uncertainties inherent in these measurements.

\begin{figure*}[htbp]
    \centering
    \includegraphics[width=\linewidth]{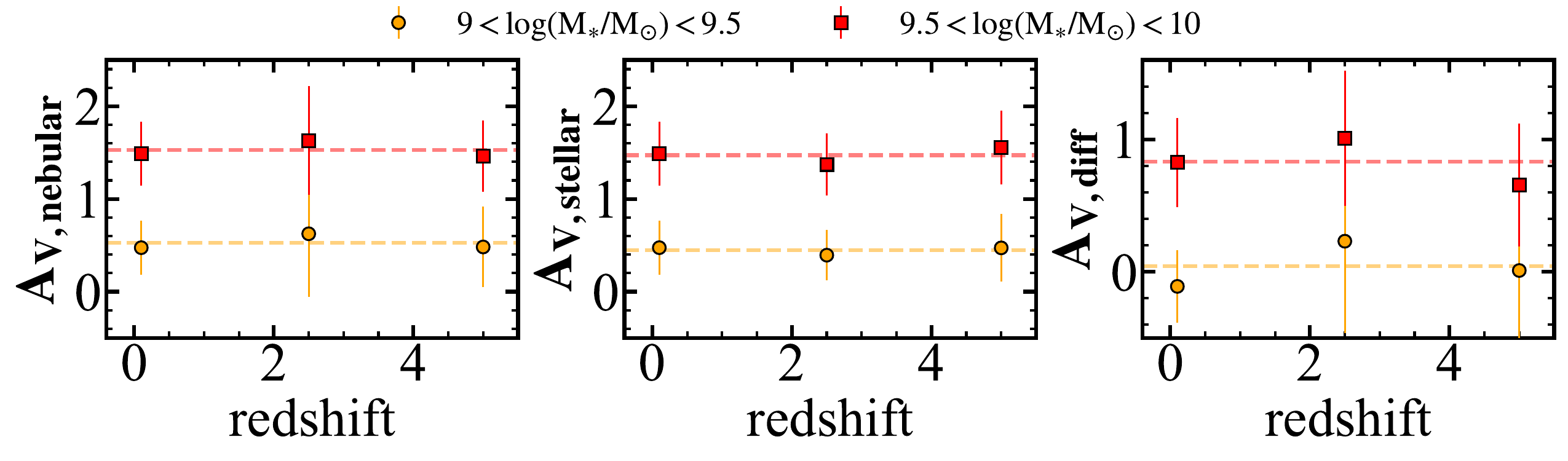}
    \caption{The redshift evolution of $A_{\rm V, nebular}$ (left), $A_{\rm V, stellar}$ (middle) , and $A_{\rm V, diff}$(right). The orange circles and red squares represent the result for galaxies with $M_\ast$ in the ranges $9 < \log(M_\ast/M_\odot) < 9.5$ and $9.5 < \log(M_\ast/M_\odot) < 10$, respectively. For clarity, the red squares are shifted upward by 0.75 mag.}
    \label{Fig:8}
\end{figure*}

\begin{figure*}[htbp]
    \centering
    \includegraphics[width=\linewidth]{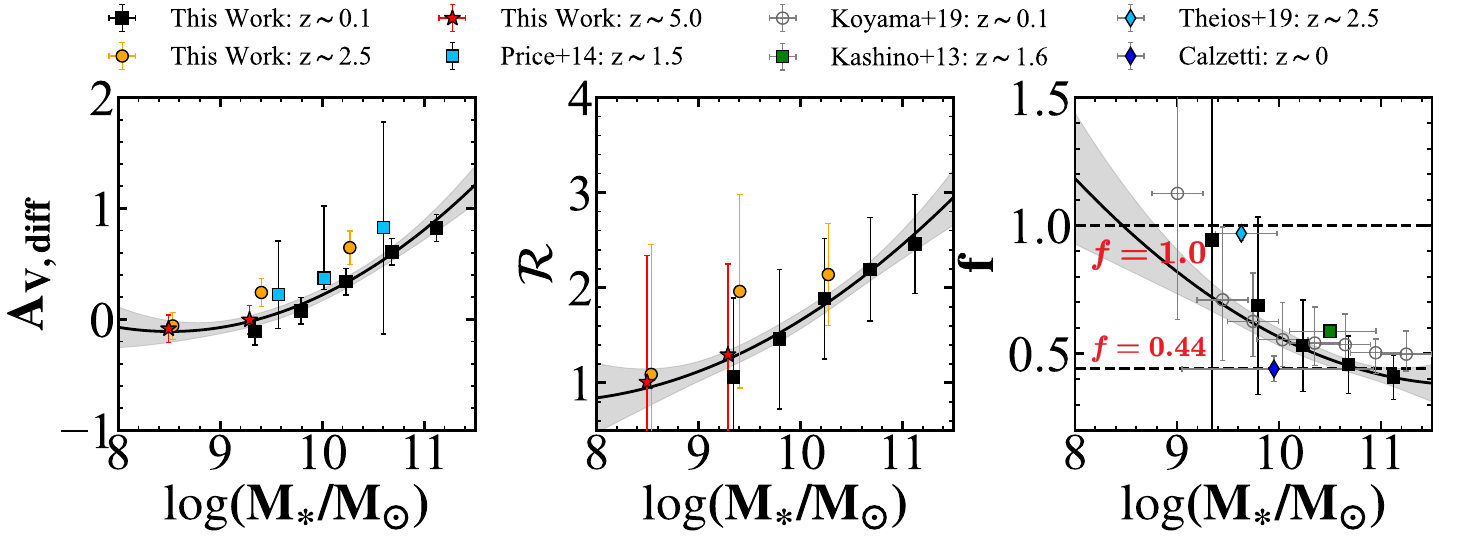}
    \caption{The difference in dust reddening between nebular and stellar components as a function of $M_\ast$ at different redshifts. The left panel presents the result of $A_{\rm V,diff}$ while the middle and right panel show the result of $\mathcal{R}$ and $f$, respectively. The black, orange, and red symbols correspond to the dust attenuation differences derived from the mean attenuation in each $M_\ast$ bin for galaxies at $0.05 < z < 0.1$, $1.5 < z < 3.5$, and $3.5 < z < 7$, respectively. The error bars are derived from the corresponding standard errors, with an additional 0.08 mag uncertainty in $A_{V,\mathrm{nebular}}$ caused by the choice of the intrinsic BD value. For clarity, we show only the $z \sim 0.1$ results in the right panel; the $z > 1.5$ results are omitted because they follow the same trend as in the middle panel. For ease of interpretation, we also indicate $f=0.44$ and 1.0 with dashed lines in the right panel. The best-fitting polynomial relation is shown as a black solid lines. The gray shaded region represents the uncertainty in the best-fitting result. For comparison, we also show the measurements from \cite{1997AJ....113..162C}, \cite{2019PASJ...71....8K}, \cite{2013ApJ...777L...8K}, \cite{2019ApJ...871..128T}, and \cite{2014ApJ...788...86P}. Overall, our results are broadly consistent with previous studies. At fixed $M_\ast$, the reddening difference exhibits little to no significant evolution with redshift. When combining all redshift bins, the $A_{\rm V, diff}$--$M_\ast$ relation flattens at the low-mass end, while a positive correlation with $M_\ast$ emerges at the high-mass end.}
    \label{Fig:9}
\end{figure*}

\begin{figure}[htbp] 
    \centering
    \includegraphics[width=\linewidth]{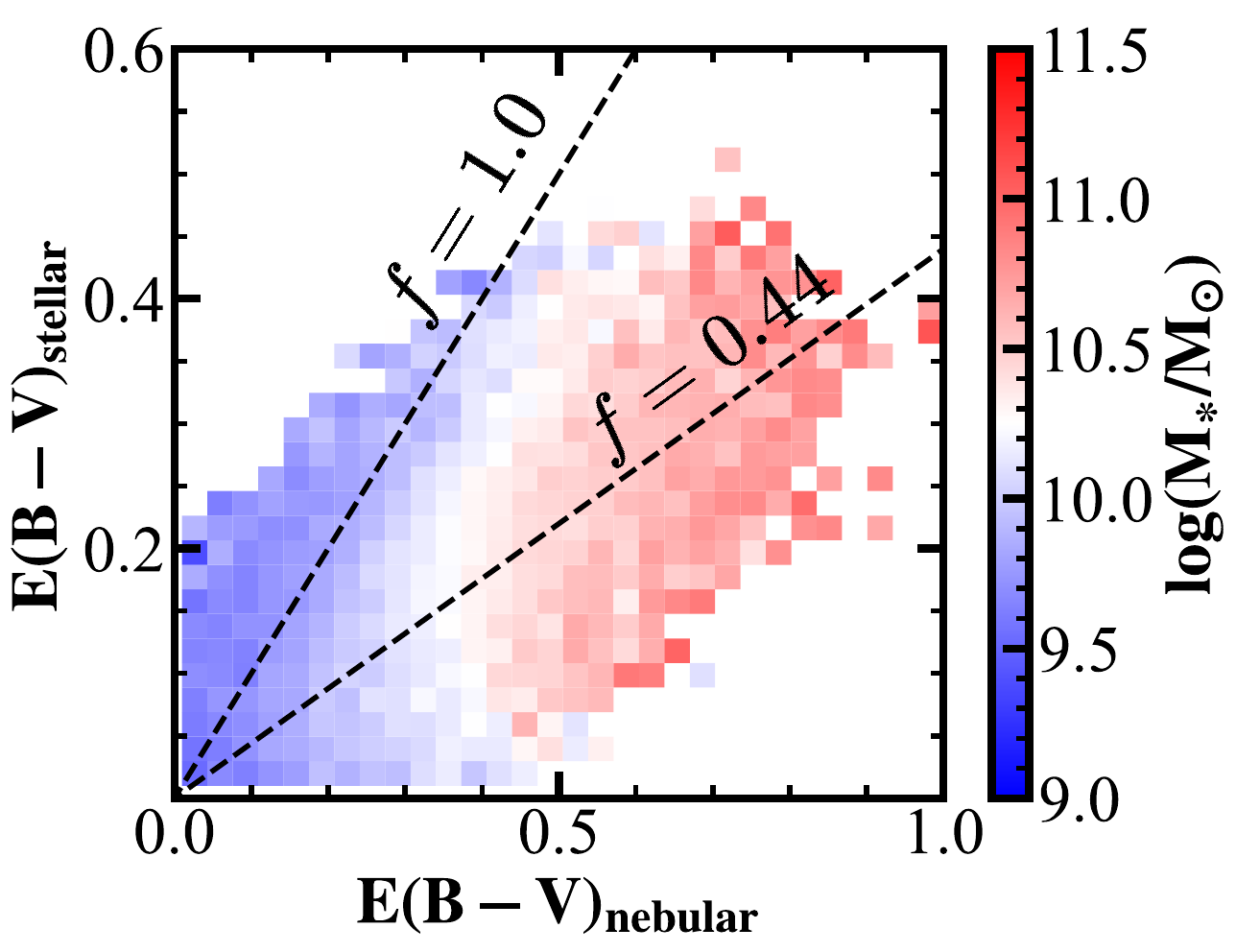}
    \caption{The comparison between $E(B-V)_{\mathrm{stellar}}$ and $E(B-V)_{\mathrm{nebular}}$, where the color scale indicates the median $M_\ast$ within each square bin. For reference, we also plot dashed lines corresponding to $f=0.44$, and 1.0 in this figure. We find a clear trend in which more massive galaxies preferentially occupy the lower-right region of the diagram, indicating a smaller result of $f$.}
    \label{Fig:10}
\end{figure}

\subsection{Redshift Evolution} \label{sec:4.3}
In the preceding subsections, we examined how $A_{V,\rm nebular}$, $A_{V,\rm stellar}$, and their difference $A_{V,\rm diff}$ depend on $M_\ast$. In this section, we further examine the redshift evolution of these quantities. As discussed in Sections~\ref{sec:4.1} and~\ref{sec:4.2}, investigating such evolution requires comparison within a consistent $M_\ast$ range. To this end, we select two mass bins, $9.0 < \log(M_*/M_\odot) < 9.5$ and $9.5 < \log(M_*/M_\odot) < 10.0$, corresponding to the overlap region between the SDSS and JADES samples. The redshift evolution of $A_{V,\rm nebular}$, $A_{V,\rm stellar}$, and $A_{V,\rm diff}$ within these two mass bins is presented in Figure~\ref{Fig:8}. For clarity of presentation, results for the $9.5 < \log(M_*/M_\odot) < 10.0$ bin are vertically offset by $0.75$ mag, and the error bars indicate the $1\sigma$ dispersion within each bin.

Figure~\ref{Fig:8} shows that, in both $M_\ast$ bins, $A_{V,\rm nebular}$, $A_{V,\rm stellar}$, and $A_{V,\rm diff}$ remain essentially unchanged with redshift, showing no statistically significant evolutionary trend. One exception is a slight elevation of $A_{V,\rm nebular}$ in the $1.5 < z < 3.5$ bin. Correspondingly, $A_{V,\rm diff}$ is also slightly elevated over the same redshift range. This offset can be explained by our adoption of distinct Case~B BD values for the two samples: $2.79$ for the JADES sample and $2.86$ for the SDSS sample (as discussed in Section~\ref{sec:3.2}), which propagates into a systematic difference of ${\sim}\,0.08$ mag in the inferred $A_{V,\rm nebular}$, as inferred from Equations \ref{eq:2} and \ref{eq:3}. Many earlier studies continued to assume the canonical intrinsic value of $2.86$ at $z < 3$.  If this standard intrinsic ratio were adopted for this redshift bin, the marginal elevation in $A_{V, \rm nebular}$ would vanish. 

Moreover, adopting a single intrinsic BD value for all galaxies at $z > 1.5$ remains a simplification. Since $T_{\rm e}$ scales with metallicity (which in turn correlates with stellar mass), using a mass-dependent intrinsic BD value would be more rigorous (e.g., \citealt{2022MNRAS.517....1S, 2024A&A...691A.305S}). However, this is beyond the scope of this work. Given that the differences between $1.5 <z < 3.5$ and the other redshift bins remain within the $1\sigma$ dispersion, we conclude that, from $z \sim 0.1$ to $z \sim 7$, $A_{V,\rm nebular}$, $A_{V,\rm stellar}$, and $A_{V,\rm diff}$ exhibit no significant redshift evolution at fixed $M_\ast$ from $z\sim0.1$ to $z\sim7$. However, we note that our sample lacks galaxies at $0.1 \lesssim z \lesssim 1.5$. In future work, we will use Subaru Prime Focus Spectrograph \citep{2016SPIE.9908E..1MT} observation to obtain a more complete picture of the redshift evolution.

Given the absence of significant redshift evolution at fixed $M_\ast$, we can combine the low-redshift and high-redshift samples to characterize the global $A_{V,\rm nebular}$--$M_*$ and $A_{V,\rm stellar}$--$M_*$ relations across cosmic time. Following the approach used in Figure~\ref{Fig:3}, we fit each relation with a polynomial function. The best-fit relations are overplotted as black solid and dashed lines for $A_{V,\rm nebular}$ and $A_{V,\rm stellar}$ in Figure~\ref{Fig:4}, respectively. The best fitting relations are:

\begin{equation}
\begin{aligned}
    A_{\rm V,nebular}
    &= 0.17\log(M_{\ast}/M_\odot)^2 \\
    &\quad -2.79 \log(M_\ast/M_\odot) + 11.37
\end{aligned}
\label{eq:7}
\end{equation}

\begin{equation}
\begin{aligned}
    A_{\rm V,stellar}
    &= 0.05\log(M_{\ast}/M_\odot)^2 \\
    &\quad - 0.63 \log(M_\ast/M_\odot) + 2.36
\end{aligned}
\label{eq:8}
\end{equation}

The potential redshift evolution of dust attenuation--$M_*$ relation has been a subject of long-standing debate. Numerous studies have demonstrated that the positive correlation between dust attenuation and $M_*$ remains largely invariant from $z \sim 0$ to $z \sim 2$. This stability has been observed in both $A_{V,\rm nebular}$ (e.g., \citealt{2013ApJ...763..145D, 2014ApJ...788...86P, 2022ApJ...926..145S}) and in $A_{V,\rm stellar}$ derived from SED fitting or infrared-to-UV star formation rate ratios (e.g., \citealt{1999ApJ...521...64M, 2016ApJ...833...72B, 2017ApJ...850..208W, 2018MNRAS.476.3991M}). Extending this to $z > 3$, recent \textit{JWST} results suggest that the $A_{V,\rm nebular}$--$M_*$ relation continues to show no significant evolution (e.g., \citealt{2023ApJ...954..157S, 2025arXiv251000235W, 2026arXiv260311338K}). The situation for $A_{V,\rm stellar}$, however, remains less certain: while some investigations suggest no evolution out to $z \sim 4$ (e.g., \citealt{2015ApJ...807..141P}), others find a trend toward reduced attenuation at fixed $M_*$ with increasing redshift (e.g., \citealt{2020MNRAS.491.4724F}), and still others report that such a decrease is confined to the high-mass regime (e.g., \citealt{2026arXiv260204765W}). In this work, we find that the galaxy dust attenuation--$M_*$ relation exhibits no significant evolution out to $z \sim 7$. However, it should be noted that the $M_\ast$ distributions of the SDSS and JADES samples differ substantially. Therefore, whether $A_{V,\mathrm{nebular}}$ and $A_{V,\mathrm{stellar}}$ evolve at fixed $M_\ast$ over a broader $M_\ast$ range still requires further investigation. In our sample, however, the high- and low-redshift galaxies lie on the same dust attenuation--$M_\ast$ sequence. More data are therefore needed to enable a more detailed analysis of the high-mass regime at high redshift.

Similarly, Figure~\ref{Fig:9} presents $A_{V,\rm diff}$ (left panel, defined in Equation \ref{eq:5}) and $\mathcal{R}$ (middle panel, defined in Equation \ref{eq:6}) as functions of $M_\ast$, with the SDSS and JADES data from all redshift bins combined. Since previous studies have primarily focused on $f$, we also present the results for $f$ (defined in Equation \ref{eq:1}) in the right panel. However, for clarity, we show only the SDSS results in the right panel. To place our results within the broader context of the literature, we have also compiled measurements from \cite{1997AJ....113..162C}, \cite{2013ApJ...777L...8K}, \citet{2014ApJ...788...86P}, \cite{2019PASJ...71....8K}, and \citet{2019ApJ...871..128T}, which are overlaid on the figure for comparison. As illustrated in Figure \ref{Fig:9}, for $A_{\rm V, diff}$, the SDSS and JADES samples collapse onto a single continuous sequence, with no systematic offset between the low-redshift and high-redshift data at fixed $M_\ast$. A consistent trend is also observed for $\mathcal{R}$ and $f$, although they exhibit substantially larger scatter. When compared with measurements from the literature, they fall on or very close to this combined sequence when evaluated at their corresponding $M_\ast$. The slight differences among results from different studies are expected, as they adopt different methodologies. This convergence of independent measurements supports the conclusion that $M_\ast$, rather than redshift, is the primary driver of the differential dust attenuation between the nebular and stellar components.

Combining the samples across all redshift bins reveals that the $A_{V,\rm diff}$--$M_*$ relation exhibits a characteristic two-regime behavior. At $\log(M_*/M_\odot) \lesssim 9$, $A_{V,\rm diff}$ remains roughly constant near zero, indicating that the nebular and stellar components are subject to comparable levels of dust obscuration in low-mass galaxies. Above this threshold, $A_{V,\rm diff}$ rises monotonically with $M_*$, reaching ${\sim}\,1$ mag at $\log(M_*/M_\odot) \sim 11$. This behavior is fully consistent with the right panel of Figure~\ref{Fig:4}: at the high-mass end, $A_{V,\rm nebular}$ increases more rapidly than $A_{V,\rm stellar}$, naturally producing the monotonic rise of $A_{V,\rm diff}$ with $M_*$. To provide a quantitative description of this trend, we fit the $A_{V,\rm diff}$--$M_*$ relation with a polynomial function following the method used in Figure~\ref{Fig:3}. The resulting best-fit relation is

\begin{equation}
\begin{aligned}
    A_{V,\rm diff}
    &= 0.15\log(M_{\ast}/M_\odot)^2 \\
    &\quad - 2.51 \log(M_\ast/M_\odot) + 10.58
\end{aligned}
\label{eq:9}
\end{equation}
which is shown as the black solid line in Figure~\ref{Fig:9}. 

The $\mathcal{R}$ (and $f$) factor exhibits a qualitatively similar behavior, remaining close to unity at low $M_\ast$, although with substantial scatter driven by the small absolute $E(B-V)$ values in this regime, and increasing (decreasing) monotonically toward higher masses, ultimately approaching the canonical \cite{1997AJ....113..162C} value of $\mathcal{R} = 2.27$ ($f=0.44$) at $\log(M_*/M_\odot) \sim 11$. We similarly fit the $\mathcal{R}$--$M_*$ and $f$--$M_\ast$ relation with a polynomial function. The resulting best-fit relation is
\begin{equation}
\begin{aligned}
    \mathcal{R}
    &= 0.13\log(M_{\ast}/M_\odot)^2 \\
    &\quad - 1.93 \log(M_\ast/M_\odot) + 7.96
\end{aligned}
\label{eq:10}
\end{equation}

\begin{equation}
\begin{aligned}
    f
    &= 0.05\log(M_{\ast}/M_\odot)^2 \\
    &\quad - 1.29 \log(M_\ast/M_\odot) + 8.00
\end{aligned}
\label{eq:11}
\end{equation}
which is overplotted as the black line in the middle and right panel of Figure~\ref{Fig:9}.

Since many prior studies focused primarily on the $f$ parameter, which exhibits substantial scatter in this work, we provide a complementary perspective on the mass dependence of $f$ in Figure~\ref{Fig:10}. This figure illustrates the relationship between the two components through a two-dimensional histogram of $E(B-V)_{\rm stellar}$ versus $E(B-V)_{\rm nebular}$. The color of each squares encodes the median $M_\ast$ of galaxies within that bin. Because the JADES sample is small in comparison to the SDSS sample, only the SDSS galaxies are shown here. For reference, the loci corresponding to $f = 1.0$ and 0.44 are overlaid as black dashed lines. As is evident from this figure, massive galaxies preferentially occupy the lower-right region of the plane, implying smaller $f$ values for high-mass systems. This trend demonstrates that $f$ is not a universal constant but is instead strongly coupled to the $M_\ast$ of the host galaxy, in agreement with the trend revealed in Figure~\ref{Fig:9}.

\section{Discussion} \label{sec:5}

\begin{figure*}[htbp]
    \centering
    \includegraphics[width=\linewidth]{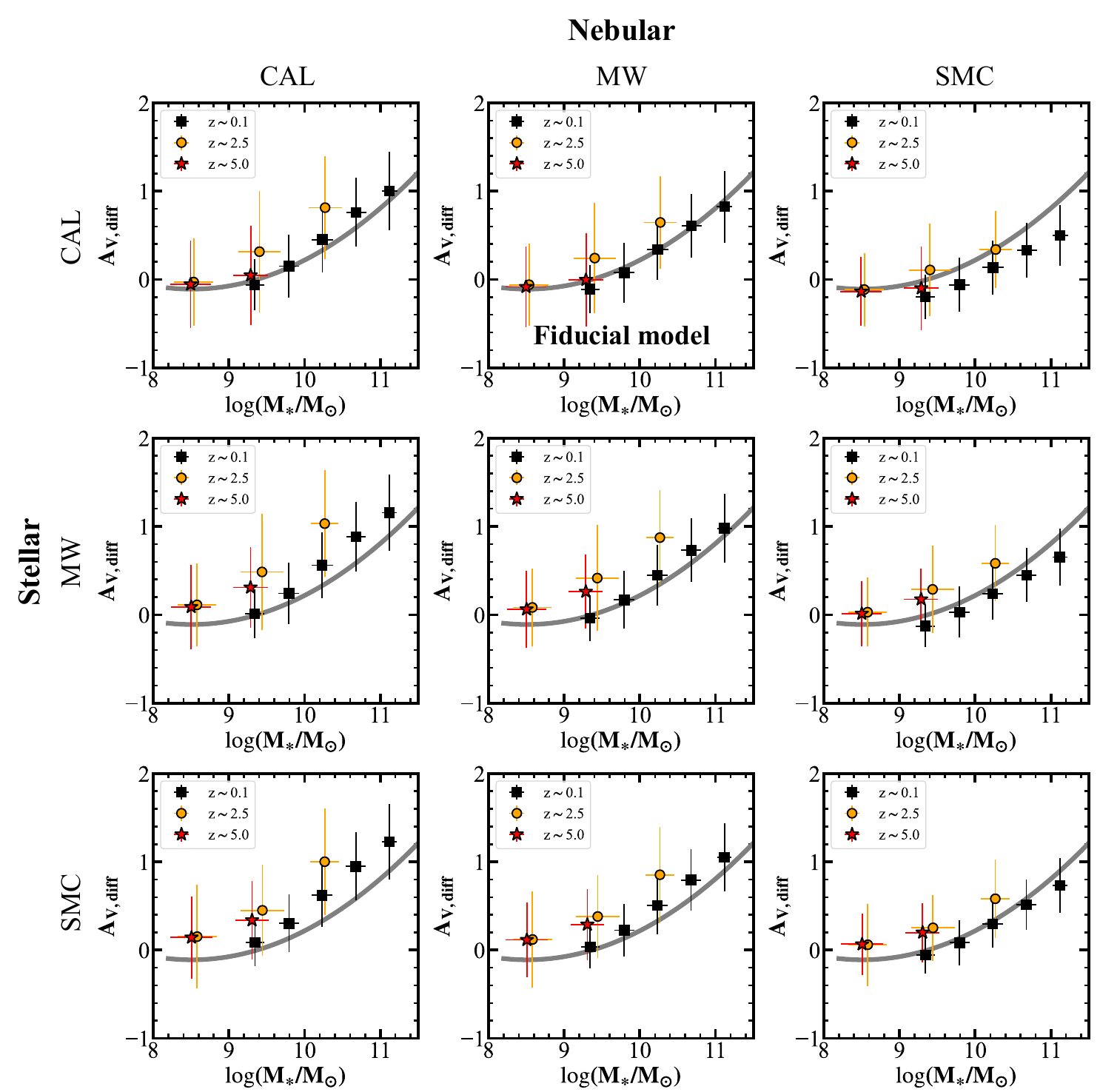}
    \caption{Similar to the left panel of Figure~\ref{Fig:9}, but showing the results obtained using different dust attenuation curves. The error bars here represent the $1 \sigma$ dispersion. The gray solid line is the best-fit result for the Fiducial model described in Equation \ref{eq:9}}
    \label{Fig:11}
\end{figure*}

\begin{figure*}[htbp]
    \centering
    \includegraphics[width=\linewidth]{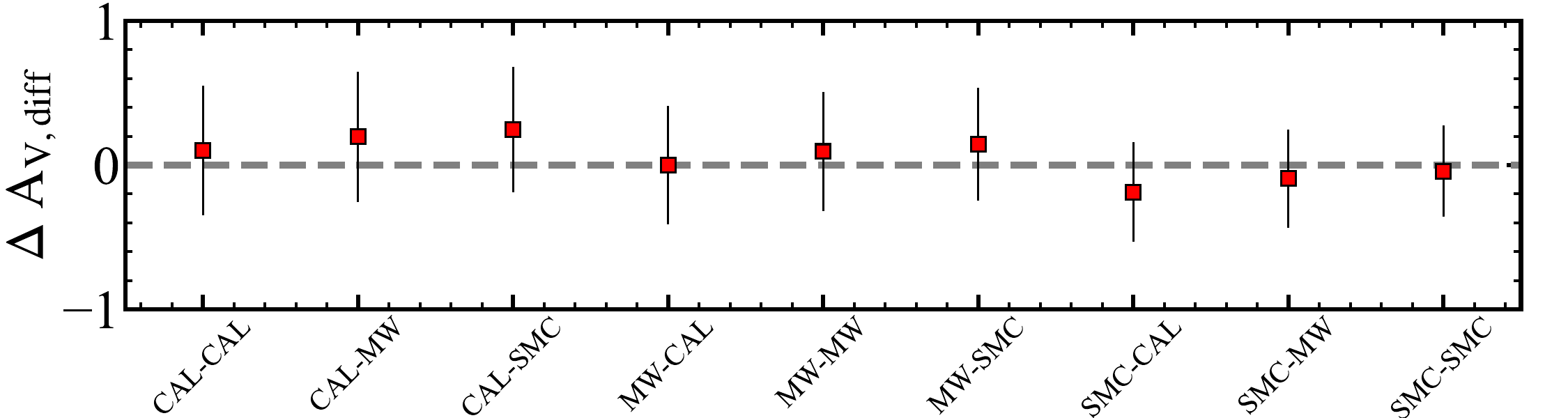}
    \caption{Average differences in $A_{V,\text{diff}}$ for various dust attenuation curve configurations relative to our fiducial model (MW-CAL). The error bars represent the corresponding $1\sigma$ dispersion. The horizontal axis indicates the specific combination of attenuation laws considered, denoted as the [nebular attenuation curve]--[stellar attenuation curve] combination.}
    \label{Fig:12}
\end{figure*}

\subsection{Results of different dust extinction curves} \label{sec:5.1}

In this work, we investigated the dependence of $A_{V,\mathrm{nebular}}$, $A_{V,\mathrm{stellar}}$, and their difference on $M_\ast$, as well as their possible redshift evolution. Compared with previous studies (e.g., \citealt{2015ApJ...806..259R, 2019PASJ...71....8K, 2025arXiv251000235W, 2026arXiv260311338K}), our work covers a broader redshift range and focuses on the evolution at fixed stellar mass, revealing a universal extinction relation that does not evolve with redshift. In the preceding sections, we adopted a CAL attenuation curve for $A_{V,\rm stellar}$ and a MW extinction curve for $A_{V,\rm nebular}$ as our fiducial configuration. Although this choice was justified in Section~\ref{sec:3.2}, numerous studies have demonstrated that the effective dust attenuation curve varies considerably among individual galaxies \citep[e.g.,][]{2018ApJ...859...11S, 2025MNRAS.539..109F, 2025NatAs...9..458M, 2026ApJ...999...15R}. To assess the sensitivity of our findings to this assumption, we repeated the analysis using all nine pairwise combinations of three widely adopted curves: CAL, MW, and SMC.

Before presenting the results, we note that systematic offsets between different configurations are expected by construction. Because the three attenuation curves differ significantly in the UV range, where the SMC curve is the steepest and the CAL curve is the shallowest, the $A_{V,\rm stellar}$ inferred from SED fitting is expected to be smallest when the SMC curve is assumed and largest when the CAL curve is assumed. For the nebular component, Equations \ref{eq:2} and \ref{eq:3} represented in Section~\ref{sec:3.2} imply $A_{V,\rm nebular,\,MW} = 0.90\,A_{V,\rm nebular,\,CAL} = 1.17\,A_{V,\rm nebular,\,SMC}$, indicating that the CAL curve yields the largest and the SMC curve the smallest $A_{V,\rm nebular}$ values. Consequently, different curve combinations produce systematic shifts in the absolute normalization of $A_{V,\rm diff}$. The relevant comparison for our analysis is therefore the relative trend with $M_\ast$ rather than the absolute level.

Figure~\ref{Fig:11} presents the $A_{V,\rm diff}$--$M_*$ relation for all nine combinations, arranged in a $3\times3$ grid in which rows correspond to the three assumed stellar attenuation curves and columns correspond to the three nebular extinction curves. For reference, the gray solid line in each panel reproduces our fiducial best-fit relation from Figure~\ref{Fig:10}. Although the absolute $A_{V,\rm diff}$ values shift systematically between panels in the manner described above, at fixed $M_\ast$, the median $A_{V,\rm diff}$ values across the three redshift bins remain consistent within the $1\sigma$ dispersion in every configuration. A positive correlation between $A_{V,\rm diff}$ and $M_*$ at the high-mass end is recovered under all nine configurations.

To quantify these systematic shifts, we computed the mean difference in $A_{V,\rm diff}$ between each alternative configuration and our fiducial setup, as shown in Figure~\ref{Fig:12}. The various combinations exhibit systematically positive or negative offsets that are entirely consistent with the expected differences in curve shapes. Notably, the absolute value of the mean offset remains within ${\sim}\,0.2$ mag for all nine pairings. This consistency confirms that the mass-dependent $A_{V,\rm diff}$--$M_*$ relation reported in Section~\ref{sec:4} is not an artifact of our specific choice of attenuation curves but represents a robust feature of the data.

\begin{figure*}[htbp]
    \centering
    \includegraphics[width=\linewidth]{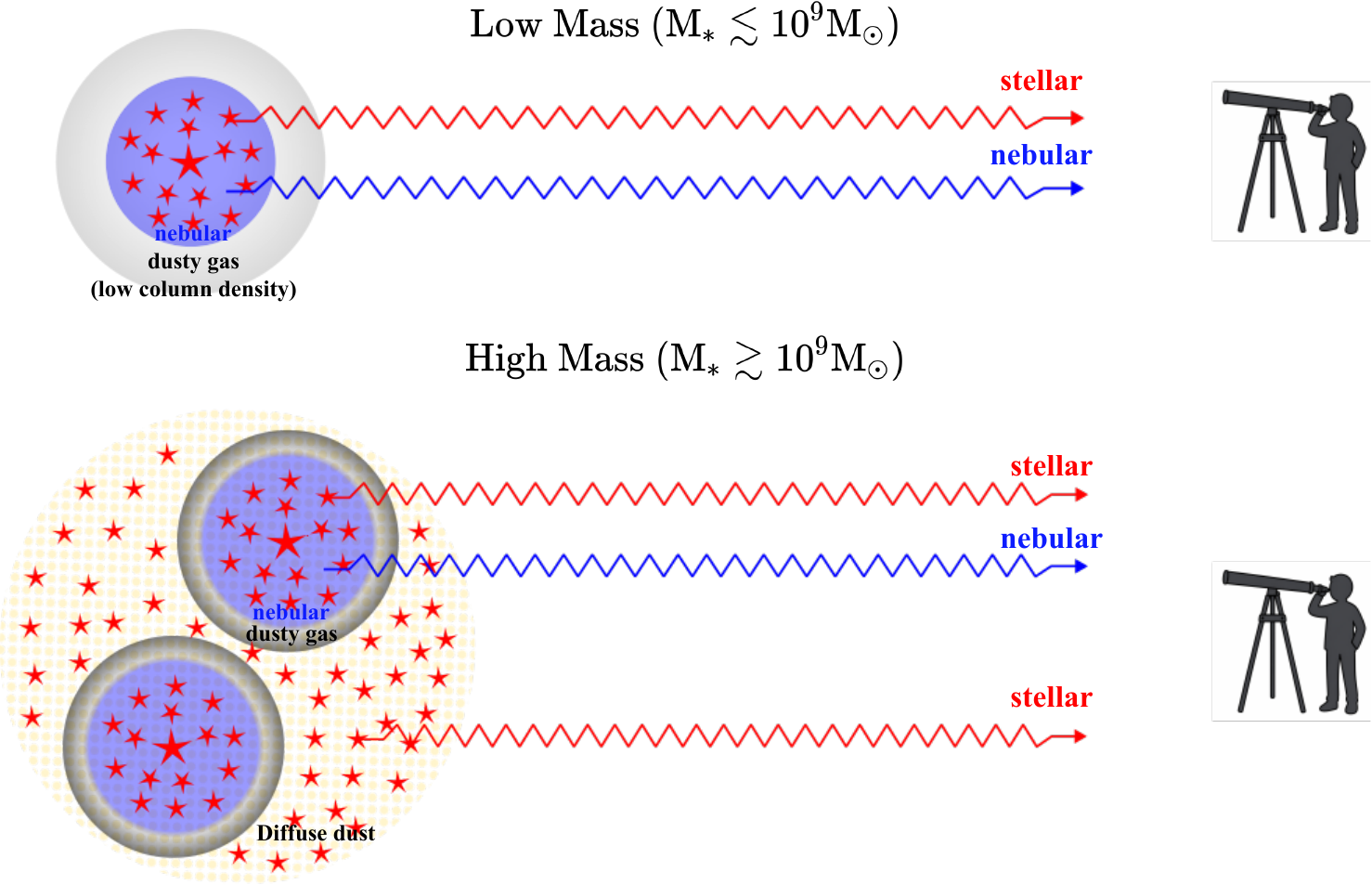}
    \caption{Schematic illustration of the two-component dust model for galaxies with low (top panel) and high (bottom panel) $M_\ast$. The gray region denotes birth cloud dust that has been dispersed by feedback processes. The light blue regions denote the nebular regions, while the light yellow regions represent the diffuse dust component within the ISM. Large red stars represent young, massive stars predominantly embedded within these birth clouds, whereas smaller red stars represent the less massive stellar populations. In low-mass galaxies, both the stellar continuum and nebular emission originate primarily from star-forming regions. Furthermore, due to the effects of stellar feedback, the birth clouds are blow apart and exhibit a more extended spatial distribution; consequently, the stellar and nebular light experience a similar degree of attenuation. In contrast, in high-mass galaxies, while nebular emission remains dominated by star-forming regions, a substantial fraction of the stellar continuum arises from older populations located outside the birth clouds. In these regions, the stellar light is primarily attenuated by the diffuse ISM. This geometric configuration leads to an increased discrepancy between $A_{V, \rm nebular}$ and $A_{V, \rm stellar}$ as a function of stellar mass.}
    \label{Fig:13}
\end{figure*}

\subsection{Physical Explanation} \label{sec:5.2}

Through the analyses presented in Section~\ref{sec:4}, we have established a coherent observational framework. At fixed $M_\ast$, both $A_{V,\rm stellar}$ and $A_{V,\rm nebular}$ exhibit no significant evolution from $z \sim 0.1$ to $z \sim 7$, and both correlate positively with $M_\ast$. These relations display a characteristic transition at $\log(M_*/M_\odot) \sim 9$: below this threshold, dust attenuation increases gradually with $M_*$, whereas above it, the slope steepens markedly. This steepening is more pronounced for $A_{V,\rm nebular}$ than for $A_{V,\rm stellar}$, resulting in an $A_{V,\rm diff}$ that is consistent with zero at $\log(M_*/M_\odot) \lesssim 9$ and increases monotonically toward higher masses. These features persist under all nine combinations of stellar and nebular attenuation curves tested in Section~\ref{sec:5.1}, indicating that they reflect intrinsic physical properties of the galaxy population rather than artifacts of methodological choices.

Because dust attenuation depends on the dust column density along each line of sight, it is governed jointly by the total dust content of a galaxy and its internal spatial distribution. The positive correlation between $A_V$ and $M_*$ follows naturally from the coupled mass--metallicity and mass--gas scaling relations \citep[e.g.,][]{2004ApJ...613..898T, 2015MNRAS.447.1610M, 2018ARA&A..56..673G, 2019MNRAS.490.1425L, 2023AJ....166..133D, 2023MNRAS.518..425C, 2023ApJ...950...84L, 2024ApJ...974..238L, 2026arXiv260118149K}: more massive galaxies host higher gas-phase metallicities, facilitating more efficient grain growth and increasing the dust-to-gas ratio (e.g., \citealt{2014A&A...563A..31R}). Simultaneously, the gas content also rises with $M_\ast$ \citep[e.g.,][]{2015MNRAS.447.1610M}, and these effects collectively lead to higher overall dust column densities in more massive systems.

The physical origin of the redshift invariance of dust attenuation at fixed $M_\ast$ remains a subtle and active topic of debate. One widely discussed interpretation is that the effective dust column density scales primarily with $M_\ast$ and is nearly independent of redshift \citep[e.g.,][]{2017ApJ...850..208W, 2022ApJ...926..145S}. Although prior studies have reported that the dust-to-stellar mass ratio in star-forming galaxies decreases with increasing redshift \citep[e.g.,][]{2022ApJ...928...68S, 2025A&A...693A.190J}, several compensating effects may offset this decline. For instance, star-forming galaxies are systematically more compact at higher redshifts at the fixed stellar mass \citep[e.g.,][]{2015ApJS..219...15S, 2024ApJ...963....9M, 2025arXiv251201684S, 2026arXiv260305045C}. Furthermore, some observations indicate that the dust-continuum sizes of high-redshift galaxies are even more concentrated than their rest-frame optical counterparts, suggesting a highly centrally concentrated dust distribution \citep[e.g.,][]{2017ApJ...850...83F, 2022MNRAS.510.3321P}. The resulting increase in dust surface density at fixed dust mass can partially counteract the reduction in the global dust-to-stellar mass ratio, thereby maintaining an effective line-of-sight column density that is approximately invariant over cosmic time. Additional factors, including a more clumpy dust geometry \citep[e.g.,][]{2000ApJ...528..799W, 2016ApJ...833..201S}, a non-unity dust covering fraction \citep[e.g.,][]{2026ApJ...999...15R}, and the potential evolution of grain size distributions \citep[e.g.,][]{2020MNRAS.491.3844A, 2023ApJ...951..100N, 2024A&A...689A..79M, 2026arXiv260406314P, 2026A&A...705A..75M}, may further modulate the integrated $A_V$ at fixed total dust content. Disentangling these various contributions requires higher-resolution multi-wavelength observations and is beyond the scope of the present work.

The transition near $\log(M_*/M_\odot)\sim9$, above which $A_{V,\mathrm{nebular}}$ rises more steeply with stellar mass, may mark the mass scale where both the dust budget and the effective dust column density begin to increase nonlinearly. For high redshift galaxies, this mass is close to the point at which the mass--metallicity relation crosses the metallicity threshold for efficient dust growth: galaxies with $M_\ast \sim10^9\,M_\odot$ typically reach $12+\log(\mathrm{O/H})\sim8.2$, corresponding to $\sim0.2$--$0.3\,Z_\odot$ (e.g., \citealt{2017MNRAS.471.3152P, 2023MNRAS.526.3610H, 2025OJAp....8E.153K}). This is comparable to the critical metallicity above which grain growth by accretion in the ISM becomes the dominant dust-production channel, leading to a rapid, order-of-magnitude increase in $M_{\rm dust}/M_*$ or $M_{\rm dust}/M_{\rm gas}$ (e.g., \citealt{2013EP&S...65..213A, 2017MNRAS.471.3152P, 2018MNRAS.478.4905A}). Below this threshold, inefficient ISM grain growth, together with strong feedback in shallow gravitational potentials, prevents dust from building up efficiently in star-forming regions, so $A_{V,\mathrm{nebular}}$ varies only weakly with $M_\ast$. However, for low-redshift galaxies, the metallicity at $M_\ast \sim 10^9 M_\odot$ is higher than $0.5Z_\odot$, already exceeding the critical metallicity (e.g., \citealt{2013ApJ...765..140A, 2020MNRAS.491..944C}). Nevertheless, we find the same result at low redshift, suggesting that the critical metallicity alone may not be sufficient to explain our results.

Additionally, structural evolution may also affect the observed dust attenuation. Several cosmological simulations indicate that, over the mass range around this threshold, galaxies with $M_\ast\gtrsim10^{8.5}\,M_\odot$ can enter a compact phase in which the stellar half-mass radius decreases with increasing $M_\ast$ (e.g., \citealt{2018MNRAS.473.4077P, 2022MNRAS.514.1921R, 2024MNRAS.534.1433S, 2026A&A...706A.125C})\footnote{In the simulations of \citet{2026A&A...706A.125C}, the half-mass radius starts to increase again at $M_\ast\gtrsim10^{9.5}\,M_\odot$, as normal inside-out growth is restored by continuous gas inflow; see also \citet{2017MNRAS.465..722F} and \citet{2023MNRAS.522.4515L}}. This compaction is driven by high central gas densities and subsequent overcooling, which trigger centrally concentrated, bursty star formation \citep[e.g.,][]{2022MNRAS.514.1921R, 2026A&A...706A.125C}. The resulting increase in gas and dust surface density enhances the column density toward \ion{H}{2} regions, naturally producing the steeper rise of $A_{V,\mathrm{nebular}}$ above the observed transition mass.


The mass dependence of $A_{V,\rm diff}$ and $\mathcal{R}$ admits a natural interpretation within the two-component dust framework. More specifically, it is set by the relative contributions of stellar continuum light from young and old populations, and is also shaped by the depth of the galaxy's gravitational potential well. This physical picture is illustrated schematically in Figure~\ref{Fig:13}.

Low-mass galaxies are characterized by higher sSFR \citep[e.g.,][]{2015A&A...581A..54T, 2018MNRAS.480.4842C}. As modeled by \cite{2004MNRAS.349..769K}, in such high-sSFR systems, the intense recent star formation ensures that young stars residing in nebular regions dominate the integrated continuum light. Because these components originate from spatially coincident regions, they probe similar dust column densities, naturally yielding $A_{V,\rm nebular} \approx A_{V,\rm stellar}$. A complementary effect operates in the same direction: feedback in low-mass galaxies repeatedly blows apart birth clouds (e.g., \citealt{2026A&A...708A.233L}). This process broadens the effective spatial distribution of birth-cloud material, causing both young and older stars to sample a mixture of birth-cloud and diffuse-ISM attenuation, again yielding $A_{V,\rm nebular} \approx A_{V,\rm stellar}$.

In contrast, in massive galaxies, the older stellar populations that contribute substantially to the integrated continuum have migrated out of their natal clouds into the diffuse ISM. Furthermore, the deeper gravitational potentials allow birth clouds to persist longer and locate in the star-forming regions. This configuration results in the classical two-component geometry: nebular emission suffers from both diffuse-ISM and birth-cloud attenuation, whereas a significant fraction of the stellar continuum is attenuated only by the diffuse ISM, thereby producing $A_{V,\rm nebular} > A_{V,\rm stellar}$ and $\mathcal{R} > 1$.

\section{Summary} \label{sec:6}

In this paper, we have presented a self-consistent census of nebular and stellar dust attenuation across cosmic time, utilizing a combined sample of 34,182 SDSS-GALEX star-forming galaxies at $0.05 < z < 0.1$ and 863 \textit{JWST}/JADES star-forming galaxies at $1.5 < z < 7$. Both samples were analyzed using a uniform \textsc{Bagpipes} SED-fitting framework to derive $A_{V, \rm stellar}$, complemented by Balmer-decrement-based $A_{V, \rm nebular}$ measurements. This homogenized treatment enables a direct comparison of the nebular-to-stellar dust attenuation across $0 < z < 7$. Our principal findings are summarized as follows:

(1) Both $A_{V, \rm nebular}$ and $A_{V, \rm stellar}$ correlate positively with $M_\ast$ across all epochs. At a fixed $M_\ast$, both $A_{V, \rm nebular}$ and $A_{V, \rm stellar}$ exhibit no significant evolution with redshift from $z \sim 7$ to $z \sim 0$, specifically, the mean dust attenuation values across the three redshift bins remain mutually consistent within the $1\sigma$ scatter.

(2) Both the $A_{V, \rm nebular}$--$M_*$ and $A_{V, \rm stellar}$--$M_*$ relations exhibit a characteristic two-regime behavior, with a transition occurring near $\log(M_*/M_\odot) \sim 9$. Below this threshold, both attenuation components increase gradually with $M_\ast$ and track each other closely. Above this mass, the relations steepen significantly. Additionally, $A_{V, \rm nebular}$ rises at a markedly higher rate than $A_{V, \rm stellar}$, leading to an increasing divergence between the two components in massive galaxies.

(3) Differential attenuation $A_{V, \rm diff}$ correlates positively with $M_\ast$ while exhibiting no statistically significant redshift evolution at a fixed $M_*$. At $\log(M_*/M_\odot) \lesssim 9$, $A_{V, \rm diff}$ remains consistent with zero, indicating that the stellar and nebular components are subject to comparable levels of dust attenuation in low-mass galaxies. Above this threshold, $A_{V, \rm diff}$ rises monotonically with $M_*$, reaching $\sim 1$~mag at $\log(M_*/M_\odot) \sim 11$. We demonstrate that this mass-dependent relation is robust within $\sim \pm 0.2$~mag across all nine combinations of the assumed stellar and nebular attenuation curves (CAL, MW, and SMC).

Taken together, these results demonstrate that the longstanding $f = 0.44$ value from \cite{1997AJ....113..162C} is not a universal constant, but is instead primarily determined by the $M_\ast$ of the galaxy. This relation extends smoothly from $f \sim 1$ at $\log(M_*/M_\odot) \lesssim 9$ to $f \sim 0.44$ at $\log(M_*/M_\odot) \sim 11$. The invariance of this relation from $z \sim 7$ to $z \sim 0$ suggests that the fundamental dust--star geometry underlying differential attenuation is established early in cosmic history and remains remarkably stable over the past 12~Gyr of galaxy evolution.

\begin{acknowledgments}
The authors thank Zesen Lin, Naveen A. Reddy, Hidenobu Yajima, and Hongxin Zhang for the valuable comments and discussions.

Funding for the Sloan Digital Sky Survey V has been provided by the Alfred P. Sloan Foundation, the Heising-Simons Foundation, the National Science Foundation, and the Participating Institutions. SDSS acknowledges support and resources from the Center for High-Performance Computing at the University of Utah. SDSS telescopes are located at Apache Point Observatory, funded by the Astrophysical Research Consortium and operated by New Mexico State University, and at Las Campanas Observatory, operated by the Carnegie Institution for Science. The SDSS web site is \url{www.sdss.org}. SDSS is managed by the Astrophysical Research Consortium for the Participating Institutions of the SDSS Collaboration, including the Carnegie Institution for Science, Chilean National Time Allocation Committee (CNTAC) ratified researchers, Caltech, the Gotham Participation Group, Harvard University, Heidelberg University, The Flatiron Institute, The Johns Hopkins University, L'Ecole polytechnique f\'{e}d\'{e}rale de Lausanne (EPFL), Leibniz-Institut f\"{u}r Astrophysik Potsdam (AIP), Max-Planck-Institut f\"{u}r Astronomie (MPIA Heidelberg), Max-Planck-Institut f\"{u}r Extraterrestrische Physik (MPE), Nanjing University, National Astronomical Observatories of China (NAOC), New Mexico State University, The Ohio State University, Pennsylvania State University, Smithsonian Astrophysical Observatory, Space Telescope Science Institute (STScI), the Stellar Astrophysics Participation Group, Universidad Nacional Aut\'{o}noma de M\'{e}xico, University of Arizona, University of Colorado Boulder, University of Illinois at Urbana-Champaign, University of Toronto, University of Utah, University of Virginia, Yale University, and Yunnan University.  

This research is based on observations made with the NASA Galaxy Evolution Explorer. GALEX is operated for NASA by the California Institute of Technology under NASA contract NAS5-98034.

This work is based on observations made with the NASA/ESA/CSA James Webb Space Telescope. The data were obtained from the Mikulski Archive for Space Telescopes at the Space Telescope Science Institute, which is operated by the Association of Universities for Research in Astronomy, Inc., under NASA contract NAS 5-03127 for JWST.  The authors acknowledge the JADES (PIs: Daniel J. Eisenstein, Nora Luetzgendorf, and Kate Isaak) for developing their observing program with a zero-exclusive-access period. All the JWST data used in this paper can be found in MAST: \url{https://dx.doi.org/10.17909/8tdj-8n28}. The English writing in this paper has been improved with the help of ChatGPT, while the software does not generate sentences from scratch.

This work was supported by Outstanding Doctoral Students Overseas Study Program of the University of Science and Technology of China. This publication is based upon work supported by the World Premier International Research Center Initiative (WPI Initiative), MEXT, Japan, and KAKENHI (25H00674) through the Japan Society for the Promotion of Science. This work was supported by the joint research program of the Institute for Cosmic Ray Research (ICRR), University of Tokyo. T.K. was supported by JSPS KAKENHI Grant Number 26KJ1232. This work was supported by the National Natural Science Foundation of China (NSFC; Grant No. 12233008), the National Key R\&D Program of China (No. 2023YFA1608100), and the Strategic Priority Research Program of the Chinese Academy of Sciences (No. XDB0550200). We also acknowledge support from the Cyrus Chun Ying Tang Foundation and the 111 Project for ``Observational and Theoretical Research on Dark Matter and Dark Energy'' (No. B23042). Y.N. acknowledges Flatiron Research Fellowhship. The Flatiron Institute is a division of the Simons Foundation.

\end{acknowledgments}

\bibliography{ref}{}

@ARTICLE{2000AJ....120.1579Y,
       author = {{York}, Donald G. and {Adelman}, J. and {Anderson}, Jr., John E. and {Anderson}, Scott F. and {Annis}, James and {Bahcall}, Neta A. and {Bakken}, J.~A. and {Barkhouser}, Robert and {Bastian}, Steven and {Berman}, Eileen and et al.},
        title = "{The Sloan Digital Sky Survey: Technical Summary}",
      journal = {\aj},
         year = 2000,
        month = sep,
       volume = {120},
       number = {3},
        pages = {1579-1587},
          doi = {10.1086/301513},
archivePrefix = {arXiv},
       eprint = {astro-ph/0006396},
 primaryClass = {astro-ph},
       adsurl = {https://ui.adsabs.harvard.edu/abs/2000AJ....120.1579Y}
}

@ARTICLE{2011ApJS..193...29A,
       author = {{Aihara}, Hiroaki and {Allende Prieto}, Carlos and {An}, Deokkeun and {Anderson}, Scott F. and {Aubourg}, {\'E}ric and {Balbinot}, Eduardo and {Beers}, Timothy C. and {Berlind}, Andreas A. and {Bickerton}, Steven J. and {Bizyaev}, Dmitry and et al.},
        title = "{The Eighth Data Release of the Sloan Digital Sky Survey: First Data from SDSS-III}",
      journal = {\apjs},
         year = 2011,
        month = apr,
       volume = {193},
       number = {2},
          eid = {29},
        pages = {29},
          doi = {10.1088/0067-0049/193/2/29},
archivePrefix = {arXiv},
       eprint = {1101.1559},
 primaryClass = {astro-ph.IM},
       adsurl = {https://ui.adsabs.harvard.edu/abs/2011ApJS..193...29A}
}

@ARTICLE{2026arXiv260401089B,
       author = {{Bakx}, Tom J.~L.~C.},
        title = "{ALMA Band 2 line survey of a $z = 3.44$ clumpy strongly-lensed submillimetre galaxy}",
      journal = {arXiv e-prints},
         year = 2026,
        month = apr,
          eid = {arXiv:2604.01089},
        pages = {arXiv:2604.01089},
          doi = {10.48550/arXiv.2604.01089},
archivePrefix = {arXiv},
       eprint = {2604.01089},
 primaryClass = {astro-ph.GA},
       adsurl = {https://ui.adsabs.harvard.edu/abs/2026arXiv260401089B}
}

@ARTICLE{2007ApJS..173..342M,
       author = {{Martin}, D. Christopher and {Wyder}, Ted K. and {Schiminovich}, David and {Barlow}, Tom A. and {Forster}, Karl and {Friedman}, Peter G. and {Morrissey}, Patrick and {Neff}, Susan G. and {Seibert}, Mark and {Small}, Todd and et al.},
        title = "{The UV-Optical Galaxy Color-Magnitude Diagram. III. Constraints on Evolution from the Blue to the Red Sequence}",
      journal = {\apjs},
         year = 2007,
        month = dec,
       volume = {173},
       number = {2},
        pages = {342-356},
          doi = {10.1086/516639},
archivePrefix = {arXiv},
       eprint = {astro-ph/0703281},
 primaryClass = {astro-ph},
       adsurl = {https://ui.adsabs.harvard.edu/abs/2007ApJS..173..342M}
}

@ARTICLE{2004MNRAS.351.1151B,
       author = {{Brinchmann}, J. and {Charlot}, S. and {White}, S.~D.~M. and {Tremonti}, C. and {Kauffmann}, G. and {Heckman}, T. and {Brinkmann}, J.},
        title = "{The physical properties of star-forming galaxies in the low-redshift Universe}",
      journal = {\mnras},
         year = 2004,
        month = jul,
       volume = {351},
       number = {4},
        pages = {1151-1179},
          doi = {10.1111/j.1365-2966.2004.07881.x},
archivePrefix = {arXiv},
       eprint = {astro-ph/0311060},
 primaryClass = {astro-ph},
       adsurl = {https://ui.adsabs.harvard.edu/abs/2004MNRAS.351.1151B}
}

@ARTICLE{2003MNRAS.341...33K,
       author = {{Kauffmann}, Guinevere and {Heckman}, Timothy M. and {White}, Simon D.~M. and {Charlot}, St{\'e}phane and {Tremonti}, Christy and {Brinchmann}, Jarle and {Bruzual}, Gustavo and {Peng}, Eric W. and {Seibert}, Mark and {Bernardi}, Mariangela and {Blanton}, Michael and {Brinkmann}, Jon and {Castander}, Francisco and {Cs{\'a}bai}, Istvan and {Fukugita}, Masataka and {Ivezic}, Zeljko and {Munn}, Jeffrey A. and {Nichol}, Robert C. and {Padmanabhan}, Nikhil and {Thakar}, Aniruddha R. and {Weinberg}, David H. and {York}, Donald},
        title = "{Stellar masses and star formation histories for {}10$^{5}$ galaxies from the Sloan Digital Sky Survey}",
      journal = {\mnras},
         year = 2003,
        month = may,
       volume = {341},
       number = {1},
        pages = {33-53},
          doi = {10.1046/j.1365-8711.2003.06291.x},
archivePrefix = {arXiv},
       eprint = {astro-ph/0204055},
 primaryClass = {astro-ph},
       adsurl = {https://ui.adsabs.harvard.edu/abs/2003MNRAS.341...33K}
}

@ARTICLE{2004ApJ...613..898T,
       author = {{Tremonti}, Christy A. and {Heckman}, Timothy M. and {Kauffmann}, Guinevere and {Brinchmann}, Jarle and {Charlot}, St{\'e}phane and {White}, Simon D.~M. and {Seibert}, Mark and {Peng}, Eric W. and {Schlegel}, David J. and {Uomoto}, Alan and {Fukugita}, Masataka and {Brinkmann}, Jon},
        title = "{The Origin of the Mass-Metallicity Relation: Insights from 53,000 Star-forming Galaxies in the Sloan Digital Sky Survey}",
      journal = {\apj},
         year = 2004,
        month = oct,
       volume = {613},
       number = {2},
        pages = {898-913},
          doi = {10.1086/423264},
archivePrefix = {arXiv},
       eprint = {astro-ph/0405537},
 primaryClass = {astro-ph},
       adsurl = {https://ui.adsabs.harvard.edu/abs/2004ApJ...613..898T}
}

@ARTICLE{2012MNRAS.422.3285P,
       author = {{Pforr}, Janine and {Maraston}, Claudia and {Tonini}, Chiara},
        title = "{Recovering galaxy stellar population properties from broad-band spectral energy distribution fitting}",
      journal = {\mnras},
         year = 2012,
        month = jun,
       volume = {422},
       number = {4},
        pages = {3285-3326},
          doi = {10.1111/j.1365-2966.2012.20848.x},
archivePrefix = {arXiv},
       eprint = {1203.3548},
 primaryClass = {astro-ph.CO},
       adsurl = {https://ui.adsabs.harvard.edu/abs/2012MNRAS.422.3285P}
}

@ARTICLE{2016ApJS..227....2S,
       author = {{Salim}, Samir and {Lee}, Janice C. and {Janowiecki}, Steven and {da Cunha}, Elisabete and {Dickinson}, Mark and {Boquien}, M{\'e}d{\'e}ric and {Burgarella}, Denis and {Salzer}, John J. and {Charlot}, St{\'e}phane},
        title = "{GALEX-SDSS-WISE Legacy Catalog (GSWLC): Star Formation Rates, Stellar Masses, and Dust Attenuations of 700,000 Low-redshift Galaxies}",
      journal = {\apjs},
         year = 2016,
        month = nov,
       volume = {227},
       number = {1},
          eid = {2},
        pages = {2},
          doi = {10.3847/0067-0049/227/1/2},
archivePrefix = {arXiv},
       eprint = {1610.00712},
 primaryClass = {astro-ph.GA},
       adsurl = {https://ui.adsabs.harvard.edu/abs/2016ApJS..227....2S}
}

@ARTICLE{2009ApJ...694.1281B,
       author = {{Budav{\'a}ri}, Tam{\'a}s and {Heinis}, S{\'e}bastien and {Szalay}, Alexander S. and {Nieto-Santisteban}, Mar{\'\i}a and {Gupchup}, Jayant and {Shiao}, Bernie and {Smith}, Myron and {Chang}, Ruixiang and {Kauffmann}, Guinevere and {Morrissey}, Patrick and et al.},
        title = "{GALEX-SDSS Catalogs for Statistical Studies}",
      journal = {\apj},
         year = 2009,
        month = apr,
       volume = {694},
       number = {2},
        pages = {1281-1292},
          doi = {10.1088/0004-637X/694/2/1281},
archivePrefix = {arXiv},
       eprint = {0904.1392},
 primaryClass = {astro-ph.CO},
       adsurl = {https://ui.adsabs.harvard.edu/abs/2009ApJ...694.1281B}
}

@ARTICLE{1981PASP...93....5B,
       author = {{Baldwin}, J.~A. and {Phillips}, M.~M. and {Terlevich}, R.},
        title = "{Classification parameters for the emission-line spectra of extragalactic objects.}",
      journal = {\pasp},
         year = 1981,
        month = feb,
       volume = {93},
        pages = {5-19},
          doi = {10.1086/130766},
       adsurl = {https://ui.adsabs.harvard.edu/abs/1981PASP...93....5B}
}

@ARTICLE{2003MNRAS.346.1055K,
       author = {{Kauffmann}, Guinevere and {Heckman}, Timothy M. and {Tremonti}, Christy and {Brinchmann}, Jarle and {Charlot}, St{\'e}phane and {White}, Simon D.~M. and {Ridgway}, Susan E. and {Brinkmann}, Jon and {Fukugita}, Masataka and {Hall}, Patrick B. and {Ivezi{\'c}}, {\v{Z}}eljko and {Richards}, Gordon T. and {Schneider}, Donald P.},
        title = "{The host galaxies of active galactic nuclei}",
      journal = {\mnras},
         year = 2003,
        month = dec,
       volume = {346},
       number = {4},
        pages = {1055-1077},
          doi = {10.1111/j.1365-2966.2003.07154.x},
archivePrefix = {arXiv},
       eprint = {astro-ph/0304239},
 primaryClass = {astro-ph},
       adsurl = {https://ui.adsabs.harvard.edu/abs/2003MNRAS.346.1055K}
}

@ARTICLE{2016MNRAS.458..184L,
       author = {{L{\'o}pez Fern{\'a}ndez}, R. and {Cid Fernandes}, R. and {Gonz{\'a}lez Delgado}, R.~M. and {Vale Asari}, N. and {P{\'e}rez}, E. and {Garc{\'\i}a-Benito}, R. and {de Amorim}, A.~L. and {Lacerda}, E.~A.~D. and {Cortijo-Ferrero}, C. and {S{\'a}nchez}, S.~F.},
        title = "{Simultaneous spectroscopic and photometric analysis of galaxies with STARLIGHT: CALIFA+GALEX}",
      journal = {\mnras},
         year = 2016,
        month = may,
       volume = {458},
       number = {1},
        pages = {184-199},
          doi = {10.1093/mnras/stw260},
archivePrefix = {arXiv},
       eprint = {1602.01123},
 primaryClass = {astro-ph.GA},
       adsurl = {https://ui.adsabs.harvard.edu/abs/2016MNRAS.458..184L}
}

@ARTICLE{2023arXiv230602465E,
       author = {{Eisenstein}, Daniel J. and {Willott}, Chris and {Alberts}, Stacey and {Arribas}, Santiago and {Bonaventura}, Nina and {Bunker}, Andrew J. and {Cameron}, Alex J. and {Carniani}, Stefano and {Charlot}, Stephane and {Curtis-Lake}, Emma and {D'Eugenio}, Francesco and {Ferruit}, Pierre and {Giardino}, Giovanna and {Hainline}, Kevin and {Hausen}, Ryan and {Jakobsen}, Peter and {Johnson}, Benjamin D. and {Maiolino}, Roberto and {Rauscher}, Bernard J. and {Rieke}, Marcia and {Rieke}, George and {Rix}, Hans-Walter and {Robertson}, Brant and {Stark}, Daniel P. and {Tacchella}, Sandro and {Williams}, Christina C. and {Willmer}, Christopher N.~A. and {Baker}, William M. and {Baum}, Stefi and {Bhatawdekar}, Rachana and {Boyett}, Kristan and {Chen}, Zuyi and {Chevallard}, Jacopo and {Circosta}, Chiara and {Curti}, Mirko and {Danhaive}, A. Lola and {DeCoursey}, Christa and {Endsley}, Ryan and {de Graaff}, Anna and {Dressler}, Alan and {Egami}, Eiichi and {Helton}, Jakob M. and {Hviding}, Raphael E. and {Ji}, Zhiyuan and {Jones}, Gareth C. and {Kumari}, Nimisha and {L{\"u}tzgendorf}, Nora and {Laseter}, Isaac and {Looser}, Tobias J. and {Lyu}, Jianwei and {Maseda}, Michael V. and {Nelson}, Erica and {Parlanti}, Eleonora and {Perna}, Michele and {Pusk{\'a}s}, D{\'a}vid and {Rawle}, Tim and {Rodr{\'\i}guez Del Pino}, Bruno and {Rujopakarn}, Wiphu and {Sandles}, Lester and {Saxena}, Aayush and {Scholtz}, Jan and {Sharpe}, Katherine and {Shivaei}, Irene and {Silcock}, Maddie S. and {Simmonds}, Charlotte and {Skarbinski}, Maya and {Smit}, Renske and {Stone}, Meredith and {Suess}, Katherine A. and {Sun}, Fengwu and {Tang}, Mengtao and {Topping}, Michael W. and {{\"U}bler}, Hannah and {Villanueva}, Natalia C. and {Wallace}, Imaan E.~B. and {Whitler}, Lily and {Witstok}, Joris and {Woodrum}, Charity},
        title = "{Overview of the JWST Advanced Deep Extragalactic Survey (JADES)}",
      journal = {arXiv e-prints},
         year = 2023,
        month = jun,
          eid = {arXiv:2306.02465},
        pages = {arXiv:2306.02465},
          doi = {10.48550/arXiv.2306.02465},
archivePrefix = {arXiv},
       eprint = {2306.02465},
 primaryClass = {astro-ph.GA},
       adsurl = {https://ui.adsabs.harvard.edu/abs/2023arXiv230602465E}
}

@ARTICLE{2024A&A...690A.288B,
       author = {{Bunker}, Andrew J. and {Cameron}, Alex J. and {Curtis-Lake}, Emma and {Jakobsen}, Peter and {Carniani}, Stefano and {Curti}, Mirko and {Witstok}, Joris and {Maiolino}, Roberto and {D'Eugenio}, Francesco and {Looser}, Tobias J. and {Willott}, Chris and {Bonaventura}, Nina and {Hainline}, Kevin and {{\"U}bler}, Hannah and {Willmer}, Christopher N.~A. and {Saxena}, Aayush and {Smit}, Renske and {Alberts}, Stacey and {Arribas}, Santiago and {Baker}, William M. and {Baum}, Stefi and {Bhatawdekar}, Rachana and {Bowler}, Rebecca A.~A. and {Boyett}, Kristan and {Charlot}, Stephane and {Chen}, Zuyi and {Chevallard}, Jacopo and {Circosta}, Chiara and {DeCoursey}, Christa and {de Graaff}, Anna and {Egami}, Eiichi and {Eisenstein}, Daniel J. and {Endsley}, Ryan and {Ferruit}, Pierre and {Giardino}, Giovanna and {Hausen}, Ryan and {Helton}, Jakob M. and {Hviding}, Raphael E. and {Ji}, Zhiyuan and {Johnson}, Benjamin D. and {Jones}, Gareth C. and {Kumari}, Nimisha and {Laseter}, Isaac and {L{\"u}tzgendorf}, Nora and {Maseda}, Michael V. and {Nelson}, Erica and {Parlanti}, Eleonora and {Perna}, Michele and {Rauscher}, Bernard J. and {Rawle}, Tim and {Rix}, Hans-Walter and {Rieke}, Marcia and {Robertson}, Brant and {Rodr{\'\i}guez Del Pino}, Bruno and {Sandles}, Lester and {Scholtz}, Jan and {Sharpe}, Katherine and {Skarbinski}, Maya and {Stark}, Daniel P. and {Sun}, Fengwu and {Tacchella}, Sandro and {Topping}, Michael W. and {Villanueva}, Natalia C. and {Wallace}, Imaan E.~B. and {Williams}, Christina C. and {Woodrum}, Charity},
        title = "{JADES NIRSpec initial data release for the Hubble Ultra Deep Field: Redshifts and line fluxes of distant galaxies from the deepest JWST Cycle 1 NIRSpec multi-object spectroscopy}",
      journal = {\aap},
         year = 2024,
        month = oct,
       volume = {690},
          eid = {A288},
        pages = {A288},
          doi = {10.1051/0004-6361/202347094},
archivePrefix = {arXiv},
       eprint = {2306.02467},
 primaryClass = {astro-ph.GA},
       adsurl = {https://ui.adsabs.harvard.edu/abs/2024A&A...690A.288B}
}

@ARTICLE{2024ApJ...964...71H,
       author = {{Hainline}, Kevin N. and {Johnson}, Benjamin D. and {Robertson}, Brant and {Tacchella}, Sandro and {Helton}, Jakob M. and {Sun}, Fengwu and {Eisenstein}, Daniel J. and {Simmonds}, Charlotte and {Topping}, Michael W. and {Whitler}, Lily and {Willmer}, Christopher N.~A. and {Rieke}, Marcia and {Suess}, Katherine A. and {Hviding}, Raphael E. and {Cameron}, Alex J. and {Alberts}, Stacey and {Baker}, William M. and {Baum}, Stefi and {Bhatawdekar}, Rachana and {Bonaventura}, Nina and {Boyett}, Kristan and {Bunker}, Andrew J. and {Carniani}, Stefano and {Charlot}, Stephane and {Chevallard}, Jacopo and {Chen}, Zuyi and {Curti}, Mirko and {Curtis-Lake}, Emma and {D'Eugenio}, Francesco and {Egami}, Eiichi and {Endsley}, Ryan and {Hausen}, Ryan and {Ji}, Zhiyuan and {Looser}, Tobias J. and {Lyu}, Jianwei and {Maiolino}, Roberto and {Nelson}, Erica and {Pusk{\'a}s}, D{\'a}vid and {Rawle}, Tim and {Sandles}, Lester and {Saxena}, Aayush and {Smit}, Renske and {Stark}, Daniel P. and {Williams}, Christina C. and {Willott}, Chris and {Witstok}, Joris},
        title = "{The Cosmos in Its Infancy: JADES Galaxy Candidates at z > 8 in GOODS-S and GOODS-N}",
      journal = {\apj},
         year = 2024,
        month = mar,
       volume = {964},
       number = {1},
          eid = {71},
        pages = {71},
          doi = {10.3847/1538-4357/ad1ee4},
archivePrefix = {arXiv},
       eprint = {2306.02468},
 primaryClass = {astro-ph.GA},
       adsurl = {https://ui.adsabs.harvard.edu/abs/2024ApJ...964...71H}
}

@ARTICLE{2023ApJS..269...16R,
       author = {{Rieke}, Marcia J. and {Robertson}, Brant and {Tacchella}, Sandro and {Hainline}, Kevin and {Johnson}, Benjamin D. and {Hausen}, Ryan and {Ji}, Zhiyuan and {Willmer}, Christopher N.~A. and {Eisenstein}, Daniel J. and {Pusk{\'a}s}, D{\'a}vid and {Alberts}, Stacey and {Arribas}, Santiago and {Baker}, William M. and {Baum}, Stefi and {Bhatawdekar}, Rachana and {Bonaventura}, Nina and {Boyett}, Kristan and {Bunker}, Andrew J. and {Cameron}, Alex J. and {Carniani}, Stefano and {Charlot}, Stephane and {Chevallard}, Jacopo and {Chen}, Zuyi and {Curti}, Mirko and {Curtis-Lake}, Emma and {Danhaive}, A. Lola and {DeCoursey}, Christa and {Dressler}, Alan and {Egami}, Eiichi and {Endsley}, Ryan and {Helton}, Jakob M. and {Hviding}, Raphael E. and {Kumari}, Nimisha and {Looser}, Tobias J. and {Lyu}, Jianwei and {Maiolino}, Roberto and {Maseda}, Michael V. and {Nelson}, Erica J. and {Rieke}, George and {Rix}, Hans-Walter and {Sandles}, Lester and {Saxena}, Aayush and {Sharpe}, Katherine and {Shivaei}, Irene and {Skarbinski}, Maya and {Smit}, Renske and {Stark}, Daniel P. and {Stone}, Meredith and {Suess}, Katherine A. and {Sun}, Fengwu and {Topping}, Michael and {{\"U}bler}, Hannah and {Villanueva}, Natalia C. and {Wallace}, Imaan E.~B. and {Williams}, Christina C. and {Willott}, Chris and {Whitler}, Lily and {Witstok}, Joris and {Woodrum}, Charity},
        title = "{JADES Initial Data Release for the Hubble Ultra Deep Field: Revealing the Faint Infrared Sky with Deep JWST NIRCam Imaging}",
      journal = {\apjs},
         year = 2023,
        month = nov,
       volume = {269},
       number = {1},
          eid = {16},
        pages = {16},
          doi = {10.3847/1538-4365/acf44d},
archivePrefix = {arXiv},
       eprint = {2306.02466},
 primaryClass = {astro-ph.GA},
       adsurl = {https://ui.adsabs.harvard.edu/abs/2023ApJS..269...16R}
}

@ARTICLE{2025ApJS..281...50E,
       author = {{Eisenstein}, Daniel J. and {Johnson}, Benjamin D. and {Robertson}, Brant and {Tacchella}, Sandro and {Hainline}, Kevin and {Jakobsen}, Peter and {Maiolino}, Roberto and {Bonaventura}, Nina and {Bunker}, Andrew J. and {Cameron}, Alex J. and {Cargile}, Phillip A. and {Curtis-Lake}, Emma and {Hausen}, Ryan and {Pusk{\'a}s}, D{\'a}vid and {Rieke}, Marcia and {Sun}, Fengwu and {Willmer}, Christopher N.~A. and {Willott}, Chris and {Alberts}, Stacey and {Arribas}, Santiago and {Baker}, William M. and {Baum}, Stefi and {Bhatawdekar}, Rachana and {Carniani}, Stefano and {Charlot}, Stephane and {Chen}, Zuyi and {Chevallard}, Jacopo and {Curti}, Mirko and {DeCoursey}, Christa and {D'Eugenio}, Francesco and {de Graaff}, Anna and {Egami}, Eiichi and {Helton}, Jakob M. and {Ji}, Zhiyuan and {Jones}, Gareth C. and {Kumari}, Nimisha and {L{\"u}tzgendorf}, Nora and {Laseter}, Isaac and {Looser}, Tobias J. and {Lyu}, Jianwei and {Maseda}, Michael V. and {Nelson}, Erica and {Parlanti}, Eleonora and {Rauscher}, Bernard J. and {Rawle}, Tim and {Rieke}, George and {Rix}, Hans-Walter and {Rujopakarn}, Wiphu and {Sandles}, Lester and {Saxena}, Aayush and {Scholtz}, Jan and {Sharpe}, Katherine and {Shivaei}, Irene and {Simmonds}, Charlotte and {Smit}, Renske and {Topping}, Michael W. and {{\"U}bler}, Hannah and {Venturi}, Giacomo and {Williams}, Christina C. and {Witstok}, Joris and {Woodrum}, Charity},
        title = "{The JADES Origins Field: A New JWST Deep Field in the JADES Second NIRCam Data Release}",
      journal = {\apjs},
         year = 2025,
        month = dec,
       volume = {281},
       number = {2},
          eid = {50},
        pages = {50},
          doi = {10.3847/1538-4365/ae1137},
archivePrefix = {arXiv},
       eprint = {2310.12340},
 primaryClass = {astro-ph.GA},
       adsurl = {https://ui.adsabs.harvard.edu/abs/2025ApJS..281...50E}
}

@ARTICLE{2025ApJS..277....4D,
       author = {{D'Eugenio}, Francesco and {Cameron}, Alex J. and {Scholtz}, Jan and {Carniani}, Stefano and {Willott}, Chris J. and {Curtis-Lake}, Emma and {Bunker}, Andrew J. and {Parlanti}, Eleonora and {Maiolino}, Roberto and {Willmer}, Christopher N.~A. and {Jakobsen}, Peter and {Robertson}, Brant E. and {Johnson}, Benjamin D. and {Tacchella}, Sandro and {Cargile}, Phillip A. and {Rawle}, Tim and {Arribas}, Santiago and {Chevallard}, Jacopo and {Curti}, Mirko and {Egami}, Eiichi and {Eisenstein}, Daniel J. and {Kumari}, Nimisha and {Looser}, Tobias J. and {Rieke}, Marcia J. and {Rodr{\'\i}guez Del Pino}, Bruno and {Saxena}, Aayush and {{\"U}bler}, Hannah and {Venturi}, Giacomo and {Witstok}, Joris and {Baker}, William M. and {Bhatawdekar}, Rachana and {Bonaventura}, Nina and {Boyett}, Kristan and {Charlot}, Stephane and {Danhaive}, A. Lola and {Hainline}, Kevin N. and {Hausen}, Ryan and {Helton}, Jakob M. and {Ji}, Xihan and {Ji}, Zhiyuan and {Jones}, Gareth C. and {Juod{\v{z}}balis}, Ignas and {Maseda}, Michael V. and {P{\'e}rez-Gonz{\'a}lez}, Pablo G. and {Perna}, Michele and {Pusk{\'a}s}, D{\'a}vid and {Shivaei}, Irene and {Silcock}, Maddie S. and {Simmonds}, Charlotte and {Smit}, Renske and {Sun}, Fengwu and {Villanueva}, Natalia C. and {Williams}, Christina C. and {Zhu}, Yongda},
        title = "{JADES Data Release 3: NIRSpec/Microshutter Assembly Spectroscopy for 4000 Galaxies in the GOODS Fields}",
      journal = {\apjs},
         year = 2025,
        month = mar,
       volume = {277},
       number = {1},
          eid = {4},
        pages = {4},
          doi = {10.3847/1538-4365/ada148},
archivePrefix = {arXiv},
       eprint = {2404.06531},
 primaryClass = {astro-ph.GA},
       adsurl = {https://ui.adsabs.harvard.edu/abs/2025ApJS..277....4D}
}

@ARTICLE{2025arXiv251001033C,
       author = {{Curtis-Lake}, Emma and {Cameron}, Alex J. and {Bunker}, Andrew J. and {Scholtz}, Jan and {Carniani}, Stefano and {Parlanti}, Eleonora and {D'Eugenio}, Francesco and {Jakobsen}, Peter and {Willmer}, Christopher N.~A. and {Arribas}, Santiago and {Baker}, William M. and {Charlot}, St{\'e}phane and {Chevallard}, Jacopo and {Circosta}, Chiara and {Curti}, Mirko and {Eisenstein}, Daniel J. and {Hainline}, Kevin and {Ji}, Zhiyuan and {Johnson}, Benjamin D. and {Jones}, Gareth C. and {Maiolino}, Roberto and {Maseda}, Michael V. and {P{\'e}rez-Gonz{\'a}lez}, Pablo G. and {Rawle}, Tim and {Rieke}, Marcia and {Rinaldi}, Pierluigi and {Robertson}, Brant and {Rodr{\'\i}gez Del Pino}, Bruno and {Saxena}, Aayush and {Shivaei}, Irene and {Smit}, Renske and {Tacchella}, Sandro and {{\"U}bler}, Hannah and {Venturi}, Giacomo and {Williams}, Christina C. and {Willott}, Chris and {Duan}, Qiao},
        title = "{JADES Data Release 4 Paper I: Sample Selection, Observing Strategy and Redshifts of the complete spectroscopic sample}",
      journal = {arXiv e-prints},
         year = 2025,
        month = oct,
          eid = {arXiv:2510.01033},
        pages = {arXiv:2510.01033},
          doi = {10.48550/arXiv.2510.01033},
archivePrefix = {arXiv},
       eprint = {2510.01033},
 primaryClass = {astro-ph.GA},
       adsurl = {https://ui.adsabs.harvard.edu/abs/2025arXiv251001033C}
}

@ARTICLE{2025arXiv251001034S,
       author = {{Scholtz}, J. and {Carniani}, S. and {Parlanti}, E. and {D'Eugenio}, F. and {Curtis-Lake}, E. and {Jakobsen}, P. and {Bunker}, A.~J. and {Cameron}, A.~J. and {Arribas}, S. and {Baker}, W.~M. and {Charlot}, S. and {Chevellard}, J. and {Circosta}, C. and {Curti}, M. and {Duan}, Q. and {Eisenstein}, D.~J. and {Hainline}, K. and {Ji}, Z. and {Johnson}, B.~D. and {Jones}, G.~C. and {Kumari}, N. and {Maiolino}, R. and {Maseda}, M.~V. and {Perna}, M. and {P{\'e}rez-Gonz{\'a}lez}, P.~G. and {Rawle}, T. and {Rieke}, M. and {Rinaldi}, P. and {Robertson}, B. and {Saxena}, A. and {Shivaei}, I. and {Silcock}, M.~S. and {Sun}, Y. and {Rodr{\'\i}guez Del Pino}, B. and {Tacchella}, S. and {{\"U}bler}, H. and {Venturi}, G. and {Williams}, C.~C. and {Willmer}, C.~N.~A. and {Willott}, C. and {Witstok}, J.},
        title = "{JADES Data Release 4 -- Paper II: Data reduction, analysis and emission-line fluxes of the complete spectroscopic sample}",
      journal = {arXiv e-prints},
         year = 2025,
        month = oct,
          eid = {arXiv:2510.01034},
        pages = {arXiv:2510.01034},
          doi = {10.48550/arXiv.2510.01034},
archivePrefix = {arXiv},
       eprint = {2510.01034},
 primaryClass = {astro-ph.GA},
       adsurl = {https://ui.adsabs.harvard.edu/abs/2025arXiv251001034S}
}

@ARTICLE{2026arXiv260115956R,
       author = {{Robertson}, Brant E. and {Johnson}, Benjamin D. and {Tacchella}, Sandro and {Eisenstein}, Daniel J. and {Hainline}, Kevin and {Alberts}, Stacey and {Arribas}, Santiago and {Baker}, William M. and {Bunker}, Andrew J. and {Cameron}, Alex J. and {Carniani}, Stefano and {Carreira}, Courtney and {Chevallard}, Jacopo and {Circosta}, Chiara and {Curtis-Lake}, Emma and {Danhaive}, A. Lola and {Duan}, Qiao and {Egami}, Eiichi and {Hausen}, Ryan and {Helton}, Jakob M. and {Ji}, Zhiyuan and {Maiolino}, Roberto and {P{\'e}rez-Gonz{\'a}lez}, Pablo G. and {Pusk{\'a}s}, D{\'a}vid and {Rieke}, Marcia and {Rinaldi}, Pierluigi and {Sun}, Fengwu and {Sun}, Yang and {{\"U}bler}, Hannah and {Trussler}, James A.~A. and {Villanueva}, Natalia C. and {Whitler}, Lily and {Williams}, Christina C. and {Willmer}, Christopher N.~A. and {Willott}, Chris and {Wu}, Zihao and {Zhu}, Yongda},
        title = "{JWST Advanced Deep Extragalactic Survey (JADES) Data Release 5: Photometric Catalog}",
      journal = {arXiv e-prints},
         year = 2026,
        month = jan,
          eid = {arXiv:2601.15956},
        pages = {arXiv:2601.15956},
          doi = {10.48550/arXiv.2601.15956},
archivePrefix = {arXiv},
       eprint = {2601.15956},
 primaryClass = {astro-ph.GA},
       adsurl = {https://ui.adsabs.harvard.edu/abs/2026arXiv260115956R}
}

@ARTICLE{2022A&A...661A..80J,
       author = {{Jakobsen}, P. and {Ferruit}, P. and {Alves de Oliveira}, C. and {Arribas}, S. and {Bagnasco}, G. and {Barho}, R. and {Beck}, T.~L. and {Birkmann}, S. and {B{\"o}ker}, T. and {Bunker}, A.~J. and et al.},
        title = "{The Near-Infrared Spectrograph (NIRSpec) on the James Webb Space Telescope. I. Overview of the instrument and its capabilities}",
      journal = {\aap},
         year = 2022,
        month = may,
       volume = {661},
          eid = {A80},
        pages = {A80},
          doi = {10.1051/0004-6361/202142663},
archivePrefix = {arXiv},
       eprint = {2202.03305},
 primaryClass = {astro-ph.IM},
       adsurl = {https://ui.adsabs.harvard.edu/abs/2022A&A...661A..80J}
}

@ARTICLE{2022A&A...661A..81F,
       author = {{Ferruit}, P. and {Jakobsen}, P. and {Giardino}, G. and {Rawle}, T. and {Alves de Oliveira}, C. and {Arribas}, S. and {Beck}, T.~L. and {Birkmann}, S. and {B{\"o}ker}, T. and {Bunker}, A.~J. and et al.},
        title = "{The Near-Infrared Spectrograph (NIRSpec) on the James Webb Space Telescope. II. Multi-object spectroscopy (MOS)}",
      journal = {\aap},
         year = 2022,
        month = may,
       volume = {661},
          eid = {A81},
        pages = {A81},
          doi = {10.1051/0004-6361/202142673},
archivePrefix = {arXiv},
       eprint = {2202.03306},
 primaryClass = {astro-ph.IM},
       adsurl = {https://ui.adsabs.harvard.edu/abs/2022A&A...661A..81F}
}

@ARTICLE{2023PASP..135c8001B,
       author = {{B{\"o}ker}, T. and {Beck}, T.~L. and {Birkmann}, S.~M. and {Giardino}, G. and {Keyes}, C. and {Kumari}, N. and {Muzerolle}, J. and {Rawle}, T. and {Zeidler}, P. and {Abul-Huda}, Y. and et al.},
        title = "{In-orbit Performance of the Near-infrared Spectrograph NIRSpec on the James Webb Space Telescope}",
      journal = {\pasp},
         year = 2023,
        month = mar,
       volume = {135},
       number = {1045},
          eid = {038001},
        pages = {038001},
          doi = {10.1088/1538-3873/acb846},
archivePrefix = {arXiv},
       eprint = {2301.13766},
 primaryClass = {astro-ph.IM},
       adsurl = {https://ui.adsabs.harvard.edu/abs/2023PASP..135c8001B}
}

@ARTICLE{2025arXiv250403551J,
       author = {{Juod{\v{z}}balis}, Ignas and {Maiolino}, Roberto and {Baker}, William M. and {Lake}, Emma Curtis and {Scholtz}, Jan and {D'Eugenio}, Francesco and {Trefoloni}, Bartolomeo and {Isobe}, Yuki and {Tacchella}, Sandro and {Bunker}, Andrew J. and et al.},
        title = "{JADES: comprehensive census of broad-line AGN from Reionization to Cosmic Noon revealed by JWST}",
      journal = {arXiv e-prints},
         year = 2025,
        month = apr,
          eid = {arXiv:2504.03551},
        pages = {arXiv:2504.03551},
          doi = {10.48550/arXiv.2504.03551},
archivePrefix = {arXiv},
       eprint = {2504.03551},
 primaryClass = {astro-ph.GA},
       adsurl = {https://ui.adsabs.harvard.edu/abs/2025arXiv250403551J}
}

@ARTICLE{2025A&A...697A.175S,
       author = {{Scholtz}, Jan and {Maiolino}, Roberto and {D'Eugenio}, Francesco and {Curtis-Lake}, Emma and {Carniani}, Stefano and {Charlot}, Stephane and {Curti}, Mirko and {Silcock}, Maddie S. and {Arribas}, Santiago and {Baker}, William and et al.},
        title = "{JADES: A large population of obscured, narrow-line active galactic nuclei at high redshift}",
      journal = {\aap},
         year = 2025,
        month = may,
       volume = {697},
          eid = {A175},
        pages = {A175},
          doi = {10.1051/0004-6361/202348804},
archivePrefix = {arXiv},
       eprint = {2311.18731},
 primaryClass = {astro-ph.GA},
       adsurl = {https://ui.adsabs.harvard.edu/abs/2025A&A...697A.175S}
}

@ARTICLE{2022ApJ...941..191L,
       author = {{Lyu}, Jianwei and {Alberts}, Stacey and {Rieke}, George H. and {Rujopakarn}, Wiphu},
        title = "{AGN Selection and Demographics in GOODS-S/HUDF from X-Ray to Radio}",
      journal = {\apj},
         year = 2022,
        month = dec,
       volume = {941},
       number = {2},
          eid = {191},
        pages = {191},
          doi = {10.3847/1538-4357/ac9e5d},
archivePrefix = {arXiv},
       eprint = {2209.06219},
 primaryClass = {astro-ph.GA},
       adsurl = {https://ui.adsabs.harvard.edu/abs/2022ApJ...941..191L}
}

@ARTICLE{2016ApJS..224...15X,
       author = {{Xue}, Y.~Q. and {Luo}, B. and {Brandt}, W.~N. and {Alexander}, D.~M. and {Bauer}, F.~E. and {Lehmer}, B.~D. and {Yang}, G.},
        title = "{The 2 Ms Chandra Deep Field-North Survey and the 250 ks Extended Chandra Deep Field-South Survey: Improved Point-source Catalogs}",
      journal = {\apjs},
         year = 2016,
        month = jun,
       volume = {224},
       number = {2},
          eid = {15},
        pages = {15},
          doi = {10.3847/0067-0049/224/2/15},
archivePrefix = {arXiv},
       eprint = {1602.06299},
 primaryClass = {astro-ph.GA},
       adsurl = {https://ui.adsabs.harvard.edu/abs/2016ApJS..224...15X}
}

@ARTICLE{2019MNRAS.483.2382W,
       author = {{Werle}, A. and {Cid Fernandes}, R. and {Vale Asari}, N. and {Bruzual}, G. and {Charlot}, S. and {Gonzalez Delgado}, R. and {Herpich}, F.~R.},
        title = "{Simultaneous analysis of SDSS spectra and GALEX photometry with STARLIGHT: method and early results}",
      journal = {\mnras},
         year = 2019,
        month = feb,
       volume = {483},
       number = {2},
        pages = {2382-2397},
          doi = {10.1093/mnras/sty3264},
archivePrefix = {arXiv},
       eprint = {1811.11255},
 primaryClass = {astro-ph.GA},
       adsurl = {https://ui.adsabs.harvard.edu/abs/2019MNRAS.483.2382W}
}

@ARTICLE{2025A&A...693A..60H,
       author = {{Heintz}, K.~E. and {Brammer}, G.~B. and {Watson}, D. and {Oesch}, P.~A. and {Keating}, L.~C. and {Hayes}, M.~J. and {Abdurro'uf} and {Arellano-C{\'o}rdova}, K.~Z. and {Carnall}, A.~C. and {Christiansen}, C.~R. and et al.},
        title = "{The JWST-PRIMAL archival survey: A JWST/NIRSpec reference sample for the physical properties and Lyman-{\ensuremath{\alpha}} absorption and emission of {\ensuremath{\sim}}600 galaxies at z = 5.0 {\ensuremath{-}} 13.4}",
      journal = {\aap},
         year = 2025,
        month = jan,
       volume = {693},
          eid = {A60},
        pages = {A60},
          doi = {10.1051/0004-6361/202450243},
archivePrefix = {arXiv},
       eprint = {2404.02211},
 primaryClass = {astro-ph.GA},
       adsurl = {https://ui.adsabs.harvard.edu/abs/2025A&A...693A..60H}
}

@ARTICLE{2019ApJ...876....3L,
       author = {{Leja}, Joel and {Carnall}, Adam C. and {Johnson}, Benjamin D. and {Conroy}, Charlie and {Speagle}, Joshua S.},
        title = "{How to Measure Galaxy Star Formation Histories. II. Nonparametric Models}",
      journal = {\apj},
         year = 2019,
        month = may,
       volume = {876},
       number = {1},
          eid = {3},
        pages = {3},
          doi = {10.3847/1538-4357/ab133c},
archivePrefix = {arXiv},
       eprint = {1811.03637},
 primaryClass = {astro-ph.GA},
       adsurl = {https://ui.adsabs.harvard.edu/abs/2019ApJ...876....3L}
}

@ARTICLE{2022ApJ...927..170T,
       author = {{Tacchella}, Sandro and {Finkelstein}, Steven L. and {Bagley}, Micaela and {Dickinson}, Mark and {Ferguson}, Henry C. and {Giavalisco}, Mauro and {Graziani}, Luca and {Grogin}, Norman A. and {Hathi}, Nimish and {Hutchison}, Taylor A. and et al.},
        title = "{On the Stellar Populations of Galaxies at z = 9-11: The Growth of Metals and Stellar Mass at Early Times}",
      journal = {\apj},
         year = 2022,
        month = mar,
       volume = {927},
       number = {2},
          eid = {170},
        pages = {170},
          doi = {10.3847/1538-4357/ac4cad},
archivePrefix = {arXiv},
       eprint = {2111.05351},
 primaryClass = {astro-ph.GA},
       adsurl = {https://ui.adsabs.harvard.edu/abs/2022ApJ...927..170T}
}

@ARTICLE{2000ApJ...533..682C,
       author = {{Calzetti}, Daniela and {Armus}, Lee and {Bohlin}, Ralph C. and {Kinney}, Anne L. and {Koornneef}, Jan and {Storchi-Bergmann}, Thaisa},
        title = "{The Dust Content and Opacity of Actively Star-forming Galaxies}",
      journal = {\apj},
         year = 2000,
        month = apr,
       volume = {533},
       number = {2},
        pages = {682-695},
          doi = {10.1086/308692},
archivePrefix = {arXiv},
       eprint = {astro-ph/9911459},
 primaryClass = {astro-ph},
       adsurl = {https://ui.adsabs.harvard.edu/abs/2000ApJ...533..682C}
}

@ARTICLE{2003ApJ...594..279G,
       author = {{Gordon}, Karl D. and {Clayton}, Geoffrey C. and {Misselt}, K.~A. and {Landolt}, Arlo U. and {Wolff}, Michael J.},
        title = "{A Quantitative Comparison of the Small Magellanic Cloud, Large Magellanic Cloud, and Milky Way Ultraviolet to Near-Infrared Extinction Curves}",
      journal = {\apj},
         year = 2003,
        month = sep,
       volume = {594},
       number = {1},
        pages = {279-293},
          doi = {10.1086/376774},
archivePrefix = {arXiv},
       eprint = {astro-ph/0305257},
 primaryClass = {astro-ph},
       adsurl = {https://ui.adsabs.harvard.edu/abs/2003ApJ...594..279G}
}

@ARTICLE{1989ApJ...345..245C,
       author = {{Cardelli}, Jason A. and {Clayton}, Geoffrey C. and {Mathis}, John S.},
        title = "{The Relationship between Infrared, Optical, and Ultraviolet Extinction}",
      journal = {\apj},
         year = 1989,
        month = oct,
       volume = {345},
        pages = {245},
          doi = {10.1086/167900},
       adsurl = {https://ui.adsabs.harvard.edu/abs/1989ApJ...345..245C}
}

@ARTICLE{2025NatAs...9..458M,
       author = {{Markov}, Vladan and {Gallerani}, Simona and {Ferrara}, Andrea and {Pallottini}, Andrea and {Parlanti}, Eleonora and {Mascia}, Fabio Di and {Sommovigo}, Laura and {Kohandel}, Mahsa},
        title = "{The evolution of dust attenuation in z {\ensuremath{\approx}} 2-12 galaxies observed by JWST}",
      journal = {Nature Astronomy},
         year = 2025,
        month = mar,
       volume = {9},
        pages = {458-468},
          doi = {10.1038/s41550-024-02426-1},
archivePrefix = {arXiv},
       eprint = {2402.05996},
 primaryClass = {astro-ph.GA},
       adsurl = {https://ui.adsabs.harvard.edu/abs/2025NatAs...9..458M}
}

@ARTICLE{2011ApJ...737..103S,
       author = {{Schlafly}, Edward F. and {Finkbeiner}, Douglas P.},
        title = "{Measuring Reddening with Sloan Digital Sky Survey Stellar Spectra and Recalibrating SFD}",
      journal = {\apj},
         year = 2011,
        month = aug,
       volume = {737},
       number = {2},
          eid = {103},
        pages = {103},
          doi = {10.1088/0004-637X/737/2/103},
archivePrefix = {arXiv},
       eprint = {1012.4804},
 primaryClass = {astro-ph.GA},
       adsurl = {https://ui.adsabs.harvard.edu/abs/2011ApJ...737..103S}
}

@ARTICLE{1994ApJ...422..158O,
       author = {{O'Donnell}, James E.},
        title = "{R v-dependent Optical and Near-Ultraviolet Extinction}",
      journal = {\apj},
         year = 1994,
        month = feb,
       volume = {422},
        pages = {158},
          doi = {10.1086/173713},
       adsurl = {https://ui.adsabs.harvard.edu/abs/1994ApJ...422..158O}
}

@ARTICLE{2022ApJ...926..145S,
       author = {{Shapley}, Alice E. and {Sanders}, Ryan L. and {Salim}, Samir and {Reddy}, Naveen A. and {Kriek}, Mariska and {Mobasher}, Bahram and {Coil}, Alison L. and {Siana}, Brian and {Price}, Sedona H. and {Shivaei}, Irene and et al.},
        title = "{The MOSFIRE Deep Evolution Field Survey: Implications of the Lack of Evolution in the Dust Attenuation-Mass Relation to z   2}",
      journal = {\apj},
         year = 2022,
        month = feb,
       volume = {926},
       number = {2},
          eid = {145},
        pages = {145},
          doi = {10.3847/1538-4357/ac4742},
archivePrefix = {arXiv},
       eprint = {2109.14630},
 primaryClass = {astro-ph.GA},
       adsurl = {https://ui.adsabs.harvard.edu/abs/2022ApJ...926..145S}
}

@ARTICLE{2001MNRAS.322..231K,
       author = {{Kroupa}, Pavel},
        title = "{On the variation of the initial mass function}",
      journal = {\mnras},
         year = 2001,
        month = apr,
       volume = {322},
       number = {2},
        pages = {231-246},
          doi = {10.1046/j.1365-8711.2001.04022.x},
archivePrefix = {arXiv},
       eprint = {astro-ph/0009005},
 primaryClass = {astro-ph},
       adsurl = {https://ui.adsabs.harvard.edu/abs/2001MNRAS.322..231K}
}

@ARTICLE{2024ApJ...977..133C,
       author = {{Clarke}, Leonardo and {Shapley}, Alice E. and {Sanders}, Ryan L. and {Topping}, Michael W. and {Brammer}, Gabriel B. and {Bento}, Trinity and {Reddy}, Naveen A. and {Kehoe}, Emily},
        title = "{The Star-forming Main Sequence in JADES and CEERS at z > 1.4: Investigating the Burstiness of Star Formation}",
      journal = {\apj},
         year = 2024,
        month = dec,
       volume = {977},
       number = {1},
          eid = {133},
        pages = {133},
          doi = {10.3847/1538-4357/ad8ba4},
archivePrefix = {arXiv},
       eprint = {2406.05178},
 primaryClass = {astro-ph.GA},
       adsurl = {https://ui.adsabs.harvard.edu/abs/2024ApJ...977..133C}
}

@ARTICLE{2019PASJ...71....8K,
       author = {{Koyama}, Yusei and {Shimakawa}, Rhythm and {Yamamura}, Issei and {Kodama}, Tadayuki and {Hayashi}, Masao},
        title = "{On the different levels of dust attenuation to nebular and stellar light in star-forming galaxies}",
      journal = {\pasj},
         year = 2019,
        month = jan,
       volume = {71},
       number = {1},
          eid = {8},
        pages = {8},
          doi = {10.1093/pasj/psy113},
archivePrefix = {arXiv},
       eprint = {1809.03715},
 primaryClass = {astro-ph.GA},
       adsurl = {https://ui.adsabs.harvard.edu/abs/2019PASJ...71....8K}
}

@ARTICLE{2026arXiv260311338K,
       author = {{Karthikeyan}, Shreya and {Clarke}, Leonardo and {Shapley}, Alice E. and {Lam}, Natalie and {Sanders}, Ryan L. and {Reddy}, Naveen A. and {Topping}, Michael W. and {Brammer}, Gabriel B.},
        title = "{Balmer Decrements and Nebular-Stellar Reddening in JADES Galaxies at $2.7<z<7$}",
      journal = {arXiv e-prints},
         year = 2026,
        month = mar,
          eid = {arXiv:2603.11338},
        pages = {arXiv:2603.11338},
          doi = {10.48550/arXiv.2603.11338},
archivePrefix = {arXiv},
       eprint = {2603.11338},
 primaryClass = {astro-ph.GA},
       adsurl = {https://ui.adsabs.harvard.edu/abs/2026arXiv260311338K}
}

@ARTICLE{2025arXiv251000235W,
       author = {{Woodrum}, Charity and {Shivaei}, Irene and {Witstok}, Joris and {Saxena}, Aayush and {Simmonds}, Charlotte and {Scholtz}, Jan and {Bhatawdekar}, Rachana and {Bunker}, Andrew J. and {Carniani}, St{\'e}fano and {Charlot}, Stephane and et al.},
        title = "{JADES: The Star Formation and Dust Attenuation Properties of Galaxies at 3<z<7}",
      journal = {arXiv e-prints},
         year = 2025,
        month = sep,
          eid = {arXiv:2510.00235},
        pages = {arXiv:2510.00235},
          doi = {10.48550/arXiv.2510.00235},
archivePrefix = {arXiv},
       eprint = {2510.00235},
 primaryClass = {astro-ph.GA},
       adsurl = {https://ui.adsabs.harvard.edu/abs/2025arXiv251000235W}
}

@ARTICLE{2015ApJ...801L..29R,
       author = {{Renzini}, Alvio and {Peng}, Ying-jie},
        title = "{An Objective Definition for the Main Sequence of Star-forming Galaxies}",
      journal = {\apjl},
         year = 2015,
        month = mar,
       volume = {801},
       number = {2},
          eid = {L29},
        pages = {L29},
          doi = {10.1088/2041-8205/801/2/L29},
archivePrefix = {arXiv},
       eprint = {1502.01027},
 primaryClass = {astro-ph.GA},
       adsurl = {https://ui.adsabs.harvard.edu/abs/2015ApJ...801L..29R}
}

@ARTICLE{2014ApJS..214...15S,
       author = {{Speagle}, J.~S. and {Steinhardt}, C.~L. and {Capak}, P.~L. and {Silverman}, J.~D.},
        title = "{A Highly Consistent Framework for the Evolution of the Star-Forming ``Main Sequence'' from z \raisebox{-0.5ex}\textasciitilde 0-6}",
      journal = {\apjs},
         year = 2014,
        month = oct,
       volume = {214},
       number = {2},
          eid = {15},
        pages = {15},
          doi = {10.1088/0067-0049/214/2/15},
archivePrefix = {arXiv},
       eprint = {1405.2041},
 primaryClass = {astro-ph.GA},
       adsurl = {https://ui.adsabs.harvard.edu/abs/2014ApJS..214...15S}
}

@ARTICLE{1997AJ....113..162C,
       author = {{Calzetti}, Daniela},
        title = "{Reddening and Star Formation in Starburst Galaxies}",
      journal = {\aj},
         year = 1997,
        month = jan,
       volume = {113},
        pages = {162-184},
          doi = {10.1086/118242},
archivePrefix = {arXiv},
       eprint = {astro-ph/9610184},
 primaryClass = {astro-ph},
       adsurl = {https://ui.adsabs.harvard.edu/abs/1997AJ....113..162C}
}

@ARTICLE{2011MNRAS.417.1760W,
       author = {{Wild}, Vivienne and {Charlot}, St{\'e}phane and {Brinchmann}, Jarle and {Heckman}, Timothy and {Vince}, Oliver and {Pacifici}, Camilla and {Chevallard}, Jacopo},
        title = "{Empirical determination of the shape of dust attenuation curves in star-forming galaxies}",
      journal = {\mnras},
         year = 2011,
        month = nov,
       volume = {417},
       number = {3},
        pages = {1760-1786},
          doi = {10.1111/j.1365-2966.2011.19367.x},
archivePrefix = {arXiv},
       eprint = {1106.1646},
 primaryClass = {astro-ph.CO},
       adsurl = {https://ui.adsabs.harvard.edu/abs/2011MNRAS.417.1760W}
}

@ARTICLE{2020ApJ...888...88L,
       author = {{Lin}, Zesen and {Kong}, Xu},
        title = "{A Variant Stellar-to-nebular Dust Attenuation Ratio on Subgalactic and Galactic Scales}",
      journal = {\apj},
         year = 2020,
        month = jan,
       volume = {888},
       number = {2},
          eid = {88},
        pages = {88},
          doi = {10.3847/1538-4357/ab5f0e},
archivePrefix = {arXiv},
       eprint = {1912.01851},
 primaryClass = {astro-ph.GA},
       adsurl = {https://ui.adsabs.harvard.edu/abs/2020ApJ...888...88L}
}

@ARTICLE{2021ApJ...917...72L,
       author = {{Li}, Niu and {Li}, Cheng and {Mo}, Houjun and {Zhou}, Shuang and {Liang}, Fu-heng and {Boquien}, M{\'e}d{\'e}ric and {Drory}, Niv and {Fern{\'a}ndez-Trincado}, Jos{\'e} G. and {Greener}, Michael and {Riffel}, Rog{\'e}rio},
        title = "{Estimating Dust Attenuation From Galactic Spectra. II. Stellar and Gas Attenuation in Star-forming and Diffuse Ionized Gas Regions in MaNGA}",
      journal = {\apj},
         year = 2021,
        month = aug,
       volume = {917},
       number = {2},
          eid = {72},
        pages = {72},
          doi = {10.3847/1538-4357/ac0973},
archivePrefix = {arXiv},
       eprint = {2103.00666},
 primaryClass = {astro-ph.GA},
       adsurl = {https://ui.adsabs.harvard.edu/abs/2021ApJ...917...72L}
}

@ARTICLE{2016ApJ...818...13B,
       author = {{Battisti}, A.~J. and {Calzetti}, D. and {Chary}, R.-R.},
        title = "{Characterizing Dust Attenuation in Local Star-forming Galaxies: UV and Optical Reddening}",
      journal = {\apj},
         year = 2016,
        month = feb,
       volume = {818},
       number = {1},
          eid = {13},
        pages = {13},
          doi = {10.3847/0004-637X/818/1/13},
archivePrefix = {arXiv},
       eprint = {1601.00208},
 primaryClass = {astro-ph.GA},
       adsurl = {https://ui.adsabs.harvard.edu/abs/2016ApJ...818...13B}
}

@ARTICLE{2019ApJ...886...28Q,
       author = {{Qin}, Jianbo and {Zheng}, Xian Zhong and {Wuyts}, Stijn and {Pan}, Zhizheng and {Ren}, Jian},
        title = "{Understanding the Discrepancy between IRX and Balmer Decrement in Tracing Galaxy Dust Attenuation}",
      journal = {\apj},
         year = 2019,
        month = nov,
       volume = {886},
       number = {1},
          eid = {28},
        pages = {28},
          doi = {10.3847/1538-4357/ab4a04},
archivePrefix = {arXiv},
       eprint = {1909.13505},
 primaryClass = {astro-ph.GA},
       adsurl = {https://ui.adsabs.harvard.edu/abs/2019ApJ...886...28Q}
}

@ARTICLE{2017ApJ...847...18Z,
       author = {{Zahid}, H. Jabran and {Kudritzki}, Rolf-Peter and {Conroy}, Charlie and {Andrews}, Brett and {Ho}, I.-Ting},
        title = "{Stellar Absorption Line Analysis of Local Star-forming Galaxies: The Relation between Stellar Mass, Metallicity, Dust Attenuation, and Star Formation Rate}",
      journal = {\apj},
         year = 2017,
        month = sep,
       volume = {847},
       number = {1},
          eid = {18},
        pages = {18},
          doi = {10.3847/1538-4357/aa88ae},
archivePrefix = {arXiv},
       eprint = {1708.07107},
 primaryClass = {astro-ph.GA},
       adsurl = {https://ui.adsabs.harvard.edu/abs/2017ApJ...847...18Z}
}

@ARTICLE{2013ApJ...777L...8K,
       author = {{Kashino}, D. and {Silverman}, J.~D. and {Rodighiero}, G. and {Renzini}, A. and {Arimoto}, N. and {Daddi}, E. and {Lilly}, S.~J. and {Sanders}, D.~B. and {Kartaltepe}, J. and {Zahid}, H.~J. and et al.},
        title = "{The FMOS-COSMOS Survey of Star-forming Galaxies at z \raisebox{-0.5ex}\textasciitilde 1.6. I. H{\ensuremath{\alpha}}-based Star Formation Rates and Dust Extinction}",
      journal = {\apjl},
         year = 2013,
        month = nov,
       volume = {777},
       number = {1},
          eid = {L8},
        pages = {L8},
          doi = {10.1088/2041-8205/777/1/L8},
archivePrefix = {arXiv},
       eprint = {1309.4774},
 primaryClass = {astro-ph.CO},
       adsurl = {https://ui.adsabs.harvard.edu/abs/2013ApJ...777L...8K}
}

@ARTICLE{2019ApJ...871..128T,
       author = {{Theios}, Rachel L. and {Steidel}, Charles C. and {Strom}, Allison L. and {Rudie}, Gwen C. and {Trainor}, Ryan F. and {Reddy}, Naveen A.},
        title = "{Dust Attenuation, Star Formation, and Metallicity in z {\ensuremath{\sim}} 2-3 Galaxies from KBSS-MOSFIRE}",
      journal = {\apj},
         year = 2019,
        month = jan,
       volume = {871},
       number = {1},
          eid = {128},
        pages = {128},
          doi = {10.3847/1538-4357/aaf386},
archivePrefix = {arXiv},
       eprint = {1805.00016},
 primaryClass = {astro-ph.GA},
       adsurl = {https://ui.adsabs.harvard.edu/abs/2019ApJ...871..128T}
}

@ARTICLE{2015ApJ...807..141P,
       author = {{Pannella}, M. and {Elbaz}, D. and {Daddi}, E. and {Dickinson}, M. and {Hwang}, H.~S. and {Schreiber}, C. and {Strazzullo}, V. and {Aussel}, H. and {Bethermin}, M. and {Buat}, V. and et al.},
        title = "{GOODS-Herschel: Star Formation, Dust Attenuation, and the FIR-radio Correlation on the Main Sequence of Star-forming Galaxies up to z ≃4}",
      journal = {\apj},
         year = 2015,
        month = jul,
       volume = {807},
       number = {2},
          eid = {141},
        pages = {141},
          doi = {10.1088/0004-637X/807/2/141},
archivePrefix = {arXiv},
       eprint = {1407.5072},
 primaryClass = {astro-ph.GA},
       adsurl = {https://ui.adsabs.harvard.edu/abs/2015ApJ...807..141P}
}

@ARTICLE{2016A&A...586A..83P,
       author = {{Puglisi}, A. and {Rodighiero}, G. and {Franceschini}, A. and {Talia}, M. and {Cimatti}, A. and {Baronchelli}, I. and {Daddi}, E. and {Renzini}, A. and {Schawinski}, K. and {Mancini}, C. and et al.},
        title = "{Dust attenuation in z \raisebox{-0.5ex}\textasciitilde 1 galaxies from Herschel and 3D-HST H{\ensuremath{\alpha}} measurements}",
      journal = {\aap},
         year = 2016,
        month = feb,
       volume = {586},
          eid = {A83},
        pages = {A83},
          doi = {10.1051/0004-6361/201526782},
archivePrefix = {arXiv},
       eprint = {1507.00005},
 primaryClass = {astro-ph.GA},
       adsurl = {https://ui.adsabs.harvard.edu/abs/2016A&A...586A..83P}
}

@ARTICLE{2014ApJ...788...86P,
       author = {{Price}, Sedona H. and {Kriek}, Mariska and {Brammer}, Gabriel B. and {Conroy}, Charlie and {F{\"o}rster Schreiber}, Natascha M. and {Franx}, Marijn and {Fumagalli}, Mattia and {Lundgren}, Britt and {Momcheva}, Ivelina and {Nelson}, Erica J. and et al.},
        title = "{Direct Measurements of Dust Attenuation in z \raisebox{-0.5ex}\textasciitilde 1.5 Star-forming Galaxies from 3D-HST: Implications for Dust Geometry and Star Formation Rates}",
      journal = {\apj},
         year = 2014,
        month = jun,
       volume = {788},
       number = {1},
          eid = {86},
        pages = {86},
          doi = {10.1088/0004-637X/788/1/86},
archivePrefix = {arXiv},
       eprint = {1310.4177},
 primaryClass = {astro-ph.CO},
       adsurl = {https://ui.adsabs.harvard.edu/abs/2014ApJ...788...86P}
}

@ARTICLE{2015ApJ...806..259R,
       author = {{Reddy}, Naveen A. and {Kriek}, Mariska and {Shapley}, Alice E. and {Freeman}, William R. and {Siana}, Brian and {Coil}, Alison L. and {Mobasher}, Bahram and {Price}, Sedona H. and {Sanders}, Ryan L. and {Shivaei}, Irene},
        title = "{The MOSDEF Survey: Measurements of Balmer Decrements and the Dust Attenuation Curve at Redshifts z \raisebox{-0.5ex}\textasciitilde 1.4-2.6}",
      journal = {\apj},
         year = 2015,
        month = jun,
       volume = {806},
       number = {2},
          eid = {259},
        pages = {259},
          doi = {10.1088/0004-637X/806/2/259},
archivePrefix = {arXiv},
       eprint = {1504.02782},
 primaryClass = {astro-ph.GA},
       adsurl = {https://ui.adsabs.harvard.edu/abs/2015ApJ...806..259R}
}

@ARTICLE{2026ApJ...997..319T,
       author = {{Tsujita}, Akiyoshi and {Fujimoto}, Seiji and {Faisst}, Andreas and {Boquien}, M{\'e}d{\'e}ric and {Li}, Juno and {Ferrara}, Andrea and {Battisti}, Andrew J. and {Dam}, Poulomi and {Aravena}, Manuel and {B{\'e}thermin}, Matthieu and {Casey}, Caitlin M. and {Cooper}, Olivia R. and {Finkelstein}, Steven L. and {Ginolfi}, Michele and {G{\'o}mez-Espinoza}, Diego A. and {Hadi}, Ali and {Herrera-Camus}, Rodrigo and {Ibar}, Edo and {Inami}, Hanae and {Jones}, Gareth C. and {Koekemoer}, Anton M. and {Kohno}, Kotaro and {Lemaux}, Brian C. and {Liu}, Zhaoxuan and {de Looze}, Ilse and {Mitsuhashi}, Ikki and {Mobasher}, Bahram and {Molina}, Juan and {Nanni}, Ambra and {Pozzi}, Francesca and {Reddy}, Naveen A. and {Relano}, Monica and {Rodighiero}, Giulia and {Romano}, Michael and {Sanders}, David B. and {Sawant}, Prasad and {Solimano}, Manuel and {Sommovigo}, Laura and {Spilker}, Justin and {Tadaki}, Ken-Ichi and {Vallini}, Livia and {Villanueva}, Vicente and {Wang}, Wuji and {Zamorani}, Giovanni and {Alpine+Cristal Collaborations}},
        title = "{The ALPINE-CRISTAL-JWST Survey: Stellar and Nebular Dust Attenuation of Main-sequence Galaxies at z {\ensuremath{\sim}} 4─6}",
      journal = {\apj},
         year = 2026,
        month = feb,
       volume = {997},
       number = {2},
          eid = {319},
        pages = {319},
          doi = {10.3847/1538-4357/ae22d8},
archivePrefix = {arXiv},
       eprint = {2510.18248},
 primaryClass = {astro-ph.GA},
       adsurl = {https://ui.adsabs.harvard.edu/abs/2026ApJ...997..319T}
}

@ARTICLE{2018ARA&A..56..673G,
       author = {{Galliano}, Fr{\'e}d{\'e}ric and {Galametz}, Maud and {Jones}, Anthony P.},
        title = "{The Interstellar Dust Properties of Nearby Galaxies}",
      journal = {\araa},
         year = 2018,
        month = sep,
       volume = {56},
        pages = {673-713},
          doi = {10.1146/annurev-astro-081817-051900},
archivePrefix = {arXiv},
       eprint = {1711.07434},
 primaryClass = {astro-ph.GA},
       adsurl = {https://ui.adsabs.harvard.edu/abs/2018ARA&A..56..673G}
}

@ARTICLE{2017MNRAS.466..105A,
       author = {{Aoyama}, Shohei and {Hou}, Kuan-Chou and {Shimizu}, Ikkoh and {Hirashita}, Hiroyuki and {Todoroki}, Keita and {Choi}, Jun-Hwan and {Nagamine}, Kentaro},
        title = "{Galaxy simulation with dust formation and destruction}",
      journal = {\mnras},
         year = 2017,
        month = apr,
       volume = {466},
       number = {1},
        pages = {105-121},
          doi = {10.1093/mnras/stw3061},
archivePrefix = {arXiv},
       eprint = {1609.07547},
 primaryClass = {astro-ph.GA},
       adsurl = {https://ui.adsabs.harvard.edu/abs/2017MNRAS.466..105A}
}

@ARTICLE{2000ApJ...539..718C,
       author = {{Charlot}, St{\'e}phane and {Fall}, S. Michael},
        title = "{A Simple Model for the Absorption of Starlight by Dust in Galaxies}",
      journal = {\apj},
         year = 2000,
        month = aug,
       volume = {539},
       number = {2},
        pages = {718-731},
          doi = {10.1086/309250},
archivePrefix = {arXiv},
       eprint = {astro-ph/0003128},
 primaryClass = {astro-ph},
       adsurl = {https://ui.adsabs.harvard.edu/abs/2000ApJ...539..718C}
}

@ARTICLE{2012ARA&A..50..531K,
       author = {{Kennicutt}, Robert C. and {Evans}, Neal J.},
        title = "{Star Formation in the Milky Way and Nearby Galaxies}",
      journal = {\araa},
         year = 2012,
        month = sep,
       volume = {50},
        pages = {531-608},
          doi = {10.1146/annurev-astro-081811-125610},
archivePrefix = {arXiv},
       eprint = {1204.3552},
 primaryClass = {astro-ph.GA},
       adsurl = {https://ui.adsabs.harvard.edu/abs/2012ARA&A..50..531K}
}

@ARTICLE{1998ARA&A..36..189K,
       author = {{Kennicutt}, Jr., Robert C.},
        title = "{Star Formation in Galaxies Along the Hubble Sequence}",
      journal = {\araa},
         year = 1998,
        month = jan,
       volume = {36},
        pages = {189-232},
          doi = {10.1146/annurev.astro.36.1.189},
archivePrefix = {arXiv},
       eprint = {astro-ph/9807187},
 primaryClass = {astro-ph},
       adsurl = {https://ui.adsabs.harvard.edu/abs/1998ARA&A..36..189K}
}

@ARTICLE{2006A&A...451..417D,
       author = {{Dole}, H. and {Lagache}, G. and {Puget}, J.-L. and {Caputi}, K.~I. and {Fern{\'a}ndez-Conde}, N. and {Le Floc'h}, E. and {Papovich}, C. and {P{\'e}rez-Gonz{\'a}lez}, P.~G. and {Rieke}, G.~H. and {Blaylock}, M.},
        title = "{The cosmic infrared background resolved by Spitzer. Contributions of mid-infrared galaxies to the far-infrared background}",
      journal = {\aap},
         year = 2006,
        month = may,
       volume = {451},
       number = {2},
        pages = {417-429},
          doi = {10.1051/0004-6361:20054446},
archivePrefix = {arXiv},
       eprint = {astro-ph/0603208},
 primaryClass = {astro-ph},
       adsurl = {https://ui.adsabs.harvard.edu/abs/2006A&A...451..417D}
}

@ARTICLE{2013ApJ...779...32V,
       author = {{Viero}, M.~P. and {Moncelsi}, L. and {Quadri}, R.~F. and {Arumugam}, V. and {Assef}, R.~J. and {B{\'e}thermin}, M. and {Bock}, J. and {Bridge}, C. and {Casey}, C.~M. and {Conley}, A. and et al.},
        title = "{HerMES: The Contribution to the Cosmic Infrared Background from Galaxies Selected by Mass and Redshift}",
      journal = {\apj},
         year = 2013,
        month = dec,
       volume = {779},
       number = {1},
          eid = {32},
        pages = {32},
          doi = {10.1088/0004-637X/779/1/32},
archivePrefix = {arXiv},
       eprint = {1304.0446},
 primaryClass = {astro-ph.CO},
       adsurl = {https://ui.adsabs.harvard.edu/abs/2013ApJ...779...32V}
}

@ARTICLE{1977ApJ...217..425M,
       author = {{Mathis}, J.~S. and {Rumpl}, W. and {Nordsieck}, K.~H.},
        title = "{The size distribution of interstellar grains.}",
      journal = {\apj},
         year = 1977,
        month = oct,
       volume = {217},
        pages = {425-433},
          doi = {10.1086/155591},
       adsurl = {https://ui.adsabs.harvard.edu/abs/1977ApJ...217..425M}
}

@ARTICLE{2013MNRAS.432..637A,
       author = {{Asano}, Ryosuke S. and {Takeuchi}, Tsutomu T. and {Hirashita}, Hiroyuki and {Nozawa}, Takaya},
        title = "{What determines the grain size distribution in galaxies?}",
      journal = {\mnras},
         year = 2013,
        month = jun,
       volume = {432},
       number = {1},
        pages = {637-652},
          doi = {10.1093/mnras/stt506},
archivePrefix = {arXiv},
       eprint = {1303.5528},
 primaryClass = {astro-ph.GA},
       adsurl = {https://ui.adsabs.harvard.edu/abs/2013MNRAS.432..637A}
}

@ARTICLE{2020MNRAS.491.3844A,
       author = {{Aoyama}, Shohei and {Hirashita}, Hiroyuki and {Nagamine}, Kentaro},
        title = "{Galaxy simulation with the evolution of grain size distribution}",
      journal = {\mnras},
         year = 2020,
        month = jan,
       volume = {491},
       number = {3},
        pages = {3844-3859},
          doi = {10.1093/mnras/stz3253},
archivePrefix = {arXiv},
       eprint = {1906.01917},
 primaryClass = {astro-ph.GA},
       adsurl = {https://ui.adsabs.harvard.edu/abs/2020MNRAS.491.3844A}
}

@ARTICLE{2001ApJ...548..296W,
       author = {{Weingartner}, Joseph C. and {Draine}, B.~T.},
        title = "{Dust Grain-Size Distributions and Extinction in the Milky Way, Large Magellanic Cloud, and Small Magellanic Cloud}",
      journal = {\apj},
         year = 2001,
        month = feb,
       volume = {548},
       number = {1},
        pages = {296-309},
          doi = {10.1086/318651},
archivePrefix = {arXiv},
       eprint = {astro-ph/0008146},
 primaryClass = {astro-ph},
       adsurl = {https://ui.adsabs.harvard.edu/abs/2001ApJ...548..296W}
}

@BOOK{2011piim.book.....D,
       author = {{Draine}, Bruce T.},
        title = "{Physics of the Interstellar and Intergalactic Medium}",
         year = 2011,
       adsurl = {https://ui.adsabs.harvard.edu/abs/2011piim.book.....D}
}

@ARTICLE{1994ApJ...429..582C,
       author = {{Calzetti}, Daniela and {Kinney}, Anne L. and {Storchi-Bergmann}, Thaisa},
        title = "{Dust Extinction of the Stellar Continua in Starburst Galaxies: The Ultraviolet and Optical Extinction Law}",
      journal = {\apj},
         year = 1994,
        month = jul,
       volume = {429},
        pages = {582},
          doi = {10.1086/174346},
       adsurl = {https://ui.adsabs.harvard.edu/abs/1994ApJ...429..582C}
}

@ARTICLE{2013MNRAS.432.2061C,
       author = {{Chevallard}, J. and {Charlot}, S. and {Wandelt}, B. and {Wild}, V.},
        title = "{Insights into the content and spatial distribution of dust from the integrated spectral properties of galaxies}",
      journal = {\mnras},
         year = 2013,
        month = jul,
       volume = {432},
       number = {3},
        pages = {2061-2091},
          doi = {10.1093/mnras/stt523},
archivePrefix = {arXiv},
       eprint = {1303.6631},
 primaryClass = {astro-ph.CO},
       adsurl = {https://ui.adsabs.harvard.edu/abs/2013MNRAS.432.2061C}
}

@ARTICLE{2013ApJ...779..135W,
       author = {{Wuyts}, Stijn and {F{\"o}rster Schreiber}, Natascha M. and {Nelson}, Erica J. and {van Dokkum}, Pieter G. and {Brammer}, Gabe and {Chang}, Yu-Yen and {Faber}, Sandra M. and {Ferguson}, Henry C. and {Franx}, Marijn and {Fumagalli}, Mattia and et al.},
        title = "{A CANDELS-3D-HST synergy: Resolved Star Formation Patterns at 0.7 < z < 1.5}",
      journal = {\apj},
         year = 2013,
        month = dec,
       volume = {779},
       number = {2},
          eid = {135},
        pages = {135},
          doi = {10.1088/0004-637X/779/2/135},
archivePrefix = {arXiv},
       eprint = {1310.5702},
 primaryClass = {astro-ph.CO},
       adsurl = {https://ui.adsabs.harvard.edu/abs/2013ApJ...779..135W}
}

@ARTICLE{2018MNRAS.480.4379C,
       author = {{Carnall}, A.~C. and {McLure}, R.~J. and {Dunlop}, J.~S. and {Dav{\'e}}, R.},
        title = "{Inferring the star formation histories of massive quiescent galaxies with BAGPIPES: evidence for multiple quenching mechanisms}",
      journal = {\mnras},
         year = 2018,
        month = nov,
       volume = {480},
       number = {4},
        pages = {4379-4401},
          doi = {10.1093/mnras/sty2169},
archivePrefix = {arXiv},
       eprint = {1712.04452},
 primaryClass = {astro-ph.GA},
       adsurl = {https://ui.adsabs.harvard.edu/abs/2018MNRAS.480.4379C}
}

@ARTICLE{2019MNRAS.490..417C,
       author = {{Carnall}, A.~C. and {McLure}, R.~J. and {Dunlop}, J.~S. and {Cullen}, F. and {McLeod}, D.~J. and {Wild}, V. and {Johnson}, B.~D. and {Appleby}, S. and {Dav{\'e}}, R. and {Amorin}, R. and {Bolzonella}, M. and {Castellano}, M. and {Cimatti}, A. and {Cucciati}, O. and {Gargiulo}, A. and {Garilli}, B. and {Marchi}, F. and {Pentericci}, L. and {Pozzetti}, L. and {Schreiber}, C. and {Talia}, M. and {Zamorani}, G.},
        title = "{The VANDELS survey: the star-formation histories of massive quiescent galaxies at 1.0 < z < 1.3}",
      journal = {\mnras},
         year = 2019,
        month = nov,
       volume = {490},
       number = {1},
        pages = {417-439},
          doi = {10.1093/mnras/stz2544},
archivePrefix = {arXiv},
       eprint = {1903.11082},
 primaryClass = {astro-ph.GA},
       adsurl = {https://ui.adsabs.harvard.edu/abs/2019MNRAS.490..417C}
}

@ARTICLE{2005PASP..117..227K,
       author = {{Kewley}, Lisa J. and {Jansen}, Rolf A. and {Geller}, Margaret J.},
        title = "{Aperture Effects on Star Formation Rate, Metallicity, and Reddening}",
      journal = {\pasp},
         year = 2005,
        month = mar,
       volume = {117},
       number = {829},
        pages = {227-244},
          doi = {10.1086/428303},
archivePrefix = {arXiv},
       eprint = {astro-ph/0501229},
 primaryClass = {astro-ph},
       adsurl = {https://ui.adsabs.harvard.edu/abs/2005PASP..117..227K}
}

@ARTICLE{2023ApJ...954..157S,
       author = {{Shapley}, Alice E. and {Sanders}, Ryan L. and {Reddy}, Naveen A. and {Topping}, Michael W. and {Brammer}, Gabriel B.},
        title = "{JWST/NIRSpec Balmer-line Measurements of Star Formation and Dust Attenuation at z   3-6}",
      journal = {\apj},
         year = 2023,
        month = sep,
       volume = {954},
       number = {2},
          eid = {157},
        pages = {157},
          doi = {10.3847/1538-4357/acea5a},
archivePrefix = {arXiv},
       eprint = {2301.03241},
 primaryClass = {astro-ph.GA},
       adsurl = {https://ui.adsabs.harvard.edu/abs/2023ApJ...954..157S}
}

@ARTICLE{2023ApJ...948...83R,
       author = {{Reddy}, Naveen A. and {Topping}, Michael W. and {Sanders}, Ryan L. and {Shapley}, Alice E. and {Brammer}, Gabriel},
        title = "{Paschen-line Constraints on Dust Attenuation and Star Formation at z   1-3 with JWST/NIRSpec}",
      journal = {\apj},
         year = 2023,
        month = may,
       volume = {948},
       number = {2},
          eid = {83},
        pages = {83},
          doi = {10.3847/1538-4357/acc869},
archivePrefix = {arXiv},
       eprint = {2301.07249},
 primaryClass = {astro-ph.GA},
       adsurl = {https://ui.adsabs.harvard.edu/abs/2023ApJ...948...83R}
}

@ARTICLE{2024A&A...691A.305S,
       author = {{Sandles}, Lester and {D'Eugenio}, Francesco and {Maiolino}, Roberto and {Looser}, Tobias J. and {Arribas}, Santiago and {Baker}, William M. and {Bonaventura}, Nina and {Bunker}, Andrew J. and {Cameron}, Alex J. and {Carniani}, Stefano and {Charlot}, Stephane and {Chevallard}, Jacopo and {Curti}, Mirko and {Curtis-Lake}, Emma and {de Graaff}, Anna and {Eisenstein}, Daniel J. and {Hainline}, Kevin and {Ji}, Zhiyuan and {Johnson}, Benjamin D. and {Jones}, Gareth C. and {Kumari}, Nimisha and {Nelson}, Erica and {Perna}, Michele and {Rawle}, Tim and {Rix}, Hans-Walter and {Robertson}, Brant and {Del Pino}, Bruno Rodr{\'\i}guez and {Scholtz}, Jan and {Shivaei}, Irene and {Smit}, Renske and {Sun}, Fengwu and {Tacchella}, Sandro and {{\"U}bler}, Hannah and {Williams}, Christina C. and {Willott}, Chris and {Witstok}, Joris},
        title = "{JADES: Balmer decrement measurements at redshifts 4 < z < 7}",
      journal = {\aap},
         year = 2024,
        month = nov,
       volume = {691},
          eid = {A305},
        pages = {A305},
          doi = {10.1051/0004-6361/202347119},
archivePrefix = {arXiv},
       eprint = {2306.03931},
 primaryClass = {astro-ph.GA},
       adsurl = {https://ui.adsabs.harvard.edu/abs/2024A&A...691A.305S}
}

@ARTICLE{2024MNRAS.534..523C,
       author = {{Cameron}, Alex J. and {Katz}, Harley and {Witten}, Callum and {Saxena}, Aayush and {Laporte}, Nicolas and {Bunker}, Andrew J.},
        title = "{Nebular dominated galaxies: insights into the stellar initial mass function at high redshift}",
      journal = {\mnras},
         year = 2024,
        month = oct,
       volume = {534},
       number = {1},
        pages = {523-543},
          doi = {10.1093/mnras/stae1547},
archivePrefix = {arXiv},
       eprint = {2311.02051},
 primaryClass = {astro-ph.GA},
       adsurl = {https://ui.adsabs.harvard.edu/abs/2024MNRAS.534..523C}
}

@BOOK{1989agna.book.....O,
       author = {{Osterbrock}, Donald E.},
        title = "{Astrophysics of gaseous nebulae and active galactic nuclei}",
         year = 1989,
       adsurl = {https://ui.adsabs.harvard.edu/abs/1989agna.book.....O}
}

@ARTICLE{2024ApJ...974..180Y,
       author = {{Yanagisawa}, Hiroto and {Ouchi}, Masami and {Nakajima}, Kimihiko and {Yajima}, Hidenobu and {Umeda}, Hiroya and {Baba}, Shunsuke and {Nakagawa}, Takao and {Nakane}, Minami and {Matsumoto}, Akinori and {Ono}, Yoshiaki and et al.},
        title = "{Balmer Decrement Anomalies in Galaxies at z {\ensuremath{\sim}} 6 Found by JWST Observations: Density-bounded Nebulae or Excited H I Clouds?}",
      journal = {\apj},
         year = 2024,
        month = oct,
       volume = {974},
       number = {2},
          eid = {180},
        pages = {180},
          doi = {10.3847/1538-4357/ad7097},
archivePrefix = {arXiv},
       eprint = {2403.20118},
 primaryClass = {astro-ph.GA},
       adsurl = {https://ui.adsabs.harvard.edu/abs/2024ApJ...974..180Y}
}

@ARTICLE{2023MNRAS.522.4515L,
       author = {{Lapiner}, Sharon and {Dekel}, Avishai and {Freundlich}, Jonathan and {Ginzburg}, Omri and {Jiang}, Fangzhou and {Kretschmer}, Michael and {Tacchella}, Sandro and {Ceverino}, Daniel and {Primack}, Joel},
        title = "{Wet compaction to a blue nugget: a critical phase in galaxy evolution}",
      journal = {\mnras},
         year = 2023,
        month = jul,
       volume = {522},
       number = {3},
        pages = {4515-4547},
          doi = {10.1093/mnras/stad1263},
archivePrefix = {arXiv},
       eprint = {2302.12234},
 primaryClass = {astro-ph.GA},
       adsurl = {https://ui.adsabs.harvard.edu/abs/2023MNRAS.522.4515L}
}

@ARTICLE{2024A&A...691A.201L,
       author = {{Lin}, Zesen and {Yan}, Renbin},
        title = "{Nebular dust attenuation with the Balmer and Paschen lines based on the MaNGA survey}",
      journal = {\aap},
         year = 2024,
        month = nov,
       volume = {691},
          eid = {A201},
        pages = {A201},
          doi = {10.1051/0004-6361/202451339},
archivePrefix = {arXiv},
       eprint = {2410.05067},
 primaryClass = {astro-ph.GA},
       adsurl = {https://ui.adsabs.harvard.edu/abs/2024A&A...691A.201L}
}

@ARTICLE{2025MNRAS.540..190M,
       author = {{McClymont}, William and {Tacchella}, Sandro and {D'Eugenio}, Francesco and {Witten}, Callum and {Ji}, Xihan and {Smith}, Aaron and {Maiolino}, Roberto and {Arribas}, Santiago and {Scholtz}, Jan and {Simmonds}, Charlotte and et al.},
        title = "{The density-bounded twilight of starbursts in the early Universe}",
      journal = {\mnras},
         year = 2025,
        month = jun,
       volume = {540},
       number = {1},
        pages = {190-203},
          doi = {10.1093/mnras/staf745},
archivePrefix = {arXiv},
       eprint = {2405.15859},
 primaryClass = {astro-ph.GA},
       adsurl = {https://ui.adsabs.harvard.edu/abs/2025MNRAS.540..190M}
}

@ARTICLE{2024MNRAS.529.3301T,
       author = {{Topping}, Michael W. and {Stark}, Daniel P. and {Senchyna}, Peter and {Plat}, Adele and {Zitrin}, Adi and {Endsley}, Ryan and {Charlot}, St{\'e}phane and {Furtak}, Lukas J. and {Maseda}, Michael V. and {Smit}, Renske and et al.},
        title = "{Metal-poor star formation at z > 6 with JWST: new insight into hard radiation fields and nitrogen enrichment on 20 pc scales}",
      journal = {\mnras},
         year = 2024,
        month = apr,
       volume = {529},
       number = {4},
        pages = {3301-3322},
          doi = {10.1093/mnras/stae682},
archivePrefix = {arXiv},
       eprint = {2401.08764},
 primaryClass = {astro-ph.GA},
       adsurl = {https://ui.adsabs.harvard.edu/abs/2024MNRAS.529.3301T}
}

@ARTICLE{2017ApJ...851...90B,
       author = {{Battisti}, A.~J. and {Calzetti}, D. and {Chary}, R.-R.},
        title = "{Characterizing Dust Attenuation in Local Star-forming Galaxies: Inclination Effects and the 2175 {\r{A}} Feature}",
      journal = {\apj},
         year = 2017,
        month = dec,
       volume = {851},
       number = {2},
          eid = {90},
        pages = {90},
          doi = {10.3847/1538-4357/aa9a43},
archivePrefix = {arXiv},
       eprint = {1711.04814},
 primaryClass = {astro-ph.GA},
       adsurl = {https://ui.adsabs.harvard.edu/abs/2017ApJ...851...90B}
}

@ARTICLE{2026arXiv260409763R,
       author = {{Rodighiero}, Giulia and {Edes Esposito}, Gaia and {Calzetti}, Daniela and {Benotto}, Pietro and {Catone}, Michele and {Cassata}, Paolo and {Gandolfi}, Giovanni and {Bisigello}, Laura and {Carniani}, Stefano and {Renzini}, Alvio and {Shivaei}, Irene and {Vulcani}, Benedetta and {Puglisi}, Annagrazia},
        title = "{A first empirical derivation of the average dust attenuation law at 2<z<7}",
      journal = {arXiv e-prints},
         year = 2026,
        month = apr,
          eid = {arXiv:2604.09763},
        pages = {arXiv:2604.09763},
          doi = {10.48550/arXiv.2604.09763},
archivePrefix = {arXiv},
       eprint = {2604.09763},
 primaryClass = {astro-ph.GA},
       adsurl = {https://ui.adsabs.harvard.edu/abs/2026arXiv260409763R}
}

@ARTICLE{2021MNRAS.506.3588R,
       author = {{Rezaee}, Saeed and {Reddy}, Naveen and {Shivaei}, Irene and {Fetherolf}, Tara and {Emami}, Najmeh and {Khostovan}, A.~A.},
        title = "{Variation of the nebular dust attenuation curve with the properties of local star-forming galaxies}",
      journal = {\mnras},
         year = 2021,
        month = sep,
       volume = {506},
       number = {3},
        pages = {3588-3595},
          doi = {10.1093/mnras/stab1885},
archivePrefix = {arXiv},
       eprint = {2105.08166},
 primaryClass = {astro-ph.GA},
       adsurl = {https://ui.adsabs.harvard.edu/abs/2021MNRAS.506.3588R}
}

@ARTICLE{2013ApJ...763..145D,
       author = {{Dom{\'\i}nguez}, A. and {Siana}, B. and {Henry}, A.~L. and {Scarlata}, C. and {Bedregal}, A.~G. and {Malkan}, M. and {Atek}, H. and {Ross}, N.~R. and {Colbert}, J.~W. and {Teplitz}, H.~I. and et al.},
        title = "{Dust Extinction from Balmer Decrements of Star-forming Galaxies at 0.75 <= z <= 1.5 with Hubble Space Telescope/Wide-Field-Camera 3 Spectroscopy from the WFC3 Infrared Spectroscopic Parallel Survey}",
      journal = {\apj},
         year = 2013,
        month = feb,
       volume = {763},
       number = {2},
          eid = {145},
        pages = {145},
          doi = {10.1088/0004-637X/763/2/145},
archivePrefix = {arXiv},
       eprint = {1206.1867},
 primaryClass = {astro-ph.CO},
       adsurl = {https://ui.adsabs.harvard.edu/abs/2013ApJ...763..145D}
}

@ARTICLE{1999ApJ...521...64M,
       author = {{Meurer}, Gerhardt R. and {Heckman}, Timothy M. and {Calzetti}, Daniela},
        title = "{Dust Absorption and the Ultraviolet Luminosity Density at z \raisebox{-0.5ex}\textasciitilde 3 as Calibrated by Local Starburst Galaxies}",
      journal = {\apj},
         year = 1999,
        month = aug,
       volume = {521},
       number = {1},
        pages = {64-80},
          doi = {10.1086/307523},
archivePrefix = {arXiv},
       eprint = {astro-ph/9903054},
 primaryClass = {astro-ph},
       adsurl = {https://ui.adsabs.harvard.edu/abs/1999ApJ...521...64M}
}

@ARTICLE{2016ApJ...833...72B,
       author = {{Bouwens}, Rychard J. and {Aravena}, Manuel and {Decarli}, Roberto and {Walter}, Fabian and {da Cunha}, Elisabete and {Labb{\'e}}, Ivo and {Bauer}, Franz E. and {Bertoldi}, Frank and {Carilli}, Chris and {Chapman}, Scott and et al.},
        title = "{ALMA Spectroscopic Survey in the Hubble Ultra Deep Field: The Infrared Excess of UV-Selected z = 2-10 Galaxies as a Function of UV-Continuum Slope and Stellar Mass}",
      journal = {\apj},
         year = 2016,
        month = dec,
       volume = {833},
       number = {1},
          eid = {72},
        pages = {72},
          doi = {10.3847/1538-4357/833/1/72},
archivePrefix = {arXiv},
       eprint = {1606.05280},
 primaryClass = {astro-ph.GA},
       adsurl = {https://ui.adsabs.harvard.edu/abs/2016ApJ...833...72B}
}

@ARTICLE{2017ApJ...850..208W,
       author = {{Whitaker}, Katherine E. and {Pope}, Alexandra and {Cybulski}, Ryan and {Casey}, Caitlin M. and {Popping}, Gerg{\"o} and {Yun}, Min S.},
        title = "{The Constant Average Relationship between Dust-obscured Star Formation and Stellar Mass from z = 0 to z = 2.5}",
      journal = {\apj},
         year = 2017,
        month = dec,
       volume = {850},
       number = {2},
          eid = {208},
        pages = {208},
          doi = {10.3847/1538-4357/aa94ce},
archivePrefix = {arXiv},
       eprint = {1710.06872},
 primaryClass = {astro-ph.GA},
       adsurl = {https://ui.adsabs.harvard.edu/abs/2017ApJ...850..208W}
}

@ARTICLE{2018MNRAS.476.3991M,
       author = {{McLure}, R.~J. and {Dunlop}, J.~S. and {Cullen}, F. and {Bourne}, N. and {Best}, P.~N. and {Khochfar}, S. and {Bowler}, R.~A.~A. and {Biggs}, A.~D. and {Geach}, J.~E. and {Scott}, D. and et al.},
        title = "{Dust attenuation in 2 < z < 3 star-forming galaxies from deep ALMA observations of the Hubble Ultra Deep Field}",
      journal = {\mnras},
         year = 2018,
        month = may,
       volume = {476},
       number = {3},
        pages = {3991-4006},
          doi = {10.1093/mnras/sty522},
archivePrefix = {arXiv},
       eprint = {1709.06102},
 primaryClass = {astro-ph.GA},
       adsurl = {https://ui.adsabs.harvard.edu/abs/2018MNRAS.476.3991M}
}

@ARTICLE{2020MNRAS.491.4724F,
       author = {{Fudamoto}, Yoshinobu and {Oesch}, P.~A. and {Magnelli}, B. and {Schinnerer}, E. and {Liu}, D. and {Lang}, P. and {Jim{\'e}nez-Andrade}, E.~F. and {Groves}, B. and {Leslie}, S. and {Sargent}, M.~T.},
        title = "{A3COSMOS: the dust attenuation of star-forming galaxies at z = 2.5-4.0 from the COSMOS-ALMA archive}",
      journal = {\mnras},
         year = 2020,
        month = feb,
       volume = {491},
       number = {4},
        pages = {4724-4734},
          doi = {10.1093/mnras/stz3248},
archivePrefix = {arXiv},
       eprint = {1910.12885},
 primaryClass = {astro-ph.GA},
       adsurl = {https://ui.adsabs.harvard.edu/abs/2020MNRAS.491.4724F}
}

@ARTICLE{2026arXiv260204765W,
       author = {{Wijesekera}, J.~V. and {Koprowski}, M.~P. and {Dunlop}, J.~S. and {Lisiecki}, K. and {McLeod}, D.~J. and {McLure}, R.~J. and {Micha{\l}owski}, M.~J. and {Solar}, M.},
        title = "{Evolution of dust attenuation in star-forming galaxies with UV slope, stellar mass, and redshift out to $z \sim 5$}",
      journal = {arXiv e-prints},
         year = 2026,
        month = feb,
          eid = {arXiv:2602.04765},
        pages = {arXiv:2602.04765},
          doi = {10.48550/arXiv.2602.04765},
archivePrefix = {arXiv},
       eprint = {2602.04765},
 primaryClass = {astro-ph.GA},
       adsurl = {https://ui.adsabs.harvard.edu/abs/2026arXiv260204765W}
}

@ARTICLE{2018ApJ...859...11S,
       author = {{Salim}, Samir and {Boquien}, M{\'e}d{\'e}ric and {Lee}, Janice C.},
        title = "{Dust Attenuation Curves in the Local Universe: Demographics and New Laws for Star-forming Galaxies and High-redshift Analogs}",
      journal = {\apj},
         year = 2018,
        month = may,
       volume = {859},
       number = {1},
          eid = {11},
        pages = {11},
          doi = {10.3847/1538-4357/aabf3c},
archivePrefix = {arXiv},
       eprint = {1804.05850},
 primaryClass = {astro-ph.GA},
       adsurl = {https://ui.adsabs.harvard.edu/abs/2018ApJ...859...11S}
}

@ARTICLE{2025MNRAS.539..109F,
       author = {{Fisher}, R. and {Bowler}, R.~A.~A. and {Stefanon}, M. and {Rowland}, L.~E. and {Algera}, H.~S.~B. and {Aravena}, M. and {Bouwens}, R. and {Dayal}, P. and {Ferrara}, A. and {Fudamoto}, Y. and {Gulis}, C. and {Hodge}, J.~A. and {Inami}, H. and {Ormerod}, K. and {Pallottini}, A. and {Phillips}, S.~G. and {Sartorio}, N.~S. and {Schouws}, S. and {Smit}, R. and {Sommovigo}, L. and {Stark}, D.~P. and {van der Werf}, P.~P.},
        title = "{REBELS-IFU: dust attenuation curves of 12 massive galaxies at z ≃ 7}",
      journal = {\mnras},
         year = 2025,
        month = may,
       volume = {539},
       number = {1},
        pages = {109-126},
          doi = {10.1093/mnras/staf485},
archivePrefix = {arXiv},
       eprint = {2501.10541},
 primaryClass = {astro-ph.GA},
       adsurl = {https://ui.adsabs.harvard.edu/abs/2025MNRAS.539..109F}
}

@ARTICLE{2026ApJ...999...15R,
       author = {{Reddy}, Naveen A. and {Shapley}, Alice E. and {Sanders}, Ryan L. and {Topping}, Michael W. and {Ellis}, Richard S. and {Pettini}, Max and {Brammer}, Gabriel and {Cullen}, Fergus and {F{\"o}rster Schreiber}, Natascha M. and {Khostovan}, Ali A. and {McLeod}, Derek J. and {McLure}, Ross J. and {Narayanan}, Desika and {Oesch}, Pascal A. and {Pahl}, Anthony J. and {Steidel}, Charles C. and {Berg}, Danielle A.},
        title = "{The AURORA Survey: Multiple Balmer and Paschen Emission Lines for Individual Star-forming Galaxies at z = 1.5─4.4. I. A Diversity of Nebular Attenuation Curves and Evidence for Non-unity Dust Covering Fractions}",
      journal = {\apj},
         year = 2026,
        month = mar,
       volume = {999},
       number = {1},
          eid = {15},
        pages = {15},
          doi = {10.3847/1538-4357/ae38da},
archivePrefix = {arXiv},
       eprint = {2506.17396},
 primaryClass = {astro-ph.GA},
       adsurl = {https://ui.adsabs.harvard.edu/abs/2026ApJ...999...15R}
}

@ARTICLE{2023MNRAS.518.6142A,
       author = {{Algera}, Hiddo S.~B. and {Inami}, Hanae and {Oesch}, Pascal A. and {Sommovigo}, Laura and {Bouwens}, Rychard J. and {Topping}, Michael W. and {Schouws}, Sander and {Stefanon}, Mauro and {Stark}, Daniel P. and {Aravena}, Manuel and {Barrufet}, Laia and {da Cunha}, Elisabete and {Dayal}, Pratika and {Endsley}, Ryan and {Ferrara}, Andrea and {Fudamoto}, Yoshinobu and {Gonzalez}, Valentino and {Graziani}, Luca and {Hodge}, Jacqueline A. and {Hygate}, Alexander P.~S. and {de Looze}, Ilse and {Nanayakkara}, Themiya and {Schneider}, Raffaella and {van der Werf}, Paul P.},
        title = "{The ALMA REBELS survey: the dust-obscured cosmic star formation rate density at redshift 7}",
      journal = {\mnras},
         year = 2023,
        month = feb,
       volume = {518},
       number = {4},
        pages = {6142-6157},
          doi = {10.1093/mnras/stac3195},
archivePrefix = {arXiv},
       eprint = {2208.08243},
 primaryClass = {astro-ph.GA},
       adsurl = {https://ui.adsabs.harvard.edu/abs/2023MNRAS.518.6142A}
}

@ARTICLE{2020ARA&A..58..529S,
       author = {{Salim}, Samir and {Narayanan}, Desika},
        title = "{The Dust Attenuation Law in Galaxies}",
      journal = {\araa},
         year = 2020,
        month = aug,
       volume = {58},
        pages = {529-575},
          doi = {10.1146/annurev-astro-032620-021933},
archivePrefix = {arXiv},
       eprint = {2001.03181},
 primaryClass = {astro-ph.GA},
       adsurl = {https://ui.adsabs.harvard.edu/abs/2020ARA&A..58..529S}
}

@ARTICLE{2009ApJ...698L.116P,
       author = {{Pannella}, M. and {Carilli}, C.~L. and {Daddi}, E. and {McCracken}, H.~J. and {Owen}, F.~N. and {Renzini}, A. and {Strazzullo}, V. and {Civano}, F. and {Koekemoer}, A.~M. and {Schinnerer}, E. and et al.},
        title = "{Star Formation and Dust Obscuration at z {\ensuremath{\approx}} 2: Galaxies at the Dawn of Downsizing}",
      journal = {\apjl},
         year = 2009,
        month = jun,
       volume = {698},
       number = {2},
        pages = {L116-L120},
          doi = {10.1088/0004-637X/698/2/L116},
archivePrefix = {arXiv},
       eprint = {0905.1674},
 primaryClass = {astro-ph.CO},
       adsurl = {https://ui.adsabs.harvard.edu/abs/2009ApJ...698L.116P}
}

@ARTICLE{2014MNRAS.439.1337O,
       author = {{Oteo}, I. and {Bongiovanni}, {\'A}. and {Magdis}, G. and {P{\'e}rez-Garc{\'\i}a}, A.~M. and {Cepa}, J. and {Dom{\'\i}nguez S{\'a}nchez}, H. and {Ederoclite}, A. and {S{\'a}nchez-Portal}, M. and {Pintos-Castro}, I.},
        title = "{The ultraviolet to far-infrared spectral energy distribution of star-forming galaxies in the redshift desert}",
      journal = {\mnras},
         year = 2014,
        month = apr,
       volume = {439},
       number = {2},
        pages = {1337-1363},
          doi = {10.1093/mnras/stt2468},
archivePrefix = {arXiv},
       eprint = {1307.0971},
 primaryClass = {astro-ph.CO},
       adsurl = {https://ui.adsabs.harvard.edu/abs/2014MNRAS.439.1337O}
}

@ARTICLE{2022ApJ...928...68S,
       author = {{Shivaei}, Irene and {Popping}, Gerg{\"o} and {Rieke}, George and {Reddy}, Naveen and {Pope}, Alexandra and {Kennicutt}, Robert and {Mobasher}, Bahram and {Coil}, Alison and {Fudamoto}, Yoshinobu and {Kriek}, Mariska and et al.},
        title = "{Infrared Spectral Energy Distributions and Dust Masses of Sub-solar Metallicity Galaxies at z   2.3}",
      journal = {\apj},
         year = 2022,
        month = mar,
       volume = {928},
       number = {1},
          eid = {68},
        pages = {68},
          doi = {10.3847/1538-4357/ac54a9},
archivePrefix = {arXiv},
       eprint = {2201.04270},
 primaryClass = {astro-ph.GA},
       adsurl = {https://ui.adsabs.harvard.edu/abs/2022ApJ...928...68S}
}

@ARTICLE{2025A&A...693A.190J,
       author = {{Jolly}, Jean-Baptiste and {Knudsen}, Kirsten and {Laporte}, Nicolas and {Guerrero}, Andrea and {Fujimoto}, Seiji and {Kohno}, Kotaro and {Kokorev}, Vasily and {Lagos}, Claudia del P. and {Schirmer}, Thi{\'e}baut-Antoine and {Bauer}, Franz and et al.},
        title = "{ALMA Lensing Cluster Survey: Dust mass measurements as a function of redshift, stellar mass, and star formation rate from z = 1 to z = 5}",
      journal = {\aap},
         year = 2025,
        month = jan,
       volume = {693},
          eid = {A190},
        pages = {A190},
          doi = {10.1051/0004-6361/202346239},
archivePrefix = {arXiv},
       eprint = {2411.11212},
 primaryClass = {astro-ph.GA},
       adsurl = {https://ui.adsabs.harvard.edu/abs/2025A&A...693A.190J}
}

@ARTICLE{2010MNRAS.409..421G,
       author = {{Garn}, Timothy and {Best}, Philip N.},
        title = "{Predicting dust extinction from the stellar mass of a galaxy}",
      journal = {\mnras},
         year = 2010,
        month = nov,
       volume = {409},
       number = {1},
        pages = {421-432},
          doi = {10.1111/j.1365-2966.2010.17321.x},
archivePrefix = {arXiv},
       eprint = {1007.1145},
 primaryClass = {astro-ph.GA},
       adsurl = {https://ui.adsabs.harvard.edu/abs/2010MNRAS.409..421G}
}

@ARTICLE{2019MNRAS.490.1425L,
       author = {{Li}, Qi and {Narayanan}, Desika and {Dav{\'e}}, Romeel},
        title = "{The dust-to-gas and dust-to-metal ratio in galaxies from z = 0 to 6}",
      journal = {\mnras},
         year = 2019,
        month = nov,
       volume = {490},
       number = {1},
        pages = {1425-1436},
          doi = {10.1093/mnras/stz2684},
archivePrefix = {arXiv},
       eprint = {1906.09277},
 primaryClass = {astro-ph.GA},
       adsurl = {https://ui.adsabs.harvard.edu/abs/2019MNRAS.490.1425L}
}

@ARTICLE{2021ApJ...914...19S,
       author = {{Sanders}, Ryan L. and {Shapley}, Alice E. and {Jones}, Tucker and {Reddy}, Naveen A. and {Kriek}, Mariska and {Siana}, Brian and {Coil}, Alison L. and {Mobasher}, Bahram and {Shivaei}, Irene and {Dav{\'e}}, Romeel and et al.},
        title = "{The MOSDEF Survey: The Evolution of the Mass-Metallicity Relation from z = 0 to z 3.3}",
      journal = {\apj},
         year = 2021,
        month = jun,
       volume = {914},
       number = {1},
          eid = {19},
        pages = {19},
          doi = {10.3847/1538-4357/abf4c1},
archivePrefix = {arXiv},
       eprint = {2009.07292},
 primaryClass = {astro-ph.GA},
       adsurl = {https://ui.adsabs.harvard.edu/abs/2021ApJ...914...19S}
}

@ARTICLE{2018ApJ...853..179T,
       author = {{Tacconi}, L.~J. and {Genzel}, R. and {Saintonge}, A. and {Combes}, F. and {Garc{\'\i}a-Burillo}, S. and {Neri}, R. and {Bolatto}, A. and {Contini}, T. and {F{\"o}rster Schreiber}, N.~M. and {Lilly}, S. and {Lutz}, D. and {Wuyts}, S. and {Accurso}, G. and {Boissier}, J. and {Boone}, F. and {Bouch{\'e}}, N. and {Bournaud}, F. and {Burkert}, A. and {Carollo}, M. and {Cooper}, M. and {Cox}, P. and {Feruglio}, C. and {Freundlich}, J. and {Herrera-Camus}, R. and {Juneau}, S. and {Lippa}, M. and {Naab}, T. and {Renzini}, A. and {Salome}, P. and {Sternberg}, A. and {Tadaki}, K. and {{\"U}bler}, H. and {Walter}, F. and {Weiner}, B. and {Weiss}, A.},
        title = "{PHIBSS: Unified Scaling Relations of Gas Depletion Time and Molecular Gas Fractions}",
      journal = {\apj},
         year = 2018,
        month = feb,
       volume = {853},
       number = {2},
          eid = {179},
        pages = {179},
          doi = {10.3847/1538-4357/aaa4b4},
archivePrefix = {arXiv},
       eprint = {1702.01140},
 primaryClass = {astro-ph.GA},
       adsurl = {https://ui.adsabs.harvard.edu/abs/2018ApJ...853..179T}
}

@ARTICLE{2025arXiv251006681C,
       author = {{Clarke}, Leonardo and {Shapley}, Alice E. and {Lam}, Natalie and {Topping}, Michael W. and {Brammer}, Gabriel B. and {Sanders}, Ryan L. and {Reddy}, Naveen A. and {Karthikeyan}, Shreya},
        title = "{The Star-forming Main Sequence and Bursty Star-formation Histories at $z>1.4$ in JADES and AURORA}",
      journal = {arXiv e-prints},
         year = 2025,
        month = oct,
          eid = {arXiv:2510.06681},
        pages = {arXiv:2510.06681},
          doi = {10.48550/arXiv.2510.06681},
archivePrefix = {arXiv},
       eprint = {2510.06681},
 primaryClass = {astro-ph.GA},
       adsurl = {https://ui.adsabs.harvard.edu/abs/2025arXiv251006681C}
}

@ARTICLE{2023MNRAS.518..425C,
       author = {{Curti}, Mirko and {D'Eugenio}, Francesco and {Carniani}, Stefano and {Maiolino}, Roberto and {Sandles}, Lester and {Witstok}, Joris and {Baker}, William M. and {Bennett}, Jake S. and {Piotrowska}, Joanna M. and {Tacchella}, Sandro and {Charlot}, Stephane and {Nakajima}, Kimihiko and {Maheson}, Gabriel and {Mannucci}, Filippo and {Amiri}, Amirnezam and {Arribas}, Santiago and {Belfiore}, Francesco and {Bonaventura}, Nina R. and {Bunker}, Andrew J. and {Chevallard}, Jacopo and {Cresci}, Giovanni and {Curtis-Lake}, Emma and {Hayden-Pawson}, Connor and {Jones}, Gareth C. and {Kumari}, Nimisha and {Laseter}, Isaac and {Looser}, Tobias J. and {Marconi}, Alessandro and {Maseda}, Michael V. and {Scholtz}, Jan and {Smit}, Renske and {{\"U}bler}, Hannah and {Wallace}, Imaan E.~B.},
        title = "{The chemical enrichment in the early Universe as probed by JWST via direct metallicity measurements at z {\ensuremath{\sim}} 8}",
      journal = {\mnras},
         year = 2023,
        month = jan,
       volume = {518},
       number = {1},
        pages = {425-438},
          doi = {10.1093/mnras/stac2737},
archivePrefix = {arXiv},
       eprint = {2207.12375},
 primaryClass = {astro-ph.GA},
       adsurl = {https://ui.adsabs.harvard.edu/abs/2023MNRAS.518..425C}
}

@ARTICLE{2015MNRAS.447.1610M,
       author = {{Maddox}, Natasha and {Hess}, Kelley M. and {Obreschkow}, Danail and {Jarvis}, M.~J. and {Blyth}, S.-L.},
        title = "{Variation of galactic cold gas reservoirs with stellar mass}",
      journal = {\mnras},
         year = 2015,
        month = feb,
       volume = {447},
       number = {2},
        pages = {1610-1617},
          doi = {10.1093/mnras/stu2532},
archivePrefix = {arXiv},
       eprint = {1412.0852},
 primaryClass = {astro-ph.GA},
       adsurl = {https://ui.adsabs.harvard.edu/abs/2015MNRAS.447.1610M}
}
\bibliographystyle{aasjournal}

\end{document}